\documentclass[a4paper,twoside,openright,12pt]{report}
\usepackage{amsfonts}
\usepackage{amsmath}
\usepackage{amssymb}
\usepackage{latexsym}
\usepackage{color}
\usepackage{colordvi}
\usepackage{epsfig}
\usepackage{array}
\usepackage{pifont}

\newcommand{\beq}{\begin{eqnarray}}
\newcommand{\eeq}{\end{eqnarray}}

\newcommand{\be}{\begin{equation}}
\newcommand{\ee}{\end{equation}}
\newcommand{\lwrsim}{\raise0.3ex\hbox{$<$\kern-0.75em\raise-1.1ex\hbox{$\sim$}}}

\def\Am#1#2#3{\widetilde A_{#1}^{#2}(#3)}

\def\C2#1#2{({\cal C}_2)_{#1}^{#2}}

\RequirePackage[a4paper,tmargin=2.75truecm,bmargin=3truecm,rmargin=2.75truecm,lmargin=2.75truecm]{geometry}%
\usepackage[T1]{fontenc}   
\usepackage{mathpazo}
\usepackage{these} 
\usepackage{setspace}
\usepackage{epsfig}
\usepackage{axodraw}
\usepackage{arydshln}
\usepackage{latexsym}
\usepackage{amsmath}
\usepackage{amssymb}
\usepackage{amsfonts}
\usepackage[english]{babel}
\newcommand{\emu}{e_{\mu}}
\newcommand{\id}{\mathbb{I}}
\newcommand{\ice}[1]{\relax}
\def\d{\partial}
\def\tr{\text{Tr}}
\def\re{\mathfrak{Re}}
\def\MFP{\mathcal{M}_{FP}}
\def\MFPlat{\mathcal{M}^{\text{lat}}_{FP}}

\newcommand{\Lqcd}{\Lambda_{\text{QCD}}}
\newcommand{\ms}{\overline{\text{MS}}}
\newcommand{\mom}{\widetilde{\text{MOM}_{c0}}}
\newcommand{\momg}{\widetilde{\text{MOM}_g}}
\newcommand{\momc}{\widetilde{\text{MOM}_c}}

\def\bea{\begin{eqnarray} }

\def\eea{\end{eqnarray}}
\titredupoly{Non-perturbative Study of QCD Correlators}
\auteurdupoly{Alexey Lokhov}
\adressepostale{Centre de Physi\textbf{}que Th\'eorique, \\ Ecole Polytechnique \\ 91128 PALAISEAU Cedex \\ FRANCE}
\adresseelectronique{lokhov@cpht.polytechnique.fr} 
\begin{document}
%
\thispagestyle{empty}
%
%
%
%
\begin{center}
%
\vspace*{-2cm}
\begin{normalsize}
{\large C}ENTRE DE PHYSIQUE TH\'EORIQUE - {\large \'E}COLE POLYTECHNIQUE \\
\end{normalsize}
\vspace*{0.5cm}
\begin{large}
{\bf TH\`ESE DE DOCTORAT DE L'\'ECOLE POLYTECHNIOUE} \\
\end{large}
\vspace*{0.5cm}
\begin{large}
Sp\'ecialit\'e: \text{\normalsize\bf PHYSIQUE TH\'EORIQUE} \\
\end{large}
\vspace*{2cm}
\begin{normalsize}
pr\'esent\'ee par\\
\end{normalsize}
\vspace*{0.3cm}
\begin{Large}
{\bf Alexey Lokhov}\\
\end{Large}
\vspace*{0.5cm}
\begin{normalsize}
pour obtenir le grade de\\
\end{normalsize}
\vspace*{1cm}
\begin{Large}
Docteur de l'\'Ecole polytechnique\\
\end{Large}
\vspace*{2cm}
\begin{normalsize}
Sujet:\\
\end{normalsize}
\vspace*{1cm}
\begin{LARGE}
\textbf{\emph{\'Etude non-perturbative des corr\'elateurs en \\ chromodynamique quantique} }\\
\end{LARGE}
\vspace*{3.7cm}
Soutenue le 8 juin 2006 devant le jury compos\'e de:\\
\vspace*{1cm}
\begin{tabular}{lll}
MM. & Ulrich Ellwanger, & rapporteur \\
~ & Georges Grunberg, & \\ 
~ & Olivier P\`ene, & \\
~ & Silvano Petrarca, & rapporteur \\
~ & Claude Roiesnel, &  directeur de th\`ese\\
~ & Andr\'e Roug\'e, & \\
\& & Jean - Bernard Zuber, & pr\'esident du jury\\
\end{tabular}
\begin{flushright}
\vspace*{0.3cm}
{\small CPHT-T 058.0706}\\
\end{flushright}

%
\end{center}

\pagestyle{myheadings}
\tableofcontents
\addcontentsline{toc}{chapter}{Notations and conventions}
\chapter*{Notations and conventions}%
%
\vspace{-1cm}
\setstretch{1.5}
\begin{small}
%
%
\begin{center}
\begin{tabular}{ll}
$a$ & lattice spacing \\
$\mathcal{A}_\mu = A^a_\mu t^a$ & gauge field
\\
$\langle A^2 \rangle$ & $A^2$-condensate
\\
$\alpha_F$ ($\alpha_G$) & infrared exponent of the ghost (gluon) propagator
\\
$\beta=\frac{2 N_c}{g_0^2}$ & bare lattice coupling
\\
$\beta(g),\beta_i$ & the renormalisation group beta function and 
\\
& its first coefficients
\\
$\gamma(g); \bar{\gamma},\bar{\gamma}_i$ & anomalous dimension; anomalous dimension 
\\
                                         & in a generic MOM  scheme and its coefficients
\\
$\emu$ & a vector in direction $\mu$ of norm $a$
\\
$\tilde{F}^{(2)ab}(p) = \delta^{ab} \frac{ \tilde{F}(p)}{p^2}$ & ghost propagator 
\\
$\tilde{G}^{(2)ab}(p) = \delta^{ab} \frac{ \tilde{G}(p)}{p^2}\left(\delta_{\mu\nu} - \frac{p_\mu p_\nu}{p^2}\right)$ 
& gluon propagator
\\
$G,\Lambda$ & Gribov region and the fundamental modular region
\\
$\Gamma_\mu^{abc}$ & three-gluon vertex
\\
$\widetilde{\Gamma}_\mu^{abc}(p,q;r)=-ig_0 f^{abc} p_\nu \widetilde{\Gamma}_{\nu\mu}(p,q;r)$ & ghost-gluon vertex, $p$ is the momentum of the 
\\
& outgoing ghost, $r$ is the gluon momentum.
\\
$g_0, g_R$ & bare coupling, renormalised coupling
\\
$h=g^2/(4\pi)^2$
\\
$L$ & size of the lattice
\\
$\MFP$, $\MFPlat$ & Faddeev - Popov operator and its discretized version
\\
$N_c$ & number of colours
\\
$p_\mu = \frac{2\pi}{L}a^{-1} n_\mu$ & relation between physical and lattice momenta
\\
$t=\ln{\frac{\mu^2}{\Lambda^2_{\text{QCD}}}}$
\\
$U(x,x+\emu)\equiv U_{\mu}(x) = $ & link variable
\\
$=e^{i g_0 a \mathcal{A}_\mu(x+\frac{a}{2}\emu)}\in SU(N_c)$ &
\end{tabular}
\end{center}
\newpage
\begin{center}
\hspace*{-2.7cm}
\begin{tabular}{ll}
$V$ $\qquad\qquad\qquad$ & volume of the lattice
\\
$Z_3$ ($\widetilde{Z_3}$) & gluon (ghost) field renormalisation factor
\\
$\left\langle 0\right. \arrowvert \bullet \arrowvert 0 \rangle$ & 
average with respect to the perturbative vacuum
\\
$\langle \bullet \rangle$ & average with respect to the non-perturbative vacuum
	\end{tabular}
\end{center}
%
%
%
%
\paragraph*{Shortenings}
\begin{center}
	\begin{tabular}{ll}
b.c $\qquad\qquad\qquad$  & best (Gribov) copy choice 
\\
f.c. & first (Gribov) copy choice
\\
ERGE & exact renormalisation group equations
\\
IR & infrared
\\
MOM & momentum substraction renormalisation scheme (see Figures~\ref{MOMasym_ET_MOMsym_GRAPH},~\ref{MOMasym_ET_MOMsym_GRAPH_ghost})
\\
RG & renormalisation group
\\
SD & Schwinger-Dyson equations
\\
ST & Slavnov-Taylor identities
\\
UV & ultraviolet
\\
v.e.v. & vacuum expectation values
\\
ZP & zero-point (kinematic configuration, $\widetilde{\rm MOM}$ renormalisation schemes)
	\end{tabular}
\end{center}
\end{small}
\singlespacing
\addcontentsline{toc}{chapter}{General introduction}
%
%
\chapter*{General introduction}
%
%
\begin{flushright}
\textit{``Of course, you can put a theory on the lattice. But then - it is a mess!"
\\
\textbf{\footnotesize Giorgio Parisi, les Houches}
}
\end{flushright}
\vspace*{1cm}
\noindent
This PhD dissertation is devoted to a non-perturbative study of QCD correlators in Landau gauge. The main
tool that we use is lattice QCD. It allows a numerical evaluation of the functional integrals
defining vacuum expectation averages of the theory i.e. Green functions.
The advantage of this method is that it gives access to the non-perturbative domain and 
exactly preserves the gauge-invariance allowing to (numerically) study QCD 
from its first principles. However, the price to pay is the appearance of diverse discretisation
artifacts like breakdown of the Lorentz invariance, a necessity to work with the Euclidean
formulation of the theory and, in practice, at finite volume.
We discuss in details the methods allowing to handle most of the artifacts. Lattice QCD has been 
successfully used in phenomenology (mass of the charm quark, $B$- and $D$-mesons physics, 
generalised parton distributions, QCD at finite temperature and its phase diagram, etc.).
But it can also be used to study the fundamental parameters (like coupling constant) 
and properties of the theory itself. This is the main goal of the present dissertation.
We concentrated our effort on the study of the two-point correlators of the pure Yang - Mills theory 
in Landau gauge, namely the gluon and the ghost propagators. 
We are particularly interested in determining the $\Lqcd$ parameter -
the fundamental scale of the pure Yang-Mills theory. It is extracted by means of 
perturbative predictions available  up to NNNLO. The related topic is the influence of
non-perturbative effects that shows up as appearance of power-corrections to the 
low-momentum behaviour of the Green functions. We shall see that these corrections are quite 
important up to energies of the order of $10$ GeV.
\vspace*{0.5cm}
\noindent

Another question that we address is the infrared behaviour of the Green functions (in Landau gauge), at momenta of order 
and below $\Lqcd$. At low energy the power-law dependence of some Green functions changes considerably,
and this is probably related to confinement. The knowledge of the infrared behaviour of the ghost and gluon 
propagators in Landau gauge is very important, because many confinement scenarii (for example the Gribov-Zwanziger scenario)
give predictions for their momentum dependence at very low energies. \emph{Ab initio} simulations on the lattice is a 
quasi-unique method for testing these predictions and the only way to challenge the underlying models for confinement.

We try to clarify the laws that govern the infrared gluondynamics in order to understand the radical nature of the changes 
of the infrared behaviour of some Green functions. Many questions arise: the Gribov ambiguity, the validity of different non-perturbative relations 
(like Schwinger - Dyson equations, Slavnov - Taylor identities and renormalisation group equations)
at low momenta, self-consistency of the lattice approach in this domain. The lattice approach allows to check the predictions 
of analytical methods because it gives access to non-perturbative correlators. Our main goal thus is to use lattice Green functions
as a non-perturbative input for different analytical relations. This allows to control the approximations that are done within
the traditional truncation methods for the non-perturbative relations. Such a mixed numerical-analytical analysis of the 
complete Schwinger - Dyson equation for the ghost propagator provided us with an interesting alternative to the widely spread claim that the 
gluon dressing function behaves like the inverse squared ghost dressing function, a claim which is at odds with lattice data.
According to our analysis the Landau gauge gluon propagator is finite and non-zero at vanishing momentum, and the power-law behaviour 
of the ghost propagator is the same as in the free case. However, as we shall see, some puzzles remain unsolved.

%
%
%
\chapter{Continuum and lattice QCD}
%
%
%
\noindent In this chapter we recall very briefly the most fundamental ideas of QCD (definitions, symmetries, 
covariant gauge fixing, Gribov ambiguity).  As we are interested in a non-perturbative calculation
of different correlators, we also introduce diverse non-perturbative relations between Green functions,
namely Schwinger-Dyson equations, Slavnov-Taylor identities and exact renormalisation group equations.
After this we shall discuss the lattice formulation of the pure Yang-Mills theories, in particular the procedure of the
Landau gauge fixing on the lattice.

%
%
\section{General features of QCD}
%
%

%
\subsection{Definitions and symmetries}
%

\noindent Nowadays there is no doubt that Quantum Chromodynamics (QCD) is the theory of the strong interaction. 
The fundamental principle of this theory is local gauge invariance. This principle, together with 
general principles of locality, Lorentz invariance and  renormalisability,  imposes important constraints 
on the form of the Lagrangian. The simplest form in Euclidean four-dimensional space reads
\begin{equation}
\label{Lqcd}
\mathcal{L}_{\text{QCD}}=-\frac{1}{4}F^{a}_{\mu\nu}F^{a\mu\nu} + 
\sum_{\psi=u,d,s,c,b,t}\bar{\psi}(i D_\mu\gamma^\mu - m_\psi)\psi
\end{equation}
with ($g_0$ is the bare coupling)
\begin{align}
& F^a_{\mu\nu}=\partial_\mu A^a_{\nu}-\partial_\nu A^a_{\mu} + g_0 f^{abc}A^b_{\mu}A^c_{\nu}, \qquad a=1..N_C^2-1
\\ &
D_\mu = \d_\mu - i g_0 t^a A^a_\mu.
\end{align}
This Lagrangian is invariant under gauge transformations of the fields
\begin{align}
\label{GaugeTransformations}
&\mathcal{A}_\mu(x) \mapsto \mathcal{A}^{(u)}_\mu(x)=u(x)\mathcal{A}_\mu(x) u^{\dagger}(x)+i[\partial_\mu u(x)]u^{\dagger}(x)
\\ &
\psi(x) \mapsto \psi^{(u)}(x) = u(x) \psi(x),
\end{align}
where $u(x)\in SU(N_c)$ and $N_C=3$ is the number of colours.

In order to quantise QCD using the functional integration formalism one has to integrate
over the quark and the gauge bosons fields. The Grassmannian integral on the quark fields is Gaussian, that
is why we discuss only the integration on the gauge boson fields $\mathcal{A}$.
The fields $\mathcal{A}^{(u)}$ and $\mathcal{A}$ in the equation (\ref{GaugeTransformations})
are related by a gauge transformation, and thus they are physically equivalent. So, in order to
quantise a gauge theory one performs an integration over gauge transformation
equivalence classes - the \textbf{orbits} of the gauge fields. This is the
Faddeev - Popov procedure. The integration on the orbits is done by choosing a representative 
element on every orbit, i.e. fixing the gauge with some relation
\begin{equation}
\label{ContinumGaugeFixingCondition}
f[\mathcal{A}] = 0.
\end{equation}
The condition $f$ should define the orbit of the field $\mathcal{A}$ in a unique way. Then the generating functional for
Green functions reads
\begin{equation}
Z[j,\overline{\omega},\omega]=\int [\mathcal{DA} \mathcal{D} \psi\mathcal{D} \overline{\psi}]\,
\Delta_{f}[\mathcal{A}]  \, \delta \left( f\left[ \mathcal{A} \right] \right) \,
e^{- \int d^4 x \mathcal{L}_{\text{QCD}} + \int d^4 x \left( A^a_\mu j^a_\mu + \overline{\omega}\psi + \overline{\psi} \omega \right)},
\end{equation}
where all loop integrals are understood to be regularised. We denote the ultraviolet 
cut-off $a^{-1}$,  $g_0 \equiv g(a^{-1})$. The \textbf{Faddeev-Popov determinant} $\Delta_{f}[\mathcal{A}]$ 
which appears in this formula is defined by means of invariant integration
\begin{equation}
\Delta_{f}[\mathcal{A}] \int \mathcal{D}u(x) \prod_x \delta\left( f\left[ \mathcal{A}^{(u)}(x) \right] \right)  = 1
\end{equation}
yielding in the general case~\footnote{when the condition (\ref{ContinumGaugeFixingCondition}) does not fix the gauge in a unique way.}
\begin{equation}
\label{Gen_def_FP}
\Delta_{f}^{-1}[\mathcal{A}] = \sum_{i:\,f\left[\mathcal{A}^{(g_i)}\right] = 0} \det{}^{-1}\frac{\delta f\left[\mathcal{A}^{(g_i)}\right]}{\delta g}.
\end{equation}
Choosing the \textbf{Landau gauge} condition
\begin{equation}
\label{LandauGauge}
f[\mathcal{A}]:\quad \d_\mu \mathcal{A}_\mu = 0
\end{equation}
and supposing for the moment that it fixes the gauge in a unique way, one obtains 
\begin{equation}
\label{Delta_FP_Landau}
\Delta_{\text{Landau}}[\mathcal{A}] =
\det{\left(\Delta + i g_0 \d_\mu \mathcal{A}_\mu \right) }
= \int [\mathcal{D}c \mathcal{D}\bar{c}] \, e^{-\int d^4x d^4y \bar{c}^a(x) \mathcal{M}^{ab}_{FP}(x,y) c^b(y) }.
\end{equation}
The spurious anticommuting fields $c$ and $\bar{c}$ belonging to the adjoint representation of 
the gauge group are called \textbf{Faddeev-Popov ghosts}, and 
\begin{equation}
\label{FP_Operator}
\mathcal{M}^{ab}_{FP}(x,y) = \left(\Delta + i g_0 \d_\mu \mathcal{A}_\mu \right)^{ab} \delta^{(4)}(x-y)
\end{equation}
is the \textbf{Faddeev-Popov operator}. The corresponding formula for the generating 
functional can be easily generalised by choosing for the gauge condition
\begin{equation}
f[\mathcal{A}]:\quad \d_\mu \mathcal{A}_\mu = \mathfrak{a}(x), \qquad\mathfrak{a}(x)\in\mathfrak{su}(N_C).
\end{equation}
In this case $\Delta_{f}$ remains the same as (\ref{Delta_FP_Landau}), and one can integrate 
on $\mathfrak{a}(x)$ with some Gaussian weight having a dispersion $\xi$. This gives for the generating functional
\begin{align}
\label{Z_continuum}
Z[j, & \bar{\omega},\omega, \bar{\sigma},\sigma] =
\int [\mathcal{DA} \mathcal{D} \psi\mathcal{D} \bar{\psi} \mathcal{D}c \mathcal{D}\bar{c}] \,
e^{- \int d^4 x \mathcal{L}_{eff}[\mathcal{A},\psi,\bar{\psi},c,\bar{c}]  + \Sigma},
\\ &
\mathcal{L}_{eff}[\mathcal{A},\psi,\bar{\psi},c,\bar{c}]  = 
\mathcal{L}_{\text{QCD}} - \frac{(\d_\mu \mathcal{A}_\mu)^2}{2\xi} 
-\bar{c}^a(x) ( \delta^{ab} \Delta + i g_0 f^{abc} A^c_\mu \d_\mu ) c^b(x)
\\ &
\Sigma = \int d^4 x \left( A_\mu j_\mu + \bar{\omega}\psi + \bar{\psi} \omega  + \bar{\sigma}c + \bar{c}\sigma \right).
\end{align}
The choice $\xi=0$ corresponds to the Landau gauge. The gauge fixing term 
in (\ref{Z_continuum}) can be expressed as a result of Gaussian integration on an auxiliary
field $B^a(x)$. This gives another form of the Lagrangian $\mathcal{L}_{eff}$:
\begin{equation}
\label{Lagrangian_BRST}
\mathcal{L}_{\text{BRST}} = \mathcal{L}_{QCD} -\frac{\xi}{2}(B^a)^2 + B^a \d_\mu A^a_\mu 
+ \bar{c}^a ( \delta^{ab} \Delta - i g_0 f^{abc} \d_\mu A^c_\mu  ) c^b.
\end{equation}
The QCD Lagrangian written in this form is invariant under \textbf{BRST transformations}. If $\lambda$ is a constant
infinitesimal Grassmann number these transformations take the form
\begin{align}
\label{BRST}
& \delta A^a_\mu = \lambda D^{ac}_\mu c^c
\nonumber
\\ &
\delta \psi = i g_0 \lambda t^a \psi
\nonumber
\\ &
\delta c^a = - \frac{1}{2} g_0 \lambda f^{abc} c^b c^c
\nonumber
\\ &
\delta \bar{c}^a = \lambda B^a
\nonumber
\\ &
\delta B^a = 0.
\end{align}
The virtue of the BRST transformation is its global nature. This simplifies a lot 
the derivation of the Slavnov-Taylor identities (direct consequence of the gauge invariance). We discuss 
this question below.

%
\subsection{The Gribov ambiguity}
%
\label{subsection_Gribov_ambiguity_general}

A serious theoretical difficulty pointed out by Gribov~\cite{Gribov:1977wm} arises when performing the 
quantisation of a non-Abelian gauge theory (in covariant gauge) in the case of  large field magnitudes. 
The reason for this is the non-uniqueness of the Landau gauge condition (\ref{LandauGauge}).
Indeed, let us find all the intersections of the gauge orbit with the 
hypersurface defined by (\ref{LandauGauge}). Imposing the Landau gauge conditions for both fields $\mathcal{A}_\mu$ 
and $\mathcal{A}^{(\overline{u})}_\mu$ in (\ref{GaugeTransformations}) we obtain the following equation for $\overline{u}(x)$:
\begin{equation}
\label{Eq_Gribov_EXACT}
\left\{
	\begin{array}{l}
	  \mathcal{A}_\mu(x) \mapsto \mathcal{A}^{(\overline{u})}_\mu(x) \\
	  \d_\mu \mathcal{A}_\mu =0\\
	  \d_\mu \mathcal{A}^{(\overline{u})}_\mu = 0
	\end{array}
\right. 
\quad \rightarrow \quad
\d_\mu \left(  \overline{u}(x) D_\mu(x) \overline{u}^\dagger(x) \right) = 0.
\end{equation}
Setting at the leading order
\begin{equation}
\overline{u} \simeq \id + i\overline{\alpha}(x),
\quad 
\overline{u}^\dagger \simeq \id - i\overline{\alpha}(x)
\qquad  
\overline{\alpha}(x)\in\mathfrak{su}(N_c)
\end{equation}
we obtain the following equation for $\overline{\alpha}$:
\begin{equation}
\label{Eq_Gribov}
\d_\mu
\left( 
	\d_\mu \overline{\alpha} + i [\mathcal{A}_\mu,\overline{\alpha} ]
\right) =0
\quad
\longrightarrow
\quad
\d_\mu D_\mu \overline{\alpha} = 0.
\end{equation}
But $\d_\mu D_\mu$ is nothing else but the Faddeev-Popov operator in the covariant gauge. Thus, any non-trivial
zero mode of the Faddeev-Popov operator generates an intersection point of the gauge orbit with the 
hypersurface (\ref{LandauGauge}). If this point is not unique, we speak about the so called \textbf{Gribov copy}.
All these secondary gauge configurations correspond to the same physical field $\mathcal{A}_\mu$, and thus
they must be removed from the functional integration measure in the partition function.
%
\begin{figure}[!h]
\begin{center}
\includegraphics[width=0.4\linewidth]{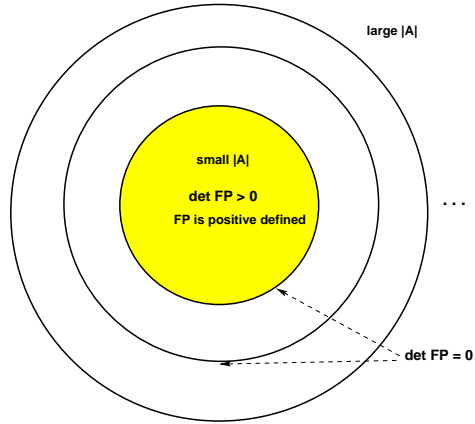}
\end{center}
\caption{\footnotesize \it The Gribov region.}
\label{GribovRegion}
\end{figure}
%

A solution to this problem is to supplement the initial gauge-fixing condition (\ref{LandauGauge}) with some 
additional requirement. Gribov's solution consists in restricting the integration measure (for the gluonic field) in (\ref{Z_continuum}) 
to the domain where (\ref{Eq_Gribov}) has a unique solution, see Fig. \ref{GribovRegion}.
This domain is called the \textbf{Gribov region}, and its boundary (where the Faddeev-Popov determinant vanishes) 
is called the \textbf{Gribov horizon}. It has been argued that some of topological solutions
like instantons lie on this boundary~\cite{Maas:2005qt}.

Inside the Gribov region all the eigenvalues of the Faddeev-Popov operator are strictly positive~\footnote{we recall that in the Euclidean formulation the Faddeev - Popov operator is hermitian.}.
Hence one can realise the Gribov quantisation by using the Minimal Landau gauge. In this gauge one integrates on 
the fields satisfying the ordinary Landau gauge and belonging to the set of local minima of the integral 
$$
\int d^4 x \left( A^a_\mu(x) \right)^2.
$$
As we shall see, this ensures that all the proper values of the Faddeev-Popov operator are positive. The Minimal Landau gauge
will be discussed in details in the subsection~\ref{subsection_lattice_gauge_fixing}.

Nowadays it is known that the Gribov quantisation prescription is not exact - there are secondary solutions to (\ref{Eq_Gribov_EXACT})
for some fileds inside the Gribov region. The domain free of them is located inside the Gribov region, and it is called the 
\textbf{fundamental modular region}~\cite{Franke_Semenov_Tyan_Shanskii},\cite{vanBaal:1991zw}.
%
\begin{figure}[h]
\begin{center}
\includegraphics[width=0.4\linewidth]{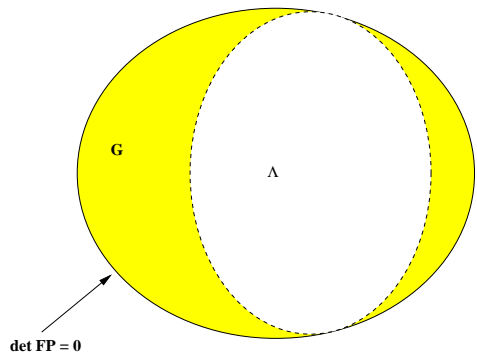}
\end{center}
\caption{\footnotesize \it Gribov region and the fundamental modular region.}
\label{GribovRegion_et_fundamental_modilar_region}
\end{figure}
%
This means that the correct quantisation prescription consists in restricting the 
integration in (\ref{Z_continuum}) to the fundamental modular region $\Lambda$ instead of
the Gribov region $G$, see figure \ref{GribovRegion_et_fundamental_modilar_region}.
However it is argued in~\cite{Zwanziger:2003cf} that the  expectation values calculated by 
integration over the Gribov region or the fundamental modular region are equal.
So the Gribov quantisation prescription would become in fact exact. We discuss this 
question in details in the section \ref{section_Gribov_copies_sur_reseau}.

%
\subsection{Schwinger-Dyson equations}
%

The Schwinger-Dyson(SD) equations is a specific class of non-perturbative equations relating Green functions and vertices. 
They can be easily derived in the functional integration formalism (for a review see \cite{Alkofer:2000wg}).
Let $Z[J]$ denote a normalised ($Z[0]=1$) generating functional (\ref{Z_continuum}) 
for the Green functions, and $W[J] = \log{Z[J]}$ - the one for connected Green functions.
Then the effective action, which is a generating functional for one-particle irreducible (1-PI) vertex functions, 
is obtained by the Legendre transformation
\begin{equation}
\label{Legendre}
\Gamma[\phi^c] = \frac{\delta W[J]}{\delta j_i}  j_i -W[J] ,
\qquad \phi^c_i  = \frac{\delta W[J]}{\delta j_i},
\end{equation}
where  $\phi_i$ denotes the generic field ($A^a_\mu, c^a,\ldots$) and $j_i$ is the corresponding source.
Then, introducing the action $S = \int d^4 x \, \mathcal{L}_{eff}$ and using the quantisation prescription based on
the integration over the fundamental modular region, the Schwinger-Dyson equations are obtained from the observation that 
\begin{equation}
\label{SD_begin}
\int\limits_\Lambda \big[\prod_j \mathcal{D}\phi_j\big]  \frac{\delta}{\delta \phi_i}\left(e^{-S + \phi \cdot j} \right) 
\equiv \int\limits_\Lambda \big[\prod_j \mathcal{D}\phi_j \big] e^{-S + \phi \cdot j} \left( - \frac{\delta S}{\delta \phi_i} + j_i\right)
= \int\limits_{\partial\Lambda_i}\big[\prod_j \mathcal{D}\phi_j\big] \, e^{-S + \phi \cdot j}
\end{equation}
Using the Zwanziger's argument~\cite{Zwanziger:2003cf} (quoted above) on the equivalence of integrations over
$\Lambda$ and $G$ we restrict the integration domain to the Gribov region. It follows from this that
the integral on the boundary vanishes because the Faddeev-Popov determinant 
present in the integration measure is equal to zero on $\partial G$ (by definition).
This allows to write the whole set of the Schwinger-Dyson equations for the Green functions in a compact form
\begin{align}
\label{SD_Tower}
\left( -\frac{\delta S[\phi]}{\delta\phi_i} \left[ \frac{\delta}{\delta j}\right]  + j_i \right)Z[J] = 0.
\end{align}
We use a generic relation between two smooth functions $f(x)$ and $w(x)$
\begin{equation}
\label{ralation_exp_derivee}
f\left(\frac{d}{dx}\right)e^{w(x)} = e^{w(x)}f\left(\frac{dw(x)}{dx} + \frac{d}{dx}\right)1,
\end{equation}
that can be applied to (\ref{SD_Tower}) and yield the equations for the functional $W$ generating
the connected Green functions
\begin{align}
 -\frac{\delta S[\phi]}{\delta\phi_i} \left[ \frac{\delta}{\delta j} + \frac{\delta W}{\delta j}\right]1 + j_i = 0.
\end{align}
Finally, performing a Legendre transformation (\ref{Legendre}), we have 
\begin{align}
- \frac{\delta S[\phi]}{\delta\phi_i} 
\left[ 
	\phi + \frac{\delta^2 W}{\delta j \delta j} \frac{\delta}{\delta \phi}
\right]1+ \frac{\delta\Gamma[\phi]}{\delta\phi_i} 
= 0,
\end{align}
corresponding to the full set of Schwinger-Dyson equations for proper functions.

As an example, we derive explicitly the SD equation for the full ghost propagator. Varying $S$ with respect
to the antighost field $\bar{c}^a$ we obtain
\begin{equation}
\left\langle 
-\frac{\delta S}{\delta \bar{c}^a(x)}  + \sigma^a(x)
\right\rangle_{[j,\bar{\sigma},\sigma]} = 0.
\end{equation}
Varying the last relation with respect to $\sigma^b(y)$ one obtains 
\begin{equation}
\left\langle \frac{\delta S}{\delta \bar{c}^a(x)} \bar{c}^b(y)\right\rangle 
= \delta^{ab} \delta^{(4)}(x-y) 
= 
\left\langle 
	\left( 
		\d_\mu D^{ac}_\mu c^c(x)
	\right)\bar{c}^b(y)
\right\rangle.
\end{equation}
Denoting the full ghost propagator as
\begin{equation}
F^{(2)ab}(x,y) = \left\langle c^a(x)  \bar{c}^b(y)\right\rangle,
\end{equation}
we obtain ( $\Delta(x,y)\equiv \delta^{(4)}(x-y) \delta$ is the Laplace operator)
\begin{equation}
\label{Ghost_SD_Green}
\delta^{(4)}(x-y) = \Delta(x,z) F^{(2)}(z,y) + 
i g_0 \d^{(x)}_\mu \left\langle \mathcal{A}_\mu(x)  c(x)  \bar{c}(y)\right\rangle.
\end{equation}
Note that this equation can be obtained in a simpler way only using the definition of the 
Faddeev-Popov operator. Indeed, the ghost correlator in the background of 
a given gauge field configuration $\mathcal{A}_\mu = A^a_\mu t^a$ is given by 
\begin{eqnarray}
\label{SD_simple_begin}
{F_{\text{1conf}}^{(2)}}\left(\mathcal{A},x,y\right) \equiv \mathcal{M}^{\text{1conf}}_{FP}(x,y)^{-1}.
\end{eqnarray}
The subscript means here that the equation is valid for a given gauge configuration. Thus one obviously has
\begin{eqnarray}
\delta^{4}(x-y) \equiv  \mathcal{M}^{\text{1conf}}_{FP}(x,z) F_{\text{1conf}}^{(2)}(\mathcal{A},z,y),
\end{eqnarray}
where a summation on $z$ is understood. Using the explicit formula (\ref{FP_Operator}) for $\MFP$ we get
\begin{eqnarray}
\label{SDghost_conf_par_conf}
\delta(x-y) = \Delta(x,z)F_{\text{1conf}}^{(2)} (z,y) +  i g_0\,\partial^{(x)}_\mu \left( \mathcal{A}_\mu(x)   F_{\text{1conf}}^{(2)}(\mathcal{A},x,y)\right),
\end{eqnarray}
valid for \emph{any} gauge field configuration $\mathcal{A}$. Performing the functional integration on $\mathcal{A}$ one
gets the mean value on gauge configurations 
\begin{eqnarray}
\delta(x-y) = \Delta(x,z)\left \langle F_{\text{1conf}}^{(2)} (z,y)\right\rangle +  i g_0 
\partial^{(x)}_\mu  \left\langle\mathcal{A}_\mu(x) F_{\text{1conf}}^{(2)}(\mathcal{A},x,y)\right\rangle.
\end{eqnarray}
Using $\left\langle F_{\text{1conf}}^{(2)} (z,y) \right\rangle \equiv F^{(2)}(z,y)$ we find the
equation (\ref{Ghost_SD_Green}). The translational invariance of the Green functions allows to replace
$\partial^{(x)}_\mu$ by $-\partial^{(y)}_\mu$. Performing the Fourier transform on $(x-y)$, we have finally
\begin{eqnarray}
\label{SD_simple_end}
1 = -p^2 {F}^{(2)}(p^2) -  i g_0 p_\mu
\left\langle\mathcal{A}_\mu(0) {F}_{\text{1conf}}^{(2)}(\mathcal{A},p)\right\rangle.
\end{eqnarray}

\noindent This derivation elucidates the trivial dependence of this equation on the functional integral weight
with which we calculate the average $\langle \bullet \rangle$ on the gauge fields $\mathcal{A}$.
The form (\ref{SDghost_conf_par_conf}) allows an explicit discussion of the Gribov copies 
dependence of the solutions. We address this question below.

\begin{figure}[!h]
\vspace{.2cm}
\begin{center}
	\includegraphics[width=.8\linewidth]{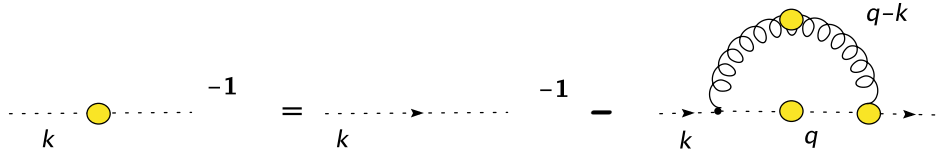}
\end{center}
\caption{\footnotesize\it Schwinger-Dyson equation for the ghost propagator in a pure gauge theory, diagrammatically.}  
\label{Ghost_SD_Graph}
\end{figure}

\begin{figure}[ht]
\vspace{.2cm}
\begin{center}
	\includegraphics[width=.8\linewidth]{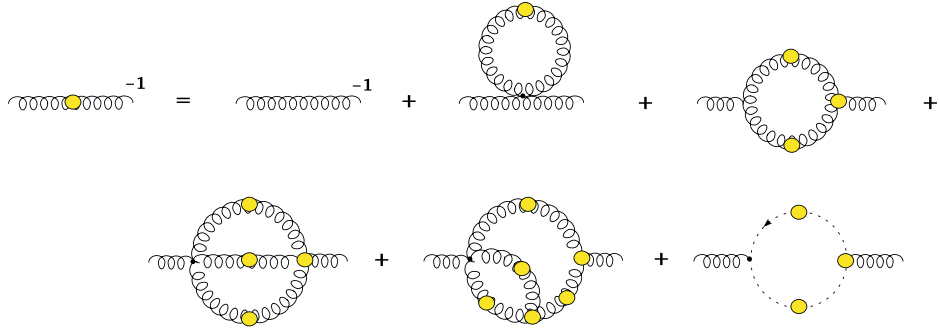}
\end{center}
\caption{\footnotesize\it Schwinger-Dyson equation for the gluon propagator in a pure gauge theory, diagrammatically.}  
\label{Gluon_SD_Graph}
\end{figure}

Performing the Legendre transformation for the three-point function in (\ref{Ghost_SD_Green}) and introducing 
the ghost-gluon vertex
\begin{equation}
\widetilde{\Gamma}_{\mu}^{abc}(-q,k;q-k) =  g_0  f^{abc} (i q_{\nu'}) \widetilde{\Gamma}_{\nu'\nu}(-q,k;q-k)
\end{equation}
and the full gluon propagator $G^{(2)ab}_{\mu\nu}(p)$, we write the Schwinger-Dyson equation 
for the ghost propagator in Fourier space
\begin{small}
\begin{align}
\label{SDghost}
\left(F^{(2)}\right)^{-1}_{ab}(k) =-\delta_{ab} k^2  
- g_0^2 f^{acd} f^{ebf} 
\int \frac{d^4q}{(2\pi)^4 }  F^{(2)}_{ce}(q)
(i q_{\nu'}) \widetilde{\Gamma}_{\nu'\nu}(-q,k;q-k) \left(i k_\mu\right) \left(G^{(2)}\right)_{\mu\nu}^{fd}(q-k),
\end{align}
\end{small}
given in a diagrammatic form at Figure \ref{Ghost_SD_Graph}.
%
%
This equation is much simpler than the one for the gluon propagator, because the last 
involves complete three- and four-gluon vertices (cf. Figure \ref{Gluon_SD_Graph}). Another virtue 
of (\ref{SDghost}) is that its \emph{form} is explicitly independent of the choice of the integration domain in 
the functional integral (\ref{SD_begin}), because the equality (\ref{SDghost_conf_par_conf}) 
holds valid for \emph{individual} gauge configurations. 
%
%

%
\subsection{Slavnov-Taylor identities}
%

The Slavnov-Taylor identities (\cite{Slavnov:1972fg},\cite{Taylor:1971ff}) are Ward 
identities in the case of a non-Abelian gauge theory. These identities follow directy
from the gauge symmetry. We write the generating functional in the case of a 
general gauge fixing condition (\ref{ContinumGaugeFixingCondition})
\begin{equation}
Z[j] = \int[\mathcal{DA}] \, \det{\MFP} \, e^{-\int d^4x \left( \mathcal{L}_{\text{QCD}}-
\frac{\left(f^a[A]\right)^2}{2\xi} + j^a_\mu A^a_\mu \right)},
\end{equation}
and use the fact that the functional integration measure is invariant under the specific gauge transformations
\begin{equation}
\label{transf_de_jauge_ST}
\begin{array}{l}
\delta A_\mu^a = D_\mu^{ab} \epsilon^b
\\
\delta f^a[\mathcal{A}] = \MFP^{ab}\epsilon^b.
\end{array}
\end{equation}
The second equation is just a general definition of the Faddeev-Popov operator. The integration measure
times the Faddeev-Popov determinant is invariant with respect to (\ref{transf_de_jauge_ST}), and hence
\begin{equation}
Z[j] = \int[\mathcal{DA}] \, \det{\MFP}\, e^{-\int d^4x \left( \mathcal{L}_{\text{QCD}}-
\frac{\left(f^a[\mathcal{A}]\right)^2}{2\xi} + j_\mu A_\mu-\frac{1}{\xi} f^a\MFP^{ab}\epsilon^{b} + j_\mu^a D_\mu^{ab}\epsilon^{b}  \right)}.
\end{equation}
Choosing a particular form for $\epsilon$
\begin{equation}
\label{epsilon_omega}
\epsilon^b = \left( \MFP^{-1} \right)^{bc} \omega^c
\end{equation}
we obtain a functional relation for the generating functional $Z$
\begin{equation}
\label{ST_general_M_moins_1}
\left( 
\frac{1}{\xi}f^a\left[\frac{\delta}{\delta j}\right] 
- \int d^4 y \, j^b_\mu(y) D_\mu^{bc}\left[\frac{\delta}{\delta j}\right] 
\left( \MFP^{-1} \right)^{ca}\left[x,y;\frac{\delta}{\delta j}\right] 
\right) Z[j] = 0.
\end{equation}
In principle, one can use the relation (\ref{ralation_exp_derivee}) in order to obtain 
an equation for the functional $W[j]$. However, this derivation would lead to a very
cumbersome expression.
In fact it is much easier to use the Slavnov-Taylor relations obtained within the
BRST formalism. The main idea of the derivation remains the same because
the BRST transformation is just a specific form of the gauge transformation (\ref{transf_de_jauge_ST}),
$\MFP^{-1}(x,y)$ being the propagator of the Faddeev-Popov ghost in the background field $\mathcal{A}$.
One has using (\ref{Lagrangian_BRST},\ref{BRST}) and the BRST invariance of the generating functional $Z[j]$
\begin{equation}
\label{generatirice_ST}
\int d^4x \int [\mathcal{D}\mathcal{A}\mathcal{D}c\mathcal{D}\bar{c}] \, e^{-\int d^4x \mathcal{L}_{eff}} \,
\left( j^a_\mu\cdot D^{ab}_\mu c^b - \frac{1}{\xi} \partial_\mu A^a_\mu \sigma^a - \bar{\sigma}^a \frac{g_0}{2} f^{abc} c^b c^c \right) = 0.
\end{equation}
This equation (\ref{generatirice_ST}) allows to obtain the Slavnov-Taylor identities for the Green functions by differentiating
with respect to the sources $j,\sigma,\bar{\sigma}$. Writing (\ref{generatirice_ST}) in terms of generating functionals 
$W$ and $\Gamma$, one obtains (see~\cite{Taylor:1976ru},~\cite{Slavnov:1972fg},~\cite{Taylor:1971ff} for details) 
the general form of Slavnov-Taylor identities between propagators and proper vertices.
The relation that we shall use in the following relates the three-gluon vertex $\Gamma_{\lambda \mu \nu} (p, q, r)$ to
the ghost-antighost-gluon vertex, and involves the complete ghost and gluon propagators.
It reads (\cite{Slavnov:1972fg},\cite{Taylor:1971ff})
\begin{equation}
\label{STid}
\begin{split}
p^\lambda\Gamma_{\lambda \mu \nu} (p, q, r) & =
\frac{p^2 F^{(2)}(p^2)}{r^2G^{(2)}(r^2)} \left(\delta_{\lambda\nu} r^2 - r_\lambda r_\nu\right) \widetilde{\Gamma}_{\lambda\mu}(r,p;q)-
\\ & -
\frac{p^2 F^{(2)}(p^2)}{q^2 G^{(2)}(q^2)} \left(\delta_{\lambda\mu} q^2 - q_\lambda q_\mu\right) \widetilde{\Gamma}_{\lambda\nu}(q,p;r).
\end{split}
\end{equation}

Some remarks regarding the non-perturbative validity of the Slavnov-Taylor identities are in order.
The above derivation is invalid when $\MFP^{-1}$ is singular 
(see (\ref{transf_de_jauge_ST}), (\ref{epsilon_omega}) and (\ref{ST_general_M_moins_1})),
i.e. for gauge fields lying on the Gribov horizon. However, 
this transformation is well defined inside the Gribov horizon. Note also that 
the general form of the Slavnov-Taylor identities does not depend on the choice of the integration 
domain inside the Gribov horizon ($\Lambda$ or $G$). Another argument in favour of 
non-perturbative validity of the Slavnov-Taylor identities may be given within 
the stochastic quantisation formalism~\cite{Zwanziger:2003cf}.

%
\subsection{Renormalisation group equations}
%

For the sake of completeness we present here another set of non-perturbative relations between the
correlators, namely the exact renormalisation group equations (or ERGE, see~\cite{Bagnuls:2000ae} for a review). 
Those are flow equations describing the variation of the effective action with the 
infrared (or ultraviolet) cut-off. The infrared cut-off is introduced by adding a special term
to the action
\begin{equation}
\Delta S_{k} = \frac{1}{2} \sum_i \int \frac{d^4 p}{(2\pi)^4} \, \phi_i(p) R_i(k,p^2) \phi_i(-p),
\end{equation}
where the momentum cut-off function $R_i$ for the field $\phi_i$ satisfies
\begin{equation}
\label{ERGE_cut_off_func}
\left\{
\begin{array}{lc}
R_i(k,p^2)\rightarrow 0 & \quad p \gtrsim k 
\\
R_i(k,p^2)\rightarrow k^2 & \quad p \lesssim k.
\end{array}
\right.
\end{equation}
The role of $\Delta S_{k}$ is to suppress quantum fluctuations with momenta below $k$. Then one may show that
the partition function satisfies the equation
\begin{equation}
\label{ERGE_Zk}
\d_k Z_k(j) = -\frac{1}{2}\int \frac{d^4p}{(2\pi)^4} \d_k R(k,p^2) \frac{\delta^2 Z_k(j)}{\delta j(p) \delta j(-p)}.
\end{equation}
The problem that arises within the formalism of the RG equations is the loss of the gauge invariance caused by
the cut-off term. However, this problem can be solved by considering a modified set of 
Slavnov-Taylor identities~\cite{Ellwanger:1996wy}. One may express the 
equation (\ref{ERGE_Zk}) in terms of the generating functional for proper vertices, that lead to 
an infinite system of partial differential equations relating different Green functions and the cut-off 
function (\ref{ERGE_cut_off_func}). We shall review some of the results for solutions of the truncated system
of such equations in the section~\ref{Review_of_today_s_analytical_results}.

%
%
\section{Lattice QCD}
%
%

%
%
%
\subsection{Lattice QCD partition function}
%
%
%

A fully non-perturbative study from the first principles of QCD phenomenon requires a direct
calculation of the functional integral of the type (\ref{Z_continuum}). These integrals can be approximately evaluated by 
means of lattice simulations. Another interest in the lattice regularisation is that it preserves the gauge invariance.
The inverse lattice spacing $a^{-1}$ plays the role of the ultraviolet cut-off, and we recover the continuum limit theory
by sending $a$ to zero. 

\begin{figure}[ht]
\begin{center}
\includegraphics[scale=0.6]{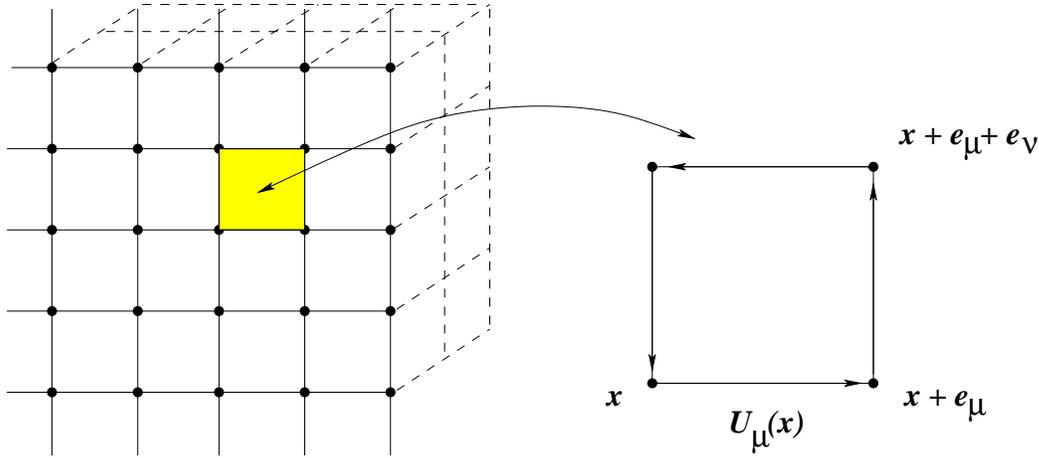}
\end{center}
\caption{\footnotesize\it Gauge fields $\mathcal{A}_\mu$ are defined on the links of the lattice ($\| \emu \| = a$), and the fermion fields $\psi (x)$ are defined on its sites.}
\label{reseau}
\end{figure}

In what follows we discuss only pure Yang-Mills theories. In practice, when doing a lattice simulation,
one considers a theory in a finite volume $V=L^4$ with (most often) periodical boundary conditions; and generates some
(quite large) number $M$ of gauge field configurations $\{C_i\}$ distributed according to the probability measure 
\begin{equation}
d\mu[\mathcal{A}]=e^{-\int d^4x \mathcal{L}_{\text{Yang-Mills}}(A)}
[\mathcal{D A}].
\end{equation}
Then one can calculate a Monte-Carlo approximation for any operator $\mathcal{O}$ 
\begin{equation}
O = \left\langle \mathcal{O}(\mathcal{A}_\nu) \right\rangle 
= \int d\mu\mathcal{O}(\mathcal{A}_\nu) \approx\frac{1}{M}\sum_{i=1}^{M}\mathcal{O}(C_i).
\end{equation}
Let us now discuss the measure $d\mu[A]$ in the discrete case. Gauge fields are defined on the
links of the Euclidean lattice (cf. Figure~\ref{reseau}), and the fundamental lattice variable for the gauge field 
is the \textbf{link variable}
\begin{equation}
U(x,x+\emu)\equiv U_{\mu}(x)= e^{i g_0 a \mathcal{A}_\mu(x+\frac{a}{2}\emu)}\in SU(N_c),
\end{equation}
where $\emu$ is a vector in direction $\mu$, $\| \emu \| = a$.
%
%
For small values of the lattice spacing $a$ we can extract the field $\mathcal{A}_\mu$ using 
an approximate formula
\begin{equation}
\label{U_to_A_conversion}
\frac{U_\mu(x) - U^\dagger_\mu(x)}{2 i a g_0} = \mathcal{A}_\nu\left( x+\frac{\emu}{2}\right) + O(a).
\end{equation}
We use this definition of the gluon field in what follows. Then one can define 
an elementary gauge invariant variable - a \textbf{plaquette}
\begin{equation}
U(p)=U_{\mu}(x) U_{\nu}(x+ae_{\mu}) U^{\dagger}_{\mu}(x+ae_{\nu})U^{\dagger}_{\nu}(x).
\end{equation}
Using this variables we can write the simplest action that converges to the pure Yang-Mills' action in the  continuum limit 
(\cite{Wilson:1974sk}, see \cite{Montvay:1994cy},\cite{Smit:2002ug} for a review):
\begin{equation}
\label{WilsonAction}
S_{g}[U_{\mu}(x)]=\beta\sum_{x}\sum_{\mu,\nu}\Big(\id-\frac{1}{N_c}\re\tr\, U(p) \Big),
\end{equation}
where the lattice bare coupling is defined by
\begin{equation}
\beta=\frac{2N_c}{g_0^2}.
\end{equation}
Thus, the partition function in the lattice formulation reads
\begin{equation}
\label{LatticePartitionFunction}
Z_{\text{lat}}[U]=\int [\mathcal{D} U_{\mu}(x)] e^{-S_g[U]}.
\end{equation}
Of course there exist an infinite number of lattice actions giving (\ref{Z_continuum}) in the continuum limit $a\rightarrow 0$.
They should all give compatible results when the lattice spacing is small enough and the renormalisation group scaling laws are verified.
In practice, one generates the gauge configurations $\{C_i\},i=1\ldots M$ according to the probability measure 
$[\mathcal{D} U_{\mu}(x)] e^{-S_g[U]}$. In all our simulations we have used the Wilson action (\ref{WilsonAction}).
Usually,  the Metropolis or Heatbath algorithms are used for the Monte-Carlo generation. For a detailed review on this topic
see \cite{Montvay:1994cy},\cite{Smit:2002ug}.

The important question that has not been discussed up to now is the removal of the ultraviolet divergences of the theory in the 
continuum limit. The regularised lattice partition function contains one parameter - the bare coupling constant $g_0$ 
(or, equivalently, the lattice spacing $a$). When calculating some physical quantity on the lattice, say, the string tension $\sigma$, we
obtain it as a function of $a$. If the ultraviolet cut-off $a^{-1}$ is large enough, the calculated quantity obeys the scaling law,
and one can compare the result for $\sigma(a)$ of the lattice simulation to the known 
experimental data $\sigma_{\text{exp}}$, and thus determine the corresponding physical value of the 
lattice spacing $a$ by solving the equation $\sigma(a)=\sigma_{\text{exp}}$.
In the unquenched case one has to calculate several physical quantities because
more non-fixed parameters are involved (the masses of quarks). 
When all free parameters of the lattice theory are fixed in the scaling region with some experimental inputs,
all further calculations~\footnote{of experimentaly observable quantities} are automatically renormalised,
and do not contain any divergences. Moreover, all the calculations are now the \emph{predictions} of the theory.
Thus, the only limitations of the numerical method \emph{a priori} are discretisation errors and
the errors on the experimental inputs. We discuss in details the systematic error of lattice simulations
in the section~\ref{ErrorOfTheCalculation}. In practice one is also often limited by the computer power.

Let us say some words about the strong coupling limit of (\ref{LatticePartitionFunction}), corresponding to the 
perturbative expansion in the $\beta$ parameter. The ordinary perturbation expansions  in gauge theories 
(in powers of $g_0$) are at most asymptotic, but power series of (\ref{LatticePartitionFunction}) in $\beta$
is proven to have a \emph{finite} radius of convergence~\cite{Osterwalder:1977pc}. Many interesting results,
like the existence of the mass gap and the area law for the Wilson loop, have been analytically proven within
this approach. However, it has been argued that the region of the strong coupling is analytically disconnected from
the weak-coupling domain, corresponding to the continuum limit of the theory~\cite{Itzykson:1980fz}.
Lattice gauge theories are believed to possess an essential singularity at some finite value $\beta_\text{rough}$, 
corresponding to an infinite order phase transition. This phenomenon is well known in statistical physics 
(roughening transition). Thus it is impossible to know whether the strong-coupling predictions are applicable
in the physically interesting weak-coupling regime. In practice, one has to perform numerical simulations 
with $\beta > \beta_\text{rough}$.

%
%
%
\subsection{Fixing the Minimal Landau gauge on the lattice}
%
%
%
\label{subsection_lattice_gauge_fixing}

The continuum gauge fixing condition (\ref{ContinumGaugeFixingCondition}) 
is modified by the discretization, so one works with its lattice version $f_L(U_\mu)=0$.
A gauge configuration $U_{\mathcal{C}_0}$ generated during the simulation process does not
satisfy the Landau gauge condition. One has to perform a gauge transformation $u(x)$ on it 
in order to move the field configuration along its gauge orbit up to an intersection with the surface $f_L(U_\mu)=0$.
But there is no need to have an explicit form of $u(x)$. Instead we perform an iterative minimisation process that
converges to the gauge fixed configuration $U^{(u)}_{\mathcal{C}}$. 
Let us first illustrate it on the example of the Landau gauge in the continuum limit. For every gauge field $\mathcal{A}$
we calculate the functional
\begin{equation}
\label{GaugeMinimisingFunctional}
F_\mathcal{A}[u(x)]=-\tr\int d^4x \left( \mathcal{A}_{\mu}^{(u)}(x) \mathcal{A}_{\mu}^{(u)}(x) \right),
\quad u(x)\in SU(N_c).
\end{equation}
Expanding it up to the second order around some element $u_0(x)$ we have ($u=e^{X}u_0$, $X\in\mathfrak{su}(N_c)$)
\begin{equation}
\label{GaugeMinimisingFunctional_1}
F_\mathcal{A}[u]= F_\mathcal{A}[u_0] - 2\int d^4x \tr\left( X \partial_\mu \mathcal{A}^{(u_0)}_{\mu } \right)
+\int d^4 x \tr \left(X^\dagger \mathcal{M}_{\text{FP}}\left[\mathcal{A}^{(u_0)}\right]  X \right)+\ldots,
\end{equation}
where the matrix $\mathcal{M}_{\text{FP}}$ in the quadratic term defines the Faddeev-Popov operator.
Obviously, if $u_0$ is a local minimum of (\ref{GaugeMinimisingFunctional}) then we have a double condition
\begin{equation}
\label{GF_minimum_condition}
\left\{
	\begin{array}{l}
		\partial_\mu \mathcal{A}^{(u_0)}_{\mu } = 0 
		\\ 
		\mathcal{M}_{\text{FP}}\left[\mathcal{A}^{(u_0)}\right] \quad \text{is positively defined.}
	\end{array}
\right.
\end{equation}
Hence, the minimisation of the functional (\ref{GaugeMinimisingFunctional}) allows not only to fix the Landau gauge,
but also to obtain a gauge configuration inside the Gribov horizon 
(because all the eigenvalues of $\mathcal{M}_{\text{FP}}$ are positive).  
On the lattice, the discretized functional (\ref{GaugeMinimisingFunctional}) reads
\begin{equation}
\label{LatticeGaugeMinimisingFunctional}
F_U[u(x)]=-\frac{1}{V}\re\tr\sum_{x,\mu}u(x)U_\mu(x)u^{\dagger}(x+\emu).
\end{equation}
Then at a local minimum $u_0$ we have a discretized Landau gauge fixing condition 
\begin{equation}
\label{LatticeLandauGaugeCondition}
\sum_{\mu}
\left[
\mathcal{A}^{(u_0)}_{\mu}\left(x+\frac{\emu}{2}\right) -
\mathcal{A}^{(u_0)}_{\mu}\left(x-\frac{\emu}{2} \right)
\right]=0
\end{equation}
that we write in a compact form $\nabla_\mu\mathcal{A}^{(u_0)}_\mu = 0$. Indeed, 
if  $u(x)=u_0(x) e^{s\, \omega(x)}$ where $\omega(x)$ is the element of the algebra $\mathfrak{su}(N_c)$
then
\begin{eqnarray}
\left.\frac{\delta}{\delta u}\right\arrowvert_{u_0}F(u) \equiv\left.\frac{d}{ds}\right\arrowvert_{s=0} F(u(s,x)) = 0,
\end{eqnarray}
and hence
\begin{eqnarray}
-\frac{1}{V}\re\tr \sum_{x,\mu}
\left(
	\left[  
		\omega(x+\emu) - \omega(x)
	\right]
 U_{\mu}(x)
 \right)=0
\qquad\forall\omega(x)\in\mathfrak{su}(N_c).
\end{eqnarray}
Thus
\begin{eqnarray}
\sum_{\mu}\left[
U_{\mu}(x) -  U^{\dagger}_{\mu}(x) - 
U_{\mu}(x-\emu)  +  U^{\dagger}_{\mu}(x-\emu)  
\right] =0 
\end{eqnarray}
and at leading order in the lattice spacing $a$ (\ref{U_to_A_conversion}) we obtain (\ref{LatticeLandauGaugeCondition}).
The second derivative (the equivalent of the second-order term in (\ref{GaugeMinimisingFunctional_1})) 
can be written as
\begin{equation}
\frac{d^2}{ds^2}F\left( u(s,x) \right)  = -\frac{1}{V} \left( \omega, \nabla_\mu \mathcal{A}^{\prime}_\mu \right)
\end{equation}
where
\begin{equation}
\label{A_PRIME}
\mathcal{A}^{\prime}_\mu = 
\frac{1}{2} 
\left( 
	-\omega(x) U^{(u)}_\mu(x) + U^{(u)}_\mu(x) \omega(x+\emu) + \omega(x+\emu) U^{(u)\dagger}_\mu(x) - 
	U^{(u)\dagger}_\mu(x) \omega(x)
\right).
\end{equation}
This defines a quadratic form 
\begin{small}
\begin{align}
\begin{split}
 \left(  \omega, \MFPlat[U] \omega  \right)  & = 
-\frac{1}{V}\re\tr
\sum_{x,\mu}\left(
\left[ 
\omega^2(x+\emu)  - 2 \omega(x+\emu)\omega(x) + \omega^2(x) 
 \right] U^{(u_0)}_\mu(x)
\right)  = 
\\ & =  -\frac{1}{2V} \tr \sum_{x,\mu}
	\Big( 
		\left(  \omega(x+\emu) -  \omega(x)  \right)^2 \left(  U^{(u_0)}_\mu(x) + U^{(u_0)\dagger}_\mu(x) \right)
		-  \\ & \hspace*{2.5cm} -  \left[\omega(x+\emu),\omega(x)\right]\left(  U^{(u_0)}_\mu(x) - U^{(u_0)\dagger}_\mu(x) \right)
	\Big).
\end{split}
\end{align}
\end{small}
The operator $\MFPlat[U]$ is the lattice version of the Faddeev-Popov operator. It reads 
\begin{small}
\begin{align}
\label{MFPlat}
\nonumber
\left(\MFPlat[U]\omega\right)^{a}(x) &= \frac{1}{V}\sum_{\mu}\biggl\{ G_{\mu}^{ab}(x)
          \left(\omega^{b}(x+\emu)-\omega^{b}(x)\right)
        - (x \leftrightarrow x-\emu) \\
 &\qquad + \frac{1}{2}f^{abc}\left[
             \omega^{b}(x+\emu)A_{\mu}^{c}\left(x+\frac{\emu}{2}\right)
           - \omega^{b}(x-\emu)A_{\mu}^{c}\left(x-\frac{\emu}{2}\right) \right]
          \biggr\},
\end{align}
\end{small}
where
\begin{align}
& G_{\mu}^{ab}(x) = - \frac{1}{2}\tr\left(\left\{t^{a},t^{b}\right\} \left(U_{\mu}(x)+U_{\mu}^{\dagger}(x)\right)\right) 
\\ &
A_{\mu}^{c}\left(x+\frac{\emu}{2}\right) = \tr\left(t^{c}\left(U_{\mu}(x)-U_{\mu}^{\dagger}(x)\right)\right).
\end{align}

For the Minimal Landau gauge fixing in our numerical simulation we have used the
Overrelaxation algorithm~\cite{Adler:1987ce} with $\omega=1.72$. We stop the iteration process of
the minimising algorithm when the following triple condition is fullfield:
\begin{align}
\label{StopParameters}
\frac{1}{V (N_c^2 -1)}\sum_{x,\mu}
\tr \left[  
\left(
\nabla_\mu\mathcal{A}^{(u^{(n)})}_\mu
\right)
\left( \nabla_\mu\mathcal{A}^{(u^{(n)})}_\mu \right)^{\dagger}\right] 
& \leq \Theta_{\max_x \arrowvert \partial_\mu A^a_\mu\arrowvert} = 10^{-18}
\nonumber
\\
\frac{1}{V (N_c^2 -1)}
\left\arrowvert
\sum_x \tr\left[ u^{(n)}(x) - \id \right]
\right\arrowvert & \leq \Theta_{\delta u } = 10^{-9}
\nonumber
\\
\forall a,t_1,t_2 \quad 
\left \arrowvert 
\frac{ \text{A}^a_0 (t_1)-\text{A}^a_0 (t_2)  }{\text{A}^a_0 (t_1)+\text{A}^a_0 (t_2) }
\right \arrowvert
& \leq \Theta_{\delta \text{A}_0 } = 10^{-7}.
\end{align}
where $u^{(n)}(x)$ is the matrix of the gauge transformation $u(x)$ at the iteration step $n$, 
and the charge 
\begin{equation}
\text{A}^a_0(t) = \int d^3 \vec{x} A^a_0(\vec{x},t)
\end{equation}
must be independent of $t$ in Landau gauge when periodical boundary conditions for the gauge 
field are used. The choice of  numerical values for the stopping parameters 
is discussed in~\cite{Lokhov:2005ra}.

%
%
%
\chapter{Lattice Green functions}
%
%
%
\label{chapter_Lattice_Green_functions}

In the functional integral formalism Green functions are defined as mean values 
of products of fields according to the functional measure,
giving as a result the vacuum expectation values of these products.
Often one is interested in the Green functions in Fourier space. 
Here we define the Fourier transformation on the lattice.

If a function $f(x)$ is defined on the \emph{sites} of the four-dimensional lattice with periodical 
boundary conditions then
\begin{align}
& \widetilde{f}(p) = a^4\sum_{ x } f(x) e^{-i p\cdot x}
\qquad p_\mu = \frac{2\pi}{aL}n_\mu,
\quad n_\mu=0,1,\ldots,L-1
\nonumber
\\ &
f(x) = \frac{1}{V}\sum_{ p  } e^{i p\cdot x }\widetilde{f}(p).
\end{align}
In the case of the variables defined on the $links$ of the lattice one should change the above formulae by
$x_\mu\rightarrow x_\mu + \frac{\emu}{2}$. In the infinite-volume limit $V\rightarrow\infty$ we have
\begin{equation}
\frac{1}{V}\sum_{p}\longrightarrow \frac{1}{(2\pi)^4}\int_{\left[-\frac{\pi}{a},\frac{\pi}{a}\right]^4}d^4 p.
\end{equation}

We recall in the first part of this chapter the main definitions regarding the lattice
two- and three-point Green functions in Landau gauge and the Momentum substraction renormalisation scheme.
Next we describe the details of the numerical calculation of the ghost propagator
and discuss different sources of errors of the lattice approach. The last part is devoted to
the Gribov ambiguity on the lattice and the influence of Gribov copies on the Green functions.

%
%
\section{Green functions in Landau gauge}
\label{GREEN_FUNCTIONS_IN_LANDAU_GAUGE}
%
%

The gluon propagator in Landau gauge may be parametrised at all values of momenta as
\begin{equation}
G^{(2)ab}_{\mu\nu}(p,-p) \equiv
\left\langle
\widetilde{A}^a_\mu(-p)  \widetilde{A}^b_\nu(p)
\right\rangle
= 
\delta^{ab}\left(  \delta_{\mu\nu} - \frac{p_\mu p_\nu}{p^2}  \right)
G^{(2)}(p^2).
\end{equation}
This implies that the scalar factor $G^{(2)}(p^2)$ may be extracted according to
\begin{equation}
\label{def_gluon_p}
G^{(2)}(p^2) = \frac{1}{3\left( N_C^2 - 1 \right)} \sum_{\mu,a} G^{(2)aa}_{\mu\mu}(p,-p), \qquad p^2 \neq 0 
\end{equation}
completed with
\begin{equation}
\label{def_gluon_0}
G^{(2)}(0) = \frac{1}{4\left( N_C^2 - 1 \right)} \sum_{\mu,a} G^{(2)aa}_{\mu\mu}(0,0), \qquad p^2  = 0.
\end{equation}
The difference in normalisations at zero (\ref{def_gluon_0}) and finite momenta (\ref{def_gluon_p}) is due to 
an additional degree of freedom related to global gauge transformations 
on a periodic finite lattice ($p^2=0 \leftrightarrow x_\mu \sim x_\mu + L$).

The ghost propagator is parametrised in the common way:
\begin{equation}
F^{(2)ab}(p,-p) \equiv
\left\langle
\widetilde{c}^a(-p)  \widetilde{\bar{c}}^b(p) \right\rangle 
=
\delta^{ab} F^{(2)}(p^2).
\end{equation}
It is not defined at $p^2=0$ because in this case the Faddeev-Popov operator is strictly equal 
to zero and thus it is not invertible.

\begin{figure}[h]
\vspace{.2cm}
\begin{center}
	\includegraphics[width=0.5\linewidth]{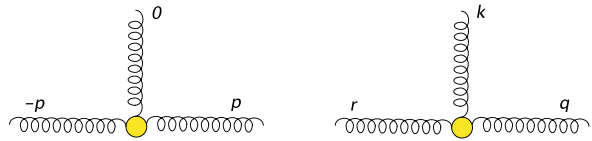} 
\end{center}
\caption{\footnotesize\it Definitions of the $\widetilde{\text{MOM}}$ scheme, $p^2=\mu^2$ (left) and $\text{MOM}$,  $q^2=r^2=k^2=\mu^2$ (right)} 
\label{MOMasym_ET_MOMsym_GRAPH}
\end{figure}

The renormalisation scheme that is widely used in order to renormalise the lattice 
Green functions is the so-called Momentum substraction scheme (MOM). The virtue of this
scheme is its non-perturbative definition. The renormalisation constants are defined by 
setting the corresponding Green functions to their tree values at some renormalisation 
point $\mu^2$. In the case of the two-point Green function the renormalisation 
constant $Z_3(\mu^2)$ of the gauge field or the one of the ghost 
field (denoted $\widetilde{Z}_3(\mu^2)$) is unambiguously defined as:
\begin{align}
\label{DEFINITION_Z3_Z3TILDE}
& Z_3(\mu^2) = \mu^2 G^{(2)}(\mu^2)
\\ & 
\widetilde{Z}_3(\mu^2) = \mu^2 F^{(2)}(\mu^2).
\end{align}
The coupling constant has also to be renormalised to complete the renormalisation of a pure 
Yang-Mills theory. It can be defined non-perturbatively by an amputation of a  three-point Green-functions 
from its external propagators. But this requires to fix the  kinematic configuration of the three-point
Green-function at the normalisation  point. On the lattice one usually uses
either a fully symmetric kinematic configuration (denoted $\text{MOM}$)  or
a zero point (ZP) kinematic configuration with one vanishing external momentum
(denoted generically  $\widetilde{\text{MOM}}$), see Figure \ref{MOMasym_ET_MOMsym_GRAPH}.

\subsubsection{Three-gluon vertex: symmetric case} 
There are only two independent tensors in Landau gauge in the case of the symmetric 
three-gluon Green function~\cite{Boucaud:1998bq}:
\begin{align}
&\mathcal{T}^{[1]}_{\mu_{1},\mu_{2},\mu_{3}}(p_1,p_2,p_3) = 
\delta_{\mu_{1}\mu_{2}}(p_1-p_2)_{\mu_{3}} + 
\delta_{\mu_{2}\mu_{3}}(p_2-p_3)_{\mu_{1}} + 
\delta_{\mu_{3}\mu_{1}}(p_3-p_1)_{\mu_{2}}
\\&
\mathcal{T}^{[2]}_{\mu_{1},\mu_{2},\mu_{3}}(p_1,p_2,p_3) = 
\frac{(p_1 - p_2)_{\mu_{3}}  (p_2 - p_3)_{\mu_{1}}  (p_3 - p_1)_{\mu_{2}}  }{p^2}.
\end{align}
Then the three-gluon Green function in $\text{MOM}$ scheme ($p_1^2 = p_2^2 = p_3^2 = \mu^2$) can be parametrised as
\begin{small}
\begin{equation}
\begin{split}
\left\langle 
\widetilde{A}^a_{\mu_{1}}(p_1) \widetilde{A}^b_{\mu_{2}}(p_2) \widetilde{A}^c_{\mu_{3}}(p_3)
\right\rangle  =
f^{abc} &
\Big[ 
G^{(3)\text{sym}}(\mu^2) \mathcal{T}^{[1]}_{\mu^{\prime}_{1},\mu^{\prime}_{2},\mu^{\prime}_{3}}(p_1,p_2,p_3)
\prod_{i=1,3} 
	\left( 
		\delta_{\mu^{\prime}_{i} \mu_{i}} - \frac{{p_i}_{\mu^{\prime}_{i}} {p_i}_{\mu_{i}} }{\mu^2}
	\right) +
\\ & + H^{(3)}(\mu^2) \mathcal{T}^{[2]}_{\mu_{1},\mu_{2},\mu_{3}}(p_1,p_2,p_3)
\Big]
\end{split}
\end{equation}
\end{small}
For a non-perturbative MOM definition of the renormalised coupling $g_R$ one need to extract 
the scalar function $G^{(3)\text{sym}}(\mu^2)$, proportional to the coupling $g_0$ at the tree order.
This is done by the following projection:
\begin{equation}
\begin{split}
G^{(3)\text{sym}}(\mu^2) = &
\left( 
\mathcal{T}^{[1]}_{\mu^{\prime}_{1},\mu^{\prime}_{2},\mu^{\prime}_{3}}(p_1,p_2,p_3)
\prod_{i=1,3} \left( 
\delta_{\mu^{\prime}_{i} \mu_{i}} - \frac{{p_i}_{\mu^{\prime}_{i}} {p_i}_{\mu_{i}} }{\mu^2}\right)
+
\frac{1}{2}\mathcal{T}^{[2]}_{\mu^{\prime}_{1},\mu^{\prime}_{2},\mu^{\prime}_{3}}(p_1,p_2,p_3)
\right)\times
\\ & \times
\frac{1}{18\mu^2}
\frac{f^{abc}}{N_c(N_c^2-1)} 
\left\langle 
	\widetilde{A}^a_{\mu_{1}}(p_1) \widetilde{A}^b_{\mu_{2}}(p_2) \widetilde{A}^c_{\mu_{3}}(p_3)
\right\rangle
\end{split}
\end{equation}
\subsubsection{Three-gluon vertex: asymmetric case}
The three-gluon Green function with one vanishing external propagator (\cite{Alles:1996ka},\cite{Boucaud:1998bq}) can be parametrised as
\begin{equation}
G^{(3)abc}_{\mu\nu\rho}(p,0,-p) \equiv
\left\langle
\widetilde{A}^a_\mu(-p)  \widetilde{A}^b_\nu(p) \widetilde{A}^c_\rho(0)
\right\rangle
=2f^{abc} p_\rho \left(  \delta_{\mu\nu} - \frac{p_\mu p_\nu}{p^2}  \right) G^{(3)\text{asym}}(p^2),
\end{equation}
and thus
\begin{equation}
G^{(3)\text{asym}}(p^2) = \frac{1}{6p^2}\frac{f^{abc}}{N_c(N_c^2-1)}\delta_{\mu\nu}p_\rho G^{(3)abc}_{\mu\nu\rho}(p,0,-p).
\end{equation}

\subsubsection{Ghost-gluon vertex} 
\begin{figure}[h]
\vspace{.2cm}
\begin{center}
\includegraphics[width=0.5\linewidth]{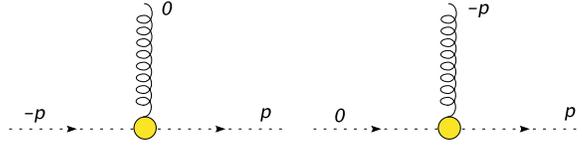} 
\end{center}
\caption{\footnotesize\it Kinematic configurations of the $\widetilde{\rm MOM}_c$, $\widetilde{\rm MOM}_{c0}$ schemes, $p^2=\mu^2$.}
\label{MOMasym_ET_MOMsym_GRAPH_ghost}
\end{figure}

Similar parametrisation may be written in the case of the ghost-ghost-gluon Green function 
(cf. Figure~\ref{MOMasym_ET_MOMsym_GRAPH_ghost}). 
%
%
But in this case one obtains two different renormalisation schemes 
$\widetilde{\rm MOM}_c$ and $\widetilde{\rm MOM}_{c0}$ corresponding to the 
zero-point kinematic configuration with vanishing momentum of, respectively, the gluon and the entering ghost.
%
%
We denote by $\widetilde{G}^{(3)}_\text{K}$ a generic scalar function extracted from a three-point function 
in a particular kinematic configuration $\text{K}$. Then the gauge coupling at the renormalisation scale $\mu^2$ is defined by
\begin{equation}
g_R^{\left(K \right)}(\mu^2) = \frac{G^{(3)}_\text{K}(p_1^2,p_2^2,p_3^2)}{G^{(2)}(p_1^2) G^{(2)}(p_2^2) G^{(2)}(p_3^2)} Z_3^{3/2}(\mu^2)
\end{equation}
in the case of three-gluon vertices, where the choice of $p_i$ determines the renormalisation scheme.
For of ghost-ghost-gluon vertices the coupling is defined by
\begin{equation}
\tilde{g}_R^{\left(K \right)}(\mu^2) = 
\frac{\widetilde{G}^{(3)}_{\text{K}}(p_1^2,p_2^2,p_3^2)}{F^{(2)}(p_1^2) F^{(2)}(p_2^2) G^{(2)}(p_3^2)} Z_3^{1/2}(\mu^2) \widetilde{Z}_3(\mu^2).
\end{equation}
where $\widetilde{G}^{(3)}_{\text{K}}(p_1^2,p_2^2,p_3^2)$ is the scalar factor of the ghost-gluon three point
function $\left\langle \widetilde{\mathcal{A}}_\mu(p_1)\widetilde{c}(p_2)\widetilde{\bar{c}}(p_3)\right\rangle $. In fact,
only the coupling in the $\widetilde{\rm MOM}_c$ scheme can be defined non-perturbatively for the reasons explained 
in the following section.

%
%
\section{Numerical calculation of the ghost Green functions in Landau gauge}
%
%

%
\subsection{Lattice implementation of the Faddeev-Popov operator}
%
\label{Lattice_implementation_Faddeev_Popov_operator}

In order to calculate numerically~\cite{Boucaud:2005np} the ghost propagator
\begin{align}
\label{eq:Ghost}
F^{(2)}(x-y)\delta^{ab} \equiv \left\langle\left({\MFPlat}^{-1}\right)^{ab}_{xy}\right\rangle
\end{align}
one uses the lattice definition (\ref{MFPlat}) of the Faddeev-Popov operator.
Most lattice implementations of the Faddeev-Popov operator use its explicit component 
form. But as we have seen the action of the Faddeev-Popov operator on a vector $\omega$ 
can also be written as a lattice divergence 
\begin{align}
\label{eq:FP2a}
\MFPlat[U]\omega = -\frac{1}{V} \nabla_\mu \mathcal{A}^{\prime}_\mu
\end{align}
where $\mathcal{A}^{\prime}$ is defined by (\ref{A_PRIME}). This form allows a very efficient lattice implementation which 
is based on the fast routines coding the group multiplication law:
\begin{footnotesize}
\begin{center}
\center{\rule{10cm}{0.5pt}}
\begin{align*}
&\#\ \omega_{in},\ \omega_{out}\ \textit{are the ghost fields.} \\
&\#\ U_\mu(x)\textit{ is the gauge configuration.} \\ 
&\textbf{type }SU(N_c)\quad U_\mu(x), dU, W, W_{+}, W_{-} \\
&\textbf{type }\mathfrak{su}(N_c)\quad \omega_{in}(x), \omega_{out}(x) \\
&\textbf{for all } x \textbf{ do}\\
&\quad dU = 0. \\
&\quad W = \omega_{in}(x) \\ 
&\quad \textbf{do}\ \mu = 1\ldots 4 \\
   &\quad\qquad W_{+} = \omega_{in}(x+\emu) \\ 
   &\quad\qquad W_{-} = \omega_{in}(x-\emu) \\ 
   &\quad\qquad dU = dU + U_{\mu}(x-\emu)\ast W + W\ast U_{\mu}(x) \\
   &\quad\qquad\quad\qquad\quad - U_{\mu}(x)\ast W_{+} 
   - W_{-}\ast U_{\mu}(x-\emu) \\
&\quad \textbf{end do} \\
&\quad \omega_{out}(x) = dU - dU^{\dagger} - \frac{1}{N_c}\tr(dU- dU^{\dagger}) \\
&\textbf{end do}
\end{align*}
\center{\rule{10cm}{0.5pt}}
\end{center}
\end{footnotesize}
%

%
\subsection{Numerical inversion of the Faddeev-Popov operator}
%
\label{Numerical_inversion_Faddeev_Popov_operator}

We invert the Faddeev-Popov operator by solving the equation 
\begin{equation}
\sum_{y,b} \MFPlat[U]^{ab}(x,y) \eta^b(y) = S^a(x),
\end{equation}
for some source $S^a(x)$ using an appropriate algorithm (for a review of algorithms see  \cite{Saad}). 
The operator $\MFPlat[U]$ has zero-modes, that is why the inversion can only be done in the 
vector subspace $K^{\perp}$ orthogonal to the kernel of the operator.
The trivial zero-modes are constant fields. If we neglect Gribov copies then 
the Faddeev-Popov operator has no other zero-modes, and thus the non-zero 
Fourier modes form a basis of $K^{\perp}$:
\begin{align}
\label{eq:ortho}
	\eta(y) = \sum_{p\neq 0} c_{p}e^{ip\cdot y}\,,\quad \forall \eta\in K^{\perp}.
\end{align}
%

\subsubsection{The inversion in one Fourier mode} 
%
Choosing the source for inversion in the form
\begin{align}
\label{source_one_mode}
   S^{a}_{p}(x) = \delta^{ab}e^{ip\cdot x}
\end{align}
and taking  the scalar product of the inverse $\MFPlat[U]^{-1}S^{a}_{p}$ with the source one obtains
\begin{align}
  \left\langle 
  	\left(S^{a}_{p}\left|\right.{\MFPlat}^{-1}S^{a}_{p}\right) 
   \right\rangle &= 
  \sum_{x,y}
  \left\langle
  \left({\MFPlat}^{-1}\right)^{aa}_{xy}
  \right\rangle e^{-ip\cdot(x-y)} \\
  \label{eq:ftp}
  &= V\,\widetilde{F}^{(2)}(p)
\end{align}
after averaging over the gauge field configurations.  This method requires one matrix 
inversion for each momentum $p$. It is suitable only when one is interested in a few 
values of the ghost propagator.

\subsubsection{The inversion in all Fourier modes} 

One can calculate the ghost propagator for all momenta $p$ doing only one matrix inversion 
noticing that 
\begin{align}
  \label{eq:delta}
  \delta_{x,y} = \frac{1}{V} + \frac{1}{V}\sum_{p\ne 0} e^{-ip\cdot(x-y)}
\end{align}
and choosing for the source:
\begin{align}
  \label{eq:zero}
  S^{a}_{0}(x) = \delta^{ab}\left(\delta_{x,0} - \frac{1}{V}\right).
\end{align}
The Fourier transform of $M^{-1}S^{a}_{0}$, averaged over the gauge
configurations, yields:
\begin{align}
  \nonumber
  \sum_{x} e^{-ip\cdot x}\left<{\MFPlat}^{-1}S^{a}_{0}\right> &=
  \sum_{x} e^{-ip\cdot x}\left<\left({\MFPlat}^{-1}\right)^{aa}_{x0}\right> - 
  \frac{1}{V}\sum_{x,y}e^{-ip\cdot x}
  \left<\left({\MFPlat}^{-1}\right)^{aa}_{xy}\right> \\
  \nonumber
  &= \sum_{x} e^{-ip\cdot x}F^{(2)}(x) -
  \frac{1}{V}\sum_{x,y}e^{-ip\cdot x}F^{(2)}(x-y) \\
  &=  \widetilde{F}^{(2)}(p) - \delta(p)\sum_{x}F^{(2)}(x),
  \label{eq:ft0}
\end{align}
where we have used the translational invariance of the ghost propagator. Therefore,
with this choice of the source, only one matrix inversion followed by one Fourier 
transformation of the solution is required to get the ghost propagator for all values of the
lattice momenta.

Because in the case of the source (\ref{eq:zero}) one inverts in all modes 
at the same time, some statistical accuracy is lost.
However, it turns out to be sufficient for our purposes.

There is one important technical point that should be mentioned. During the inversion
process it is mandatory to check, whatever the choice of sources, that rounding
errors during the inversion do not destroy the condition that
the solution is still orthogonal to the kernel of the Faddeev - Popov operator:
\begin{align}
  \label{eq:kernel}
  \sum_{x}\left({\MFPlat}^{-1}S\right)(x) = 0
\end{align}
Indeed, if the zero-mode component of the solution grows beyond some
threshold during the inversion of the Faddeev-Popov operator on a
gauge configuration, then this component starts to increase
exponentially, and a sizeable bias is produced in other components as
well. We have observed this phenomenon occasionally, about one gauge
configuration every few hundreds, when using the componentwise implementation of the
lattice Faddeev-Popov operator based on (\ref{MFPlat}). However, we have never 
observed sizeable deviations from (\ref{eq:kernel}) using the efficient 
implementation of the Faddeev-Popov operator exposed in 
the subsection~\ref{Lattice_implementation_Faddeev_Popov_operator}.

%
%
\subsection{Calculation of the ghost-gluon vertex}
%
%

In order to calculate the ghost-gluon three-point function in $\widetilde{\rm MOM}_{c}$ and 
$\widetilde{\rm MOM}_{c0}$ renormalisation schemes one has to calculate the corresponding
ghost two-point function:
\begin{equation}
\label{ghost_gluon_Green_function}
\left\langle \widetilde{\mathcal{A}}_\mu(p_1)  \underbrace{   \widetilde{c}(p_2)  \widetilde{\overline{c}}(p_3)   } \right\rangle 
\quad \substack{\text{lattice} \\ \longrightarrow} \quad
\sum_{\text{conf. }i} \widetilde{\mathcal{A}}_\mu (p_1) \widetilde{\left( \MFPlat[U_i]^{-1} \right)}(p_2,p_3)
\end{equation}
It is quite easy to calculate this Green function in the $\widetilde{\rm MOM}_{c}$
kinematic configuration, because in this case $p_2=-p,p_3=p$ and $\widetilde{\left(\MFPlat[U_i]^{-1}\right)}(-p,p)$ 
is just a ghost propagator in the background gluon field defined by $U_i$.

The situation changes when considering the kinematic configurations like $\widetilde{\rm MOM}_{c0}$, 
when the momentum of the entering (or of the outgoing) ghost is set to zero. In this case the 
inversion of the Faddeev-Popov operator has to be performed with the source (\ref{source_one_mode}).
In other words we try to solve the equation
\begin{equation}
\MFPlat[U]_{xy} \eta_{yz} = e^{i p\cdot (x-z)}.
\end{equation}
The vector $\overline{\eta}_y = \frac{1}{V}\sum_z \eta_{yz} e^{i p\cdot z}$ is the solution of 
\begin{equation}
\MFPlat[U]_{xy} \overline{\eta}_y = e^{i p\cdot x },
\end{equation}
and 
\begin{equation}
\overline{\eta}_y = \frac{1}{V}\sum_z \eta_{yz} e^{i p\cdot z} = 
\frac{1}{V} \sum_z c(y)\overline{c}(z)e^{i p\cdot z} \equiv c(y) \widetilde{\overline{c}}(p).
\end{equation}
Doing a summation on $y$ we obtain a Fourier transform of the field $c(y)$:
\begin{equation}
\frac{1}{V}\sum_{y }\overline{\eta}_y = \widetilde{c}(0) \widetilde{\overline{c}}(p).
\end{equation}
But the last equation expresses the orthogonality condition (\ref{eq:kernel}). Thus we find
$\widetilde{c}(0) \widetilde{\overline{c}}(p) = 0$ in this case. That means that (\ref{ghost_gluon_Green_function})
is also zero, and the vertex function cannot be directly extracted (on the lattice) in the kinematic configuration 
with vanishing ghost momentum $\widetilde{\rm MOM}_{c0}$, but only in the $\widetilde{\rm MOM}_{c}$ scheme.

%
%
%
\section{Errors of the calculation}
%
%
\label{ErrorOfTheCalculation}

There are three main sources of errors when calculating Green functions on the lattice: the statistical errors of the
Monte-Carlo method, the systematic bias coming from the space-time discretisation and finally the error due 
to the gauge fixing (influence of  lattice Gribov copies). The last is discussed in the following section.

%
\subsection{Estimating the statistical error}
%
\label{Jackknife_subsection}

The gauge-field configurations produced via the Monte-Carlo generation process (see \cite{Montvay:1994cy} for a review)
are not completely decorrelated. However, the residual correlations may be neglected.
Nevertheless, all data points (as function of momentum $p$) of a Green function are
calculated on the same set of gauge configurations $\{C_i\}$,  and in this sense they 
are not independent. This problem arises when calculating quantities involving functions 
of mean values, like the coupling constant in a MOM scheme. In order to take in account 
the bias induced by this correlation one uses a special method 
(called \textbf{Jackknife}~\cite{Jackknife1},~\cite{Jackknife2}) of computation of the error.
Generally speaking this method is a standard bootstrap method (of the estimation of the 
variance in the case of a non-Gaussian distribution) based on a resampling with 
replacement from the original sample. We start with a Monte-Carlo sample of size $M$.
Our purpose is to calculate the error on the estimation of the mean of this sample.
We divide it into $\left[M/m\right]$ groups of $m$ elements:
\begin{equation}
\Big[\mathcal{O}(C_1), \ldots ,\mathcal{O}(C_m) \Big] 
\quad
\Big[\mathcal{O}(C_{m+1}), \ldots ,\mathcal{O}(C_{2m})\Big] 
\quad \ldots\quad 
\Big[\ldots ,\mathcal{O}(C_M) \Big].
\end{equation}
Next one defines the partial averages
\begin{equation}
a_k=\frac{\sum_{i=1}^{M}\mathcal{O}(C_i) - \sum_{i=km}^{(k+1)m}\mathcal{O}(C_i)}{M - m}, \quad k=1 \ldots 
\widetilde{M}=\left[\frac{M}{m} \right],
\end{equation}
and finally obtains the following expression for the error:
\begin{equation}
\Delta_{\text{jackknife}}\langle\mathcal{O}\rangle = 
\sqrt{ \frac{ \widetilde{M}-1}{\widetilde{M}} 
\left(\sum_{k=1}^{\widetilde{M}}a^2_k - 
\frac{\left(\sum_{k=1}^{\widetilde{M}}a_k\right)^2}{\widetilde{M}}\right) }.
\end{equation}
This analytical expression differs from the standard formula for the dispersion of the mean 
value by an additional factor $\sim\widetilde{M}$.

%
\subsection{Handling the discretisation errors}
%
\label{subsection_discretisation_errors}

Because of discretisation of the space-time, lattice theory~\footnote{we suppose that the lattice is hypercubic.} looses the rotational
symmetry $SO(4)$ inherited from the Lorenz invariance in Minkowski space.
This symmetry is replaced by a discret isometry group 
$H_4=S_4 \ltimes P_4$ (semiproduct of the permutation and reflection groups) having
$4!\cdot 2^4 = 384$ elements. A generic scalar function $\widehat{G}(p)$ extracted form Green 
functions is thus invariant along the orbit $O(p)$ generated by the action of the group $H_4$
on the components of the lattice momentum $p\equiv\frac{2\pi}{La}\times(n_{1},n_{2},n_{3},n_{4})$. 
It may be proven in the theory of group invariants that each orbit is characterised 
by four group invariants
\begin{equation}
p^{[n]} = a^{n}\sum_{\mu}p_{\mu}^{n}, \quad n = 2, 4, 6, 8
\end{equation}
One may average on the gauge orbit of the $H_4$ group in order to increase the statistics:
\begin{align}
\label{generic_green}
  a^{2}G_{L}(p^{[2]},p^{[4]},p^{[6]},p^{[8]}) =
  \frac{1}{\text{card}\,O(p)} \sum_{p\in O(p)} \widehat{G}(p),
\end{align}
where $\text{card}\,O(p)$ is the number of elements in the orbit  $O(p)$. The resulting average is a function 
of the four invariants $p^{[n]}$. But in the continuum limit any scalar function 
is a function of the rotational invariant $p^{[2]}$. We will explain how it is possible 
to remove the dependence on the three other invariants with the example
of the free lattice gluon propagator:
\begin{align}
  \label{eq:free}
  G_{0}(p) = \frac{1}{\sum_{\mu}\widehat{p}_{\mu}^2}, 
  \quad\mathrm{where}
  \quad\widehat{p}_{\mu} =
  \frac{2}{a}\sin\left(\frac{ap_{\mu}}{2}\right).
\end{align}
If all the components of the lattice momentum verify the condition $ap_{\mu}\ll 1$, 
then $\sum_{\mu}\widehat{p}_{\mu}^2 \simeq p^2-\frac{1}{12}a^2p^{[4]}$, and thus one has up to
terms of order $\sim a^4$
\begin{equation}
G_{0}(p) =\frac{1}{p^2}\left( 1 + \frac{1}{12}\frac{a^2 p^{[4]}}{p^2} + O(a^4)\right) .
\end{equation}
So, taking the continuum limit $a\rightarrow 0$ is equivalent to taking the limit $p^{[4]}\rightarrow 0$. We apply this idea to  
(\ref{generic_green}). Making the reasonable hypothesis of regularity near the continuum limit, we expand:
\begin{align}
  \label{eq:invariants}
  G_{L}(p^{[2]},p^{[4]},p^{[6]},p^{[8]}) \approx 
  G_{L}(p^{[2]},0,0,0) + p^{[4]}\frac{\partial G_{L}}
  {\partial p^{[4]}}(p^{[2]},0,0,0) + \cdots
\end{align}
and $G_{L}(p^{[2]},0,0,0)$ is nothing but the scalar factor in the continuum in a finite volume, up to lattice artifacts which do not
break $O(4)$ invariance. When several orbits exist with the same $p^{2}$, we can remove an important part of the hypercubic artifacts by
extrapolating the lattice data towards $G_{L}(p^{[2]},0,0,0)$ using a linear regression with respect to $p^{[4]}$ 
(the other invariants are of higher order in $a$). 

%

\begin{figure}[ht]
\begin{center}
\includegraphics[width=15cm]{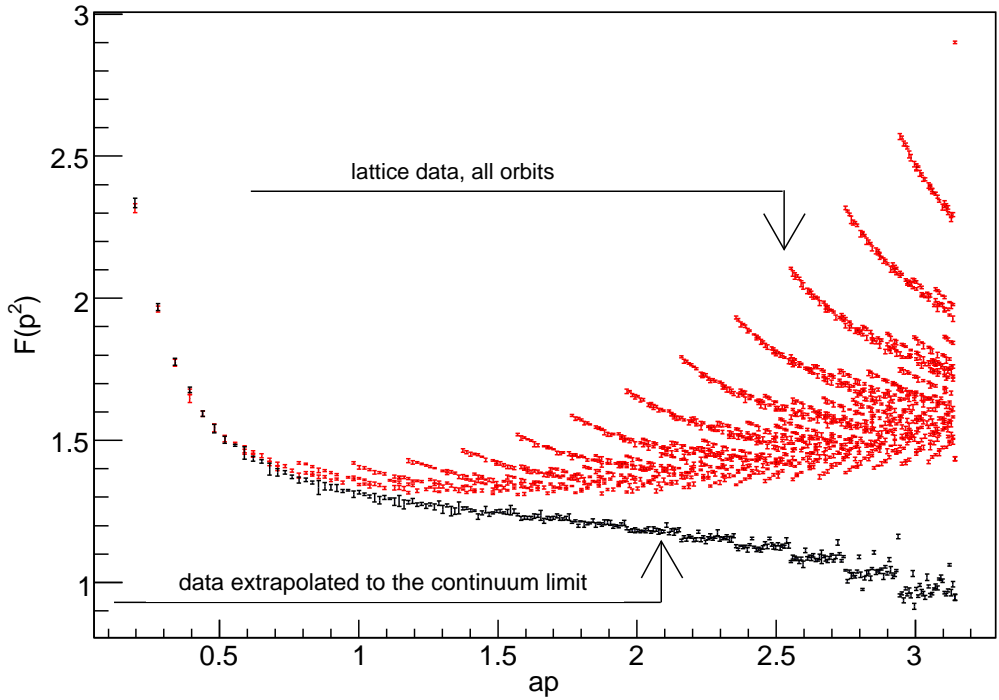}
\end{center}
\caption{\footnotesize\it Example of the extrapolation in $p^{[4]}$ for the ghost scalar factor $F(p)$ in the case of  the $SU(3)$ gauge group for
the lattice volume $32^4$ and at $\beta=6.4$. Red round dots correspond to the bare data, and blue squares to the 
extrapolated one. In practice we do not consider the moments above $ap=\frac{\pi}{2}$.}
\label{Orbits_Extrapolation}
\end{figure}
%

There are vectors $p$ that have only one orbit. In order to include them in the data analysis 
one should interpolate the slopes obtained in the extrapolation of  (\ref{eq:invariants}).
This can be done either numerically or by assuming a functional dependence of
the slope with respect to $p^{2}$. In principle the $p^2$ dependence of the slope may be
calculated using lattice perturbation theory. A dimensional analysis suggests 
the form $\sim\frac{c_1}{{\left(p^2\right)}^2}$. We have used the function
\begin{align}
  \label{eq:slope}
  \frac{\partial G_{L}}
  {\partial p^{[4]}}(p^{[2]},0,0,0) &=
 \frac{1}{\left(p^{[2]}\right)^{2}}\left( c_{1}+ c_{2}p^{[2]}\right)
\end{align}
with two fit parameters in order to fit the slopes.  The validity of the exposed method is 
qualitatively checked by the smoothness of the resulting curve. At Figure~\ref{Orbits_Extrapolation}
we present an example of removing the hypercubic artifacts. We see that the method works very well, even
at large values of $ap$. In practice we do not consider the momenta above $ap=\frac{\pi}{2}$.

%
\section{Gribov ambiguity and lattice Green functions}
%
\label{section_Gribov_copies_sur_reseau}

\begin{figure}[h]
\begin{center}
\includegraphics[width=0.5\linewidth]{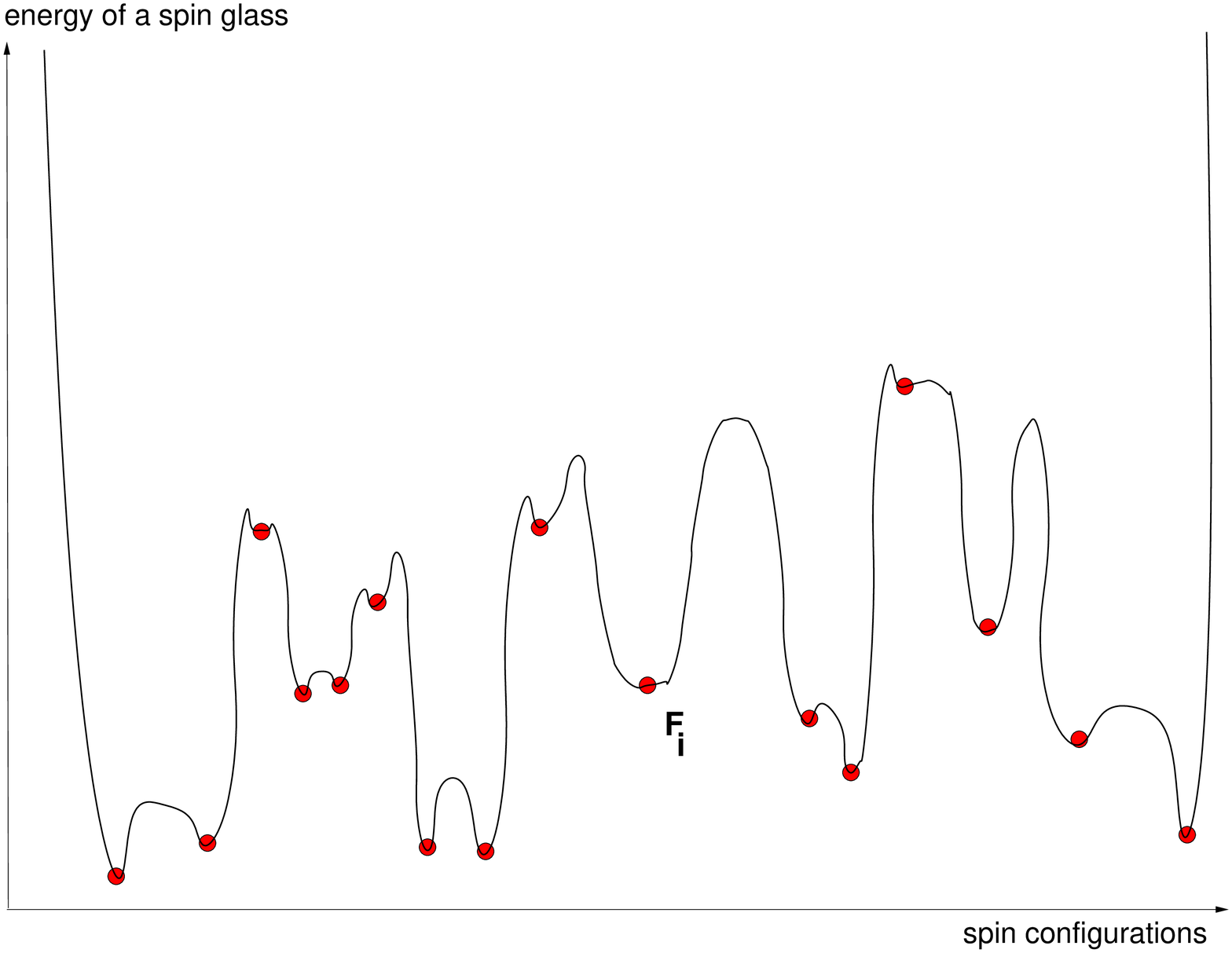}
\caption{\footnotesize\it Landscape of minima for a generic spin glass. Circles mark the configurations
corresponding to metastable states. Their number grow like $\exp{\left( \text{const} \cdot N \right)}$
as function of the number of spins $N$.}
\label{spinglass_figure}
\end{center}
\end{figure}

We have discussed in the previous sections different uncertainties introduced by the discretisation
of space-time. Another bias comes from the gauge fixing (see section~\ref{subsection_Gribov_ambiguity_general}).
As we have already mentioned, the Minimal Landau gauge fixing quantisation is equivalent to
realising the Gribov quantisation prescription. Thus lattice Green functions are calculated
within this prescription. However, the gauge is not fixed in a unique way. 
The Gribov ambiguity on the lattice shows up by the non-uniqueness of 
the minimum (\ref{GF_minimum_condition}) of the functional (\ref{LatticeGaugeMinimisingFunctional}).
Indeed, the functional (\ref{LatticeGaugeMinimisingFunctional}) has a form similar to 
the energy of a spin glass which is known to have exponentially-many metastable 
states (cf. Figure~\ref{spinglass_figure}). All these lattice Gribov copies are situated 
inside the Gribov horizon. On a finite lattice one can (in principle) fix the gauge in a unique way.
%
%
However this is very expensive in terms of computer time. Nevertheless one has to understand
the influence of the choice of the minimum of the functional (\ref{LatticeGaugeMinimisingFunctional})
on the mean values yielding Green functions. For this purpose we studied the landscape of
minima of the gauge-fixing functional~\cite{Lokhov:2005ra}. In the following subsection we 
define a specific probability to find a Gribov copy, as function of the physical momentum. 
In the next-to-the-following subsection we discuss the influence of these copies on the
Green functions, and the Zwanziger's conjecture on the equivalence of the integration
over the Gribov region and the fundamental modular region in the infinite volume limit.

%
\subsection{The landscape of minima of the gauge-fixing functional}
%
\label{subsection_landscape_of_minima_of_the_gauge_fixing_functional}

Let us consider the landscape of the functional $F_U$. One of its
characteristics is  the distribution of values at minima $F_{\text{min}}$ of
$F_U$. We know that for small magnitudes  of the gauge field all the link
matrices $U_\mu(x)$ (they play the role of couplings between the ``spin" 
variables) are close to the identity matrix, and thus the minimum is
unique.  Their number increases when the bare lattice coupling $\beta$
decreases, because the typical  magnitude of the phase of $U_\mu(x)$ grows
in this case and thus link  matrices move farther from the identity matrix.
The number of minima also increases with the number of links (at fixed
$\beta$) because in this case there are more degrees of freedom in the
system.
%
\begin{figure}[!ht]
\begin{center}
\begin{tabular}{lr}
\includegraphics[width=0.5\linewidth]{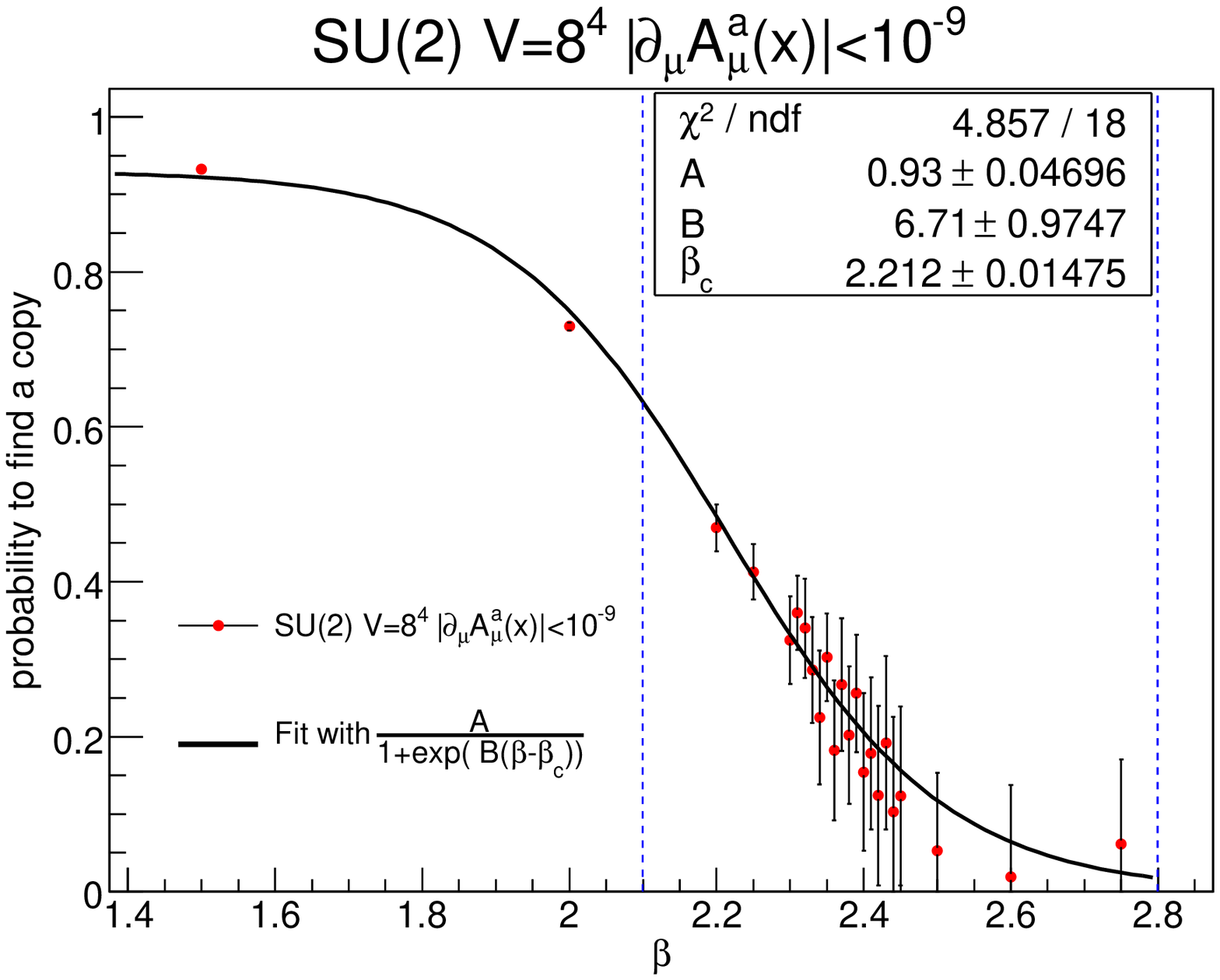}
&
\includegraphics[width=0.5\linewidth]{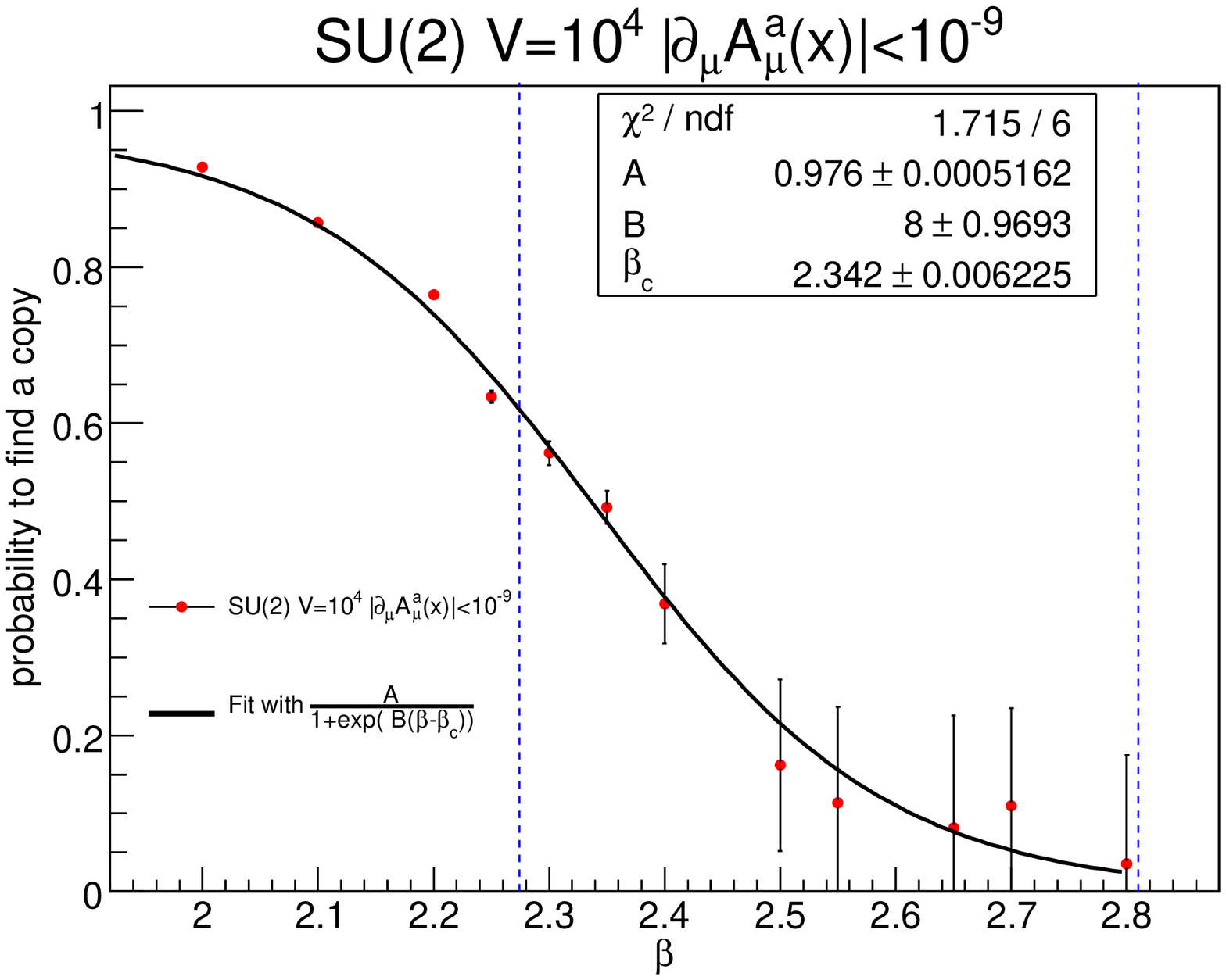}
\\
\includegraphics[width=0.5\linewidth]{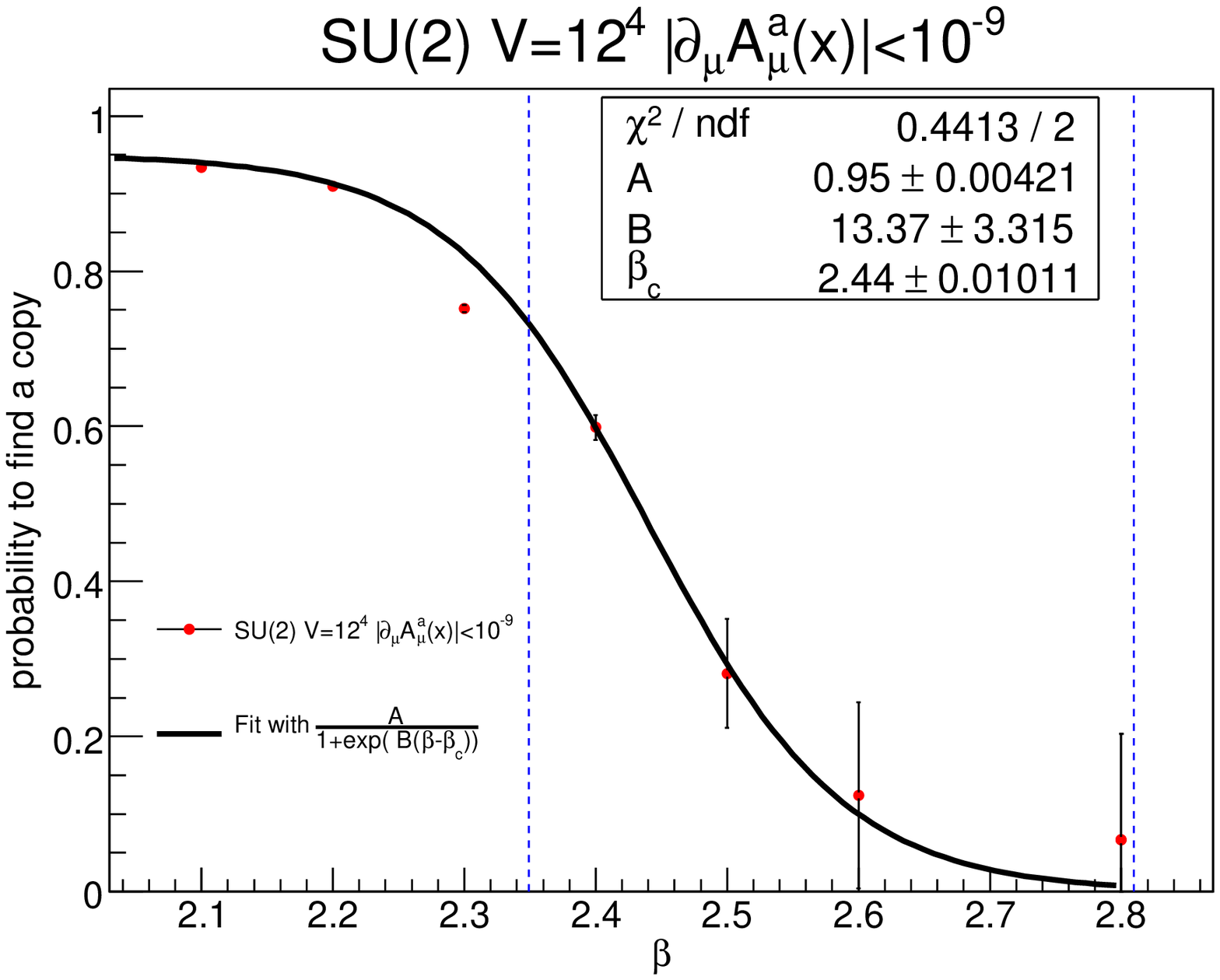}
&
\includegraphics[width=0.5\linewidth]{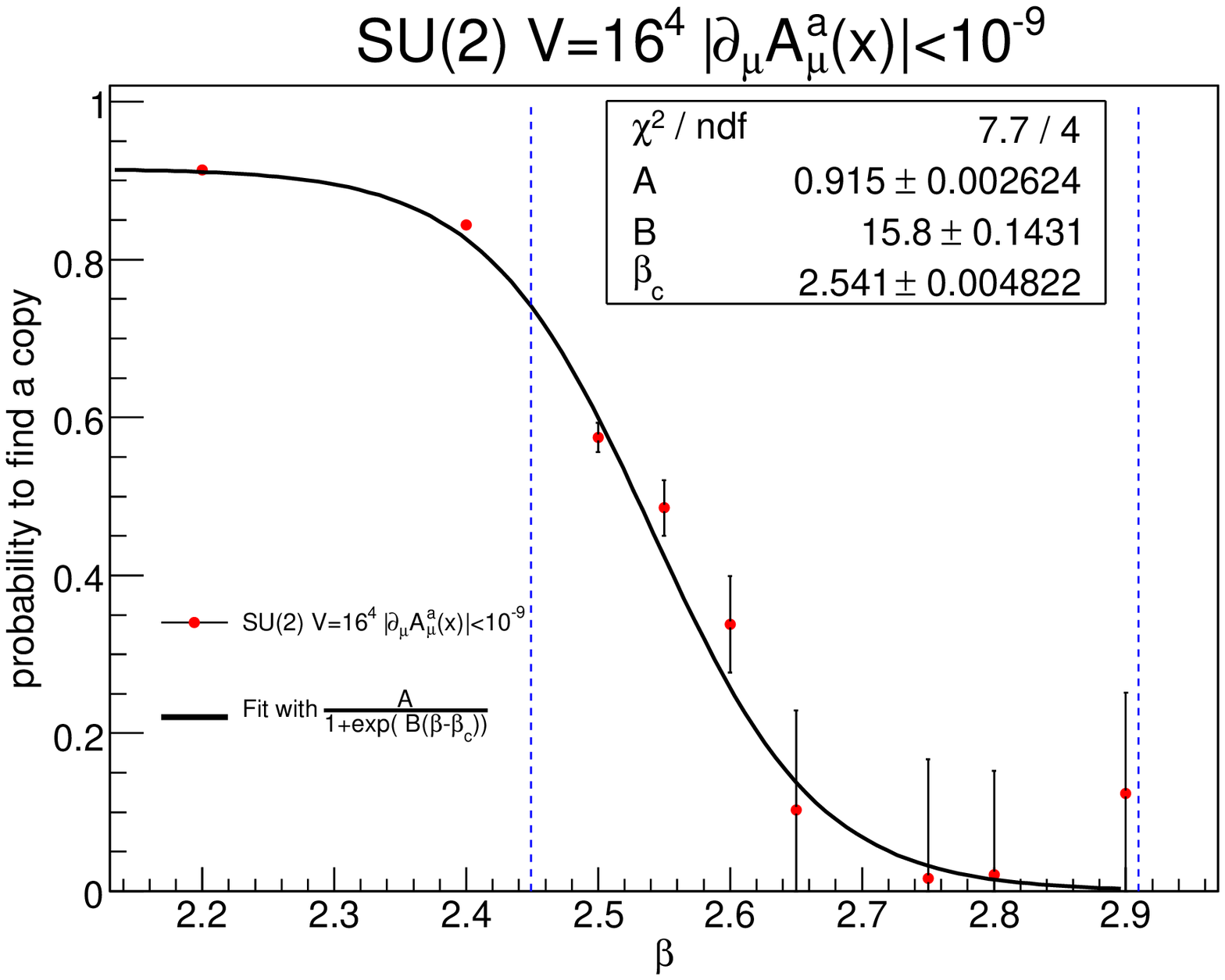}
\end{tabular}
\caption{\footnotesize\it Probability (averaged over gauge orbits) to find a secondary minimum as a function 
of $\beta$ at volumes $V=8^4,10^4,12^4,16^4$. Solid line represents a fit with an empirical formula 
(\ref{formule_empirique_proba}). The vertical dashed lines delimit the window of each fit.
}
\label{PROBA_GRIBOV}
\end{center}
\end{figure}
%

We can define a probability to find a secondary minimum, as a function of the $\beta$ parameter. 
For each orbit we fix the gauge $N_\text{GF}$ times, each gauge fixing starts after a (periodic) random gauge transformation 
of the initial field configuration. We thus obtain a distribution of minima $F_\text{min}$.
This distribution gives us the number of minima $N(F^{i})$ as a function of 
the value of $F^{i} \equiv F^{i}_{\text{min}} $. The relative frequency of a minimum $F^i$ is defined by 
\begin{equation}
\omega_i = \frac{N(F^{i})}{\sum_{i} N(F^{i}) },
\end{equation}
where $\sum_{i} N(F^{i})  = N_\text{GF}$. Then the weighted mean number of copies per value of $F_\text{min}$ is 
given by
\begin{equation}
\overline N = \sum_i \omega_i N(F^{i}).
\end{equation}
This allows us to define a probability to find a secondary minimum when
fixing the Minimal Landau gauge for a given gauge field configuration~:
\begin{equation}
p_{\text{1conf}} = 1 - \frac{\overline N}{\sum_{i} N(F^{i}) }.
\end{equation}
If one finds the same value of $F_\text{min}$ for all $N_{\text{GF}}$ tries then this probability is zero.
On the contrary, if all $F_i$ are different then $p_{\text{1conf}}$ is close to one.

Having the probability to find a secondary minimum when fixing the gauge for
\emph{one particular} configuration we can calculate the Monte-Carlo
\emph{average} $\left\langle \bullet \right\rangle$ on gauge orbits, i.e. on 
``spin couplings'' $U_\mu(x)$. We obtain finally the overall probability
to find a secondary minimum during a numerical simulation at given $\beta$:
\begin{equation}
\label{proba_copy}
P(\beta)\equiv  \left\langle p_{\text{1conf}} \right\rangle_{\{U_\mu(x) \}} = 
1 - \left\langle \frac{\overline N}{\sum_{i} N(F^{i}) } \right\rangle_{\{U_\mu(x) \}}.
\end{equation}

We have performed simulations in the case of $SU(2)$ latice gauge theory at
volumes $V=\{8^4, 10^4, 12^4, 16^4\}$  for $\beta$ varying from $1.4$ to
$2.9$. For each value of $\beta$ we generated $100$ independent Monte-Carlo 
gauge configurations, and we fixed the gauge $N_\text{GF} = 100$ times for
every configuration. Between each gauge fixing a random gauge transformation
of the initial gauge configuration was performed, and the minimising
algorithm stops  when the triple condition (\ref{StopParameters}) is
satisfied. Examples of the resulting probability to find a secondary minimum are presented at Figure~\ref{PROBA_GRIBOV}.

As expected, the probability is small when $\beta$ is large, and it is close
to one when $\beta$ is small. The dispersion was calculated using the
Jackknife method (discussed in the subsection~\ref{Jackknife_subsection}). 
The physical meaning of this dispersion is the following:
when the error is small, all gauge configurations have a similar number of
secondary minima. On the contrary, this dispersion is large if there are
some exceptional gauge configurations having a different number of copies.
At small $\beta$ almost all gauge configurations have many secondary minima,
that is why the dispersion of the probability is small. At large $\beta$
almost all gauge configurations  have a unique minimum, but some of them can
have copies. This may considerably increase the dispersion of the
probability. The appearance  of exceptional gauge configuration possessing a
large density of close-to-zero eigenvalues of the Faddeev-Popov operator has
been recently reported~\cite{Sternbeck:2005vs}. Probably these fields are
related  with those having a  lot of secondary minima at large $\beta$'s,
and this correlation deserves a separate study.

We can fit our data (Figure~\ref{PROBA_GRIBOV}) for the probability (\ref{proba_copy}) 
with an empirical formula
\begin{equation}
\label{formule_empirique_proba}
P(\beta) = \frac{A}{1+e^{B(\beta - \beta_c)}}
\end{equation}
in order to define a characteristic coupling $\beta_c$ when the probability
to find a copy decreases considerably. One can define $\beta_c$ as
corresponding to the semi-heights of the probability function $P(\beta)$. At
this value of $\beta$ an equally probable secondary attractor of the
functional $F_U$ appears. The fit has been performed for the points between
dashed lines at Figure~\ref{PROBA_GRIBOV}, and the
results for the fit parameters are given ibidem and in Table \ref{lambda_c}.
We see that $\beta_c$ depends  on the volume of the lattice. Let us check
whether these values correspond to some physical scale. According to works
\cite{Bloch:2003sk},\cite{Fingberg:1992ju} one has the following expression
for the string tension $\sigma$ for $\beta\ge 2.3$:
\begin{equation}
\label{Spacing}
[\sigma a^2] (\beta) \simeq
 e^{-\frac{4\pi^2}{\beta_0}\beta + \frac{2\beta_1}{\beta_0^2}\log{\left( \frac{4\pi^2}{\beta_0}\beta \right)} + \frac{4\pi^2}{\beta_0}\frac{d}{\beta} + c}
\end{equation}
with $c= 4.38(9)$ and $d=1.66(4)$~\footnote{in this cited formula we kept the original convention for the RG
$\beta$-function, namely $\beta_0=11/3N_C,\,\beta_1=17/3N_C^2$. They are different from the one we use in the following.}.
Using this formula,  we define a characteristic scale corresponding to the critical values $\beta_c$ from Figure~\ref{PROBA_GRIBOV} :
$$
\lambda_c = a(\beta_c)\cdot L
$$
in the string tension units, $L$ is the length of the lattice. In the last column of the Table \ref{lambda_c} we summarise the results.
\begin{table}[h]
\centering
\begin{tabular}{l|l|c|c}
\hline\hline 
$L$           &  $\beta_c$  & $\chi^2/\text{ndf}$ & $\lambda_c$, in units of $1/\sqrt{\sigma}$ 
\\ \hline
$8$           &  $2.221(14)$ & $0.27$ & $3.85(31)$ 
\\ \hdashline[0.4pt/1pt]
$10$          & $2.342(6)$   & $0.28$ & $3.20(21)$ 
\\ 
$12$          & $2.44(1)$    & $0.22$ & $2.78(20)$
\\ 
$16$          & $2.541(5)$   & $1.92$ & $2.68(17) $
\\
\hline\hline 
\end{tabular}
\caption{\footnotesize\it The characteristic length defining the appearance of
secondary minima. The errors for $\lambda_c$  include errors for $d$ and $c$
parameters, and the fitted error for $\beta_c$}
\label{lambda_c}
\end{table}
We see that for the values of $\beta_c$ in the scaling regime, when the
formula (\ref{Spacing}) is applicable ($\beta \ge 2.3$), we obtain compatible
values for the physical length $\lambda_c$. 

This suggests that in the case of $SU(2)$ gauge theory lattice Gribov 
copies appear when the physical size of the lattice exceeds 
a critical value around $2.75/\sqrt{\sigma}$. At first
approximation $\lambda_c$ is scale  invariant, but a slight dependence
in the lattice spacing remains.

In principle, the parameter $\beta_c$ can be calculated with good precision.
One should do it in the case of $SU(3)$ gauge  group, because the dependence
of the lattice spacing on the bare coupling is softer, and the scaling of
the theory  has been better studied than in the case of the $SU(2)$ theory.

A natural question that arises after the study of the distribution of
$F_\text{min}$ is whether the gauge configurations having the same value of
$F_\text{min}$ are equivalent, i.e. they differ only by a global gauge
transformation.
This can be checked by calculating the two-point gluonic correlation function on the gauge
configuration. Indeed, according to the lattice definition of the gauge field that we used (\ref{U_to_A_conversion}),
\begin{equation}
G^{(2)}_{\text{1 conf}}(x-y) \propto
\tr  \left[ \left(U_\mu(x) - U_\mu^\dagger(x)\right) \cdot \left(U_\nu(y) -
 U_\nu^\dagger(y) \right) \right].
\end{equation}
Applying a global gauge transformation $U_\mu(x)\rightarrow V U_\mu(x)
V^\dagger$ we see that the  gluon propagator remains unchanged. This is also
the case of the ghost propagator scalar function. We have checked
numerically that the values of the gluon and the ghost propagators in
Fourier space  are the same for gauge configurations having the same
$F_\text{min}$. Thus we conclude that gauge configurations having the same $F_\text{min}$
are in fact equivalent~\cite{Lokhov:2005ra}.

%
\subsection{Lattice Green functions and the Gribov ambiguity}
%
\label{subsection_Lattice_Green_functions_and_Gribov}

As we have mentioned in the subsection~\ref{subsection_Gribov_ambiguity_general}, 
there is a conjecture~\cite{Zwanziger:2003cf} saying that in the infinite volume limit the 
expectation values calculated by integration over the Gribov region or 
the fundamental modular region become equal. Let us briefly recall the argument given in ~\cite{Zwanziger:2003cf}.
Let $B_\mu$ denote a field lying on the Gribov horizon:
\begin{equation}
\left\{
\begin{array}{l}
\d_\mu B_\mu = 0 \\
\MFP(B)\omega_0=0,\qquad\omega_0(x)\neq\text{const},\quad\omega_0(x)\in\mathfrak{su}(N_c).
\end{array}
\right.
\end{equation}
We write $F_B(t,\omega)$ for the functional (\ref{GaugeMinimisingFunctional}), with $u(x)=\exp{t\omega(x)}$. On the boundary one has
\begin{equation}
\label{F_B_estimations}
\left\{
\begin{array}{l}
F_B^\prime(0,\omega_0) = 0 \\
F_B^{\prime\prime}(0,\omega_0) = 0 \\
F_B^{\prime\prime\prime}(0,\omega_0) \neq 0 \propto \sqrt{V} \quad  \\
F_B^{\prime\prime\prime\prime}(0,\omega_0) = \frac{3}{2} \int d^4x \left[ d_\mu(\omega_0^2) \right]^2 \propto V
\end{array}
\right.
\end{equation}
So for small variations of $t$ one has for the functional (\ref{GaugeMinimisingFunctional})
\begin{equation}
F_B(t) = F_B(0) + \frac{1}{3!}F_B^{\prime\prime\prime}(0) t^3 + \frac{1}{4!}F_B^{\prime\prime\prime\prime}(0) t^4,
\end{equation}
and the last expression is minimised for $t=\bar{t}\equiv -3 F_B^{\prime\prime\prime}(0)/F_B^{\prime\prime\prime\prime}(0)$.
The value that is achieved at this secondary minimum inside the Gribov horizon is 
\begin{equation}
F_B(\bar{t}) = F_B(0) - \frac{9}{8} \frac{\left[ F_B^{\prime\prime\prime}(0) \right]^4}{\left[F_B^{\prime\prime\prime\prime}(0 \right]^3}.
\end{equation}
Using the estimations (\ref{F_B_estimations}) and the fact that the third order derivative appears in an even power,
one sees that the secondary minimum is \emph{lower} than the one on the boundary, and the 
difference between them \emph{decreases} with the volume $V$. The corresponding gauge 
configuration can be written as
\begin{equation}
\label{B_modifiee}
B_\mu(x,\bar{t}) = B_\mu(x,0) + \bar{t}\left[ D_\mu(B)\omega_0 \right](x) = 
B_\mu(x,0) -3 \frac{F_B^{\prime\prime\prime}(0)}{F_B^{\prime\prime\prime\prime}(0)}\left[ D_\mu(B)\omega_0 \right](x).
\end{equation}
When the lattice volume $V$ is large, the integration (over $G$ or $\Lambda$) on the gauge configurations is dominated by a
small shell near the boundary, because the dimension of the configuration space is large ($2dV(N_C^2-1)$).
According to the above argument the configurations near the boundaries $\partial G$
and $\partial \Lambda$ draw together. So, having (\ref{B_modifiee}) for a typical 
gauge field configuration, it is natural to suppose
that in the infinite volume limit the average calculated by integration over the domains $G$ or $\Lambda$ become equal.
However at a finite lattice this is clearly not the case, that is  why it is very important to know what 
is the influence of lattice Gribov copies on the Green functions. This question that has already been considered by
different authors (\cite{Cucchieri:1997dx},\cite{Bakeev:2003rr},\cite{Sternbeck:2004qk},%
\cite{Sternbeck:2004xr},\cite{Silva:2004bv},\cite{Sternbeck:2005tk},\cite{Sternbeck:2005vs}
). To check the dependence of Green functions on the procedure of the choice of the minimum
we adopted the same strategy as in above citations~: for every of 
the $100$ gauge configurations used to compute Green functions the gauge was
fixed $100$ times (a periodic random gauge transformation is done after each gauge fixing). 
The Monte-Carlo average  was computed with respect to
the ``first copy'' (fc) found by the minimisation algorithm and the ``best
copy'' (bc), having the smallest value of $F_\text{min}$. We have calculated
the gluon and  the ghost propagators, and also the three-gluon Green
functions in symmetric and  asymmetric kinematic configurations. The
simulations have been performed in the case of the $SU(2)$ group 
on lattices of volumes  $8^4$ and $16^4$ for $\beta=2.1, 2.2, 2.3$. 
According to the results of the subsection~\ref{subsection_landscape_of_minima_of_the_gauge_fixing_functional},
at these values of $\beta$ we are sure to have lattice Gribov copies.
We conclude~\cite{Lokhov:2005ra} that no systematic effect could be found for gluonic two- and 
three-point Green functions, the Monte-Carlo average values in the cases of (fc) 
and (bc) being compatible within the statistical errors. However the ghost 
propagator is quite sensitive to the choice of the minimum - in the case 
of (bc) the infrared divergence is lessened. 
\begin{table}
\centering
\begin{footnotesize}
\begin{tabular}{ccccc}
\hline\hline
$\beta$ & $L$ & $n^2$ & $F^{(2)}_{\text{fc}}(p^2)-F^{(2)}_{\text{bc}}(p^2)$ & $\frac{F^{(2)}_{\text{fc}}(p^2)-F^{(2)}_{\text{bc}}(p^2)}{F^{(2)}_{\text{bc}}}$
\\ \hline
$2.1$ & $8~$ &$1$ & $0.211$ & $0.045$ 
\\
      &      &    &    $\vee$     & 
\\ 
$2.1$ & $16$ &$4_{~n^{[4]}=16}$ & $0.145$ & $0.033$ 
\\ \hdashline[0.4pt/1pt]
$2.2$ & $8~$ &$1$ & $0.078$ & $0.019$ 
\\ 
      &      &    &    $\vee$     & 
\\
$2.2$ & $16$ &$4_{~n^{[4]}=16}$ & $0.023$ & $0.006$ 
\\ \hdashline[0.4pt/1pt]
$2.3$ & $8~$ &$1$ & $0.086$ & $0.024$ 
\\ 
      &      &    &    $\wedge$     & 
\\
$2.3$ & $16$ &$4_{~n^{[4]}=16}$ & $0.114$ & $0.034$ 
\\ \hline\hline
\end{tabular}
\caption{\footnotesize\it Volume dependence of the ghost propagators~\cite{Lokhov:2005ra}, $p_\mu=\frac{2\pi}{aL}n_\mu$ }
\label{bc_fc}
\end{footnotesize}
\end{table}
This dependence has been found to decrease slowly with the volume~\cite{Sternbeck:2005tk}.
The results of the subsection~\ref{subsection_landscape_of_minima_of_the_gauge_fixing_functional} 
indicate that the convergence can happen only beyond the critical volume $\lambda_c^4$.
To check this we compare the fc/bc values of the ghost propagator, at one 
physical value of the momentum, for the orbit $n^2=1$ on
a $8^4$ lattice and the orbit $n^2=4, n^{[4]}=16$ on the $16^4$
lattice~\footnote{Remember that the momentum in physical units is equal to 
$2\pi\,n /(La)$ in our notations.} at the same $\beta$ (see Table~\ref{bc_fc}).
It happens indeed that the decrease is observed only at $\beta=2.1$ and $2.2$,
in accordance with Table~\ref{lambda_c}. However, these values of $\beta$ are
not in the scaling regime, and thus a study on larger lattices would be
welcome. It is not surprising that the ghost propagator depends on the bc/fc
choice: the (bc) corresponds to the  fields further from the Gribov horizon where
the Faddeev-Popov operator has a zero mode, whence the  inverse Faddeev-Popov
operator (ghost propagator) is expected to be smaller as observed.  The
correlation between the bc/fc choice and the gluon propagator is not so direct.
\begin{table}
\centering
\begin{footnotesize}
\begin{tabular}{cc|cc}
\hline \hline\\
$\beta$ & $L$ & $\overline{\langle F_\text{min} \rangle}_{\{U\}}$ & $\delta\overline{\langle F_\text{min} \rangle}_{\{U\}}$
\\ \hline
$2.2$ & $8~$ & $-0.8236$ & $0.003744$ 
\\ 
      & $10$ & $-0.8262$ & $0.002367$ 
\\ 
      & $12$ & $-0.8272$ & $0.001377$ 
\\ 
      & $16$ & $-0.8279$ & $0.000802$ 
\\ \hdashline[0.4pt/1pt]
$2.4$ & $8~$ & $-0.8642$ & $0.005270$ 
\\ 
      & $12$ & $-0.8669$ & $0.002739$ 
\\ 
      & $12$ & $-0.8686$ & $0.001849$ 
\\ 
      & $16$ & $-0.8702$ & $0.001003$ 
\\ \hline\hline
\end{tabular}
\caption{\footnotesize\it Volume dependence of the Monte-Carlo+gauge orbit mean value at minima $F_\text{min}$ and the dispersion of this mean.} 
\label{Fmin_mean_volume_dep}
\end{footnotesize}
\end{table}
Another quantity is obviously strongly correlated to the bc/fc choice:
the value of $F_\text{min}$. We tested the volume dependence of the 
Monte-Carlo+gauge orbit mean value of the quantity $F_\text{min}$  (see
Tab.\ref{Fmin_mean_volume_dep}). According to the argument given
in~\cite{Zwanziger:2003cf}, all minima  become degenerate in the infinite
volume limit, and closer to the absolute minimum (in the fundamental modular
region).  We see from the Table \ref{Fmin_mean_volume_dep} that their 
average value and dispersion decrease with  the volume at fixed $\beta$, 
in agreement with~\cite{Zwanziger:2003cf}. 

\vspace*{1cm}
\noindent We finish this section with a brief summary of our results~\cite{Lokhov:2005ra} regarding the Gribov
ambiguity on the lattice:
\begin{enumerate}
\item 
Lattice Gribov copies appear and their number grows very fast when the
physical  size of the lattice exceeds some critical value ($\approx 2.75/\sqrt{\sigma}$ in
the case of the $SU(2)$ theory). This result is fairly independent of the lattice spacing.

\item The configurations lying on the same gauge orbit and having the same
$F_\text{min}$ are equivalent, up to a global gauge transformation, and yield the same Green functions.
Those corresponding to minima of $F_{U}$ with different values of $F_\text{min}$ differ 
by a non-trivial gauge transformation, and thus they are not equivalent.

\item We confirm the result (\cite{Cucchieri:1997dx},\cite{Bakeev:2003rr},\cite{Sternbeck:2004qk},%
\cite{Sternbeck:2004xr},\cite{Silva:2004bv},\cite{Sternbeck:2005tk},\cite{Sternbeck:2005vs})
that the divergence of the ghost propagator is lessened when choosing the
``best copy" (corresponding to the choice of the gauge configuration having
the  smallest value of $F_{U}$). We also showed that gluonic Green functions
calculated in the  ``first copy'' and ``best copy'' schemes are compatible
within the statistical error, no systematic  effect was found (with periodic 
gauge tranformations).

\item We found that the influence of Gribov copies on the ghost propagator
decreases with the volume when the physical lattice size is larger  than the
critical length discussed above. We also show that  the quantity $F_\text{min}$
decreases when the volume increases. These two points are in agreement with the Zwanzigers's
argument~\cite{Zwanziger:2003cf} on the equality of the averages over the Gribov region and the
fundamental modular region.

\end{enumerate}
%
%
%
\chapter{The ultraviolet behaviour of  Green functions}
%
%
%
\label{chapter_UV_behaviour}

The Lagrangian of the Pure Yang-Mills theory in a four-dimensional space-time does not contain any dimensional 
parameters susceptible to fix an energy scale for dimensionless quantities. However, the spectrum of the 
corresponding quantum theory contains massive states (glueballs).  As a matter of fact, the quantum theory 
possesses a finite energy scale called $\Lqcd$, which is generated by the quantisation process followed by 
the renormalisation. All dimensionful physical quantities are expressed as multiples of powers of this scale, 
and thus it should be a renormalisation group invariant:
\begin{align}
\mu \frac{d}{d\mu}\Lqcd \left(\mu, g(\mu^2) \right) =0
\quad \rightarrow \quad
\left[  \mu \frac{\d}{\d \mu} + \beta \left( g(\mu^2) \right) \frac{\d}{\d g} \right] \Lqcd \left(\mu, g(\mu^2) \right) = 0,
\end{align}
where $\beta\left( g(\mu^2) \right)$ is the renormalisation group beta function and $\mu$ is the renormalisation scale. 
The solution of the above equation reads
\begin{equation}
\label{Lqcd_peturbative_integral}
\Lqcd \left(\mu, g(\mu^2) \right) = \mu \exp{\left( \displaystyle - \int^{g(\mu^2)}_{g_1} \frac{ d g^{\prime} }{\beta\left(g^{\prime}\right) } \right) },
\end{equation}
where $g_1$ is an arbitrary integration constant. $\Lqcd$ is a renormalisation scheme-dependent quantity,
although it is a renormalisation group invariant within one particular scheme. 
So it is not a real physical quantity. Still, its value is important for estimating the 
lowest bound of the domain of validity of perturbation theory. 
Knowing several first coefficients of the $\beta$-function we find from
the equation (\ref{Lqcd_peturbative_integral}):
\begin{equation}
\label{Lambda_pert_generic}
\Lqcd \left(\mu, g(\mu^2) \right) = \mu 
\exp\left[ \frac{1}{2\beta_0}\left( \frac{1}{g_1^2} - \frac{1}{g^2(\mu^2)} \right)
+\frac{\beta_1}{2\beta_0^2}\log\frac{g_1^2}{g^2(\mu^2)}
\right] + O(g^2).
\end{equation}
We see that there is an essential singularity when $g^2(\mu^2)\rightarrow 0$, and thus a perturbative
calculation of related quantities (for example, the string tension $\sqrt{\sigma} = c_\sigma \Lqcd$) 
is impossible. In the following sections we describe the method of calculation of $\Lqcd$
from lattice Green functions in Landau gauge. We start with a review of the purely perturbative results for Green
functions. The momentum range available on the lattice is situated at rather low energies where 
the non-perturbative power corrections are not negligible. The section~\ref{Green_OPE_section} is 
devoted to the estimation of the dominant power corrections. At the end of this chapter
we present the results of analysis of our lattice data.

%
\section{$\Lqcd$ and perturbative expressions for Green functions}
%

Different scalar factors of Green functions depend on the $\Lqcd$ parameter discussed in the previous 
subsection. These scalar factors can be calculated non-perturbatively in lattice simulations, and one can extract
$\Lqcd$ by fitting the lattice data in the ultraviolet domain to the corresponding 
perturbative formulae. Here we make a review of available perturbative Landau gauge
calculations for the ghost and gluon propagators in the MOM schemes.

If $\Gamma^{(N)}_R(p_i,g_R^2,\mu^2)$ is a renormalised proper vertex in Landau gauge, then the corresponding proper bare 
vertex function is independent of the renormalisation point $\mu$. This fact is reflected by the
Callan-Symanzik equation for the renormalised function:
\beq
\label{CALLAN_SYMANZIK}
\left( \frac{\d}{\d \ln{\mu^2} } + \beta\left( g_R(\mu^2) \right) \frac{\d}{\d g} - \frac{N}{2} \gamma\left(g_R(\mu^2)\right) \right)\Gamma^{(N)}_R(p_i,g_R^2,\mu^2) = 0
\eeq
where $\gamma\left(g_R(\mu^2)\right)$ is the anomalous dimension. 
In the Momentum subtraction schemes, the renormalisation conditions are defined by 
setting some of the two- and three-point functions to their tree-level values at the 
renormalisation point. Then (\ref{CALLAN_SYMANZIK}) simplifies to 
\beq
\label{gamma_3}
\lim_{a^{-1} \to \infty} \frac{d\ln(Z_{3,{\rm MOM}}(p^2=\mu^2,a^{-1})}{d\ln{\mu^2}} 
=\gamma_{3,{\rm MOM}}(g_{\rm MOM})   
\eeq
in the case of two-point Green functions, where $Z_3(\mu^2)$ is defined in (\ref{DEFINITION_Z3_Z3TILDE}),
$a^{-1}$ stands for the ultraviolet regularisation and $\gamma_{3,{\rm MOM}}(g_{\rm MOM})$ is the anomalous dimension.
A similar expression can be written for the ghost propagator renormalisation factor $\widetilde{Z_3}$.
As we have already seen in the section~\ref{GREEN_FUNCTIONS_IN_LANDAU_GAUGE},
there is an infinite number of MOM schemes differing by kinematic configurations at the substraction point.
We limit ourselves to the configurations defined by the subtraction of the transverse part 
of the three-gluon vertex ($\widetilde{\rm MOM}$) and that of the ghost-gluon vertex 
with vanishing gluon momentum ($\widetilde{\rm MOM}_{c}$) and vanishing incoming ghost momentum ($\widetilde{\rm MOM}_{c0}$),
discussed in the section~\ref{GREEN_FUNCTIONS_IN_LANDAU_GAUGE}.

Both anomalous dimensions for ghost and gluon propagators have been recently computed (\cite{Chetyrkin:2000dq},
\cite{Chetyrkin:2004mf}) in the $\ms$ scheme. The result at four-loop order reads
\begin{align}
\label{LnZ}  
  \begin{split}
 \frac{d\ln(Z_{3,\text{MOM}})}{d \ln \mu^{2}} &= 
 \frac{13}{2}\,h_{\ms} + \frac{3727}{24}\,h^{2}_{\ms} + 
 \left(\frac{2127823}{288} - \frac{9747}{16}\zeta_{3}\right) h^{3}_{\ms} \\
 &+ \left(\frac{3011547563}{6912} - \frac{18987543}{256}\zeta_{3} - 
   \frac{1431945}{64}\zeta_{5}\right) h^{4}_{\ms} 
  \nonumber 
  \end{split}
  \\
  \begin{split}
 \frac{d\ln(\widetilde{Z}_{3,\text{MOM}})}{d \ln \mu^{2}} &= 
 \frac{9}{4}\,h_{\ms} + \frac{813}{16}\,h^{2}_{\ms} + 
 \left(\frac{157303}{64} - \frac{5697}{32}\zeta_{3}\right) h^{3}_{\ms} \\
 &+ \left(\frac{219384137}{1536} - \frac{9207729}{512}\zeta_{3} - 
  \frac{221535}{32}\zeta_{5}\right) h^{4}_{\ms}   
  \end{split}
\end{align}  
where $h=g^2/(4\pi)^2$. In order to obtain the coefficients of the anomalous dimensions (\ref{gamma_3}) in 
a MOM scheme one has to express the above expressions in terms of the corresponding
coupling. We use the results of the article~\cite{Chetyrkin:2000dq} where the three-loop perturbative substraction 
of all the three-vertices appearing in the QCD Lagrangian for kinematic configurations with one 
vanishing momentum are given. In Landau gauge and in the pure Yang-Mills case one has the following 
relations between the couplings in different MOM schemes and $h_{\ms}$:
\begin{align}
\begin{split}
h_{\momg} = & h_{\ms} + \frac{70}{3} h^2_{\ms} + 
\left( \frac {51627}{576} - \frac{153}{4} \zeta(3) \right) h^3_{\ms} +
\\ & + \left(\frac{304676635}{6912} - \frac{299961}{64}\zeta_{3} -
                     \frac{81825}{64}\zeta_{5} \right)h^{4}_{\ms} \nonumber 
\end{split}
\\
\begin{split}
h_{\momc} = & h_{\ms} + \frac{223}{12} h^2_{\ms} + 
\left( \frac {918819}{1296} - \frac{351}{8} \zeta(3) \right) h^3_{\ms} + 
\\ & + \left(\frac{29551181}{864} - \frac{137199}{32}\zeta_{3} -
                    \frac{74295}{64}\zeta_{5} \right)h^{4}_{\ms}
\nonumber 
\end{split}
\\
\begin{split}
h_{\mom} = & h_{\ms} + \frac{169}{12} h^2_{\ms} + 
\left( \frac {76063}{144} - \frac{153}{4} \zeta(3) \right) h^3_{\ms} + 
\\ & + \left( \frac{42074947}{1728} - \frac{35385}{8}\zeta(3) - \frac{66765}{65}\zeta(5)  \right)
h^4_{\ms}.
\label{hmom}
\end{split}
\end{align}
Thus, inverting (\ref{hmom}) and substituting in (\ref{LnZ}), we obtain the anomalous dimensions  of the 
gluon and ghost propagator in the three above-mentioned renormalisation schemes. 
In a MOM scheme, the equations (\ref{LnZ}) may be integrated as functions of $h$ (cf. \ref{gamma_3}) \footnote{We omit the 
index specifying the renormalisation scheme both for $h$ and $\Lambda_{\text{QCD}}$ in the following formulae}:
\begin{align}
  \begin{split}
     \ln\left(\frac{Z_{\Gamma,\text{MOM}}}{Z_{0}}\right) & = 
     \log(h)\,\frac{\overline{\gamma}_{0}}{\beta_{0}} + 
     h\,\frac{\left(\beta_{0}\,\overline{\gamma}_{1}-\beta_{1}\,\overline{\gamma}_{0}\right)}{\beta_{0}^{2}} \\
     &+ h^{2}\,\frac{\left(\beta_{0}^{2}\,\overline{\gamma}_{2}-\beta_{0}\,\beta_{1}\,\overline{\gamma}_{1}-(\beta_{0}\,\beta_{2}-\beta_{1}^{2})\,\overline{\gamma}_{0}\right)}{2\beta_{0}^{3}} +  \nonumber
     \end{split}
\end{align}
\begin{align}
  \label{eq:Zmom}
  \begin{split}
     &+ h^{3}\,\bigl(\beta_{0}^{3}\,\overline{\gamma}_{3}-\beta_{0}^{2}\,\beta_{1}\,\overline{\gamma}_{2}+(\beta_{0}\,\beta_{1}^{2}-\beta_{0}^{2}\,\beta_{2})\,\overline{\gamma}_{1} \\
     &\qquad +(-\beta_{0}^{2}\,\beta_{3}+2\,\beta_{0}\,\beta_{1}\,\beta_{2}-\beta_{1}^{3})\,\overline{\gamma}_{0}\bigr)\frac{1}{3\beta_{0}^{4}} + \ldots
     \end{split}
\end{align}
where $\overline{\gamma}_i$ are the expansion coefficients of the anomalous dimension in a generic MOM type scheme and $Z_0$ 
is an integration constant.  The knowledge of the $\beta$-function 
\beq
\label{RENORMALISATION_GROUP_BETA}
\beta(h) \ = \frac{d h}{d\ln{\mu^2}}  = - \sum_{i=0}^{n} \beta_{i} \ h^{i+2} \ + \ 
{\cal O}\left(h^{n+3}\right) \ 
\eeq
at some order $n$ allows to calculate the momentum dependence of $h$. 
At four-loop order one has
\begin{align}
  \label{betainvert}
  \begin{split}
      h(t) &= \frac{1}{\beta_{0}t}
      \left(1 - \frac{\beta_{1}}{\beta_{0}^{2}}\frac{\log(t)}{t}
     + \frac{\beta_{1}^{2}}{\beta_{0}^{4}}
       \frac{1}{t^{2}}\left(\left(\log(t)-\frac{1}{2}\right)^{2}
     + \frac{\beta_{2}\beta_{0}}{\beta_{1}^{2}}-\frac{5}{4}\right)\right)+ \\
     &+ \frac{1}{(\beta_{0}t)^{4}}
 \left(\frac{\beta_{3}}{2\beta_{0}}+
   \frac{1}{2}\left(\frac{\beta_{1}}{\beta_{0}}\right)^{3}
   \left(-2\log^{3}(t)+5\log^{2}(t)+
\left(4-6\frac{\beta_{2}\beta_{0}}{\beta_{1}^{2}}\right)\log(t)-1\right)\right),
     \end{split}
\end{align}
where $t=\ln{\frac{\mu^2}{\Lambda^2_{\text{QCD}}}}$.  The last equation together with (\ref{eq:Zmom} ) 
allows us to write the ghost and gluon propagators as functions of the momentum and $\Lqcd$. The numerical 
coefficients for the $\beta$-function in (\ref{RENORMALISATION_GROUP_BETA}) are summarised in the Table~\ref{betacoefs}:
\begin{table}[!h]
\centering
\begin{tabular}{c|ccc}
\hline \hline
 & $\widetilde{MOM}$ & $\widetilde{MOM}_c$ & $\widetilde{MOM}_{c0}$
\\ \hline
$\beta_0$ & \multicolumn{3}{c}{$11$}
\\ \hdashline[0.4pt/1pt]
$\beta_1$ & \multicolumn{3}{c}{$102$} 
\\ \hdashline[0.4pt/1pt]
$\beta_2$ & $2412.16$ & $2952.73$ & $3040.48$
\\ \hdashline[0.4pt/1pt]
$\beta_3$ & $84353.8$ & $101484$ & $100541$
\\ \hline \hline
\end{tabular}
\caption{\footnotesize\it The numerical coefficients for the $\beta$-function for different MOM schemes~\cite{vanRitbergen:1997va}}
\label{betacoefs}
\end{table}
%

%
%
\section{OPE for the Green functions and dominant power corrections}
%
%
\label{Green_OPE_section}

The momentum dependence of the QCD Green functions at low energies is modified by non-perturbative effects.
These effects show up by presence of power-corrections to logarithmic series or, in other words,
by non-zero values of corresponding condensates. For example, such a non-perturbative object as instanton 
has a weight $\propto \exp{-\frac{8\pi^2}{g^2(p^2)}}$, giving at leading order a 
power correction $\propto \frac{1}{p^2}$. It is argued in (\cite{Boucaud:2000ey},\cite{Boucaud:2001qz}) 
that non-perturbative lattice gluonic two- and three-point functions include such 
contributions up to quite large energies of around $10$~GeV. For a systematic study of $\Lqcd$ one has to
know the influence of power corrections on the Green functions.

A powerfull tool to study the dependence of Green functions on the non-pert\-urb\-ati\-ve condensates is the 
Operator Product Expansion (OPE)~\cite{Wilson:1969zs}. This method is applicable to the problems having a specific energy
hierarchy, or two very different characteristic energy scales. For example, in QCD it may be applied to the study
of the influence of some background semi-classical field configurations. We recall here the idea of this method 
on the example of a two-point correlation function of a generic field $\phi$
\begin{equation}
G(x) = 
\left\langle
      \phi\left(\frac{x}{2}\right) 
      \phi\left(-\frac{x}{2}\right)
\right\rangle.
\end{equation}
It is postulated that when $x\rightarrow 0 $ the product of the fields may be expanded as
\begin{equation}
 \phi\left(\frac{x}{2}\right)  \phi\left(-\frac{x}{2}\right) = 
 \sum_{n=0}^{\infty}\sum_{i} w^{n}_{i}(x) \mathcal{O}^{[2n]}_i (0),
\end{equation}
where the second sum is performed on all local operators $\mathcal{O}^{[2n]}_i$ of mass dimension $2n$ having the same quantum 
number than the l.h.s.  The OPE suggests that all the features of the short-distance behaviour are stored in the Wilson coefficients 
\begin{equation}
w^{n}_{i}(x) \sim \left( x^2 \right)^{(n-1)} \times \Big[ \text{series in } \alpha_s \Big] ,
\end{equation}
that can be calculated in perturbation theory. In Fourier space they behave as
\begin{equation}
\widetilde{w}^{n}_{i}(p) \sim \left( \frac{1}{p^2} \right)^{(n+1)} \times \Big[ \text{series in } \alpha_s \Big],
\end{equation}
and thus 
\begin{equation}
\label{OPE_example}
\widetilde{G}(p) = \sum_{n=0}^{\infty}\sum_{i} \widetilde{v}^{n}_{i}\left(\alpha_s, \log{\left(p^2 / \mu^2\right) ,a^{-1}} \right)
 \frac{ \left\langle \mathcal{O}^{[2n]}_i \right\rangle } {\left( p^2 \right) ^{n+1}},
\end{equation}
where the coefficients $\widetilde{v}^{n}_{i}$ are computed in perturbation 
theory, and $\langle \mathcal{O}^{[2n]}_i \rangle$ are \textbf{vacuum condensates}. 
At $n=0$, corresponding to the trivial basic operator $\id$, we find an ordinary 
perturbative series for $\widetilde{G}$. But other condensates may lead to the appearance 
of non-perturbative power corrections. Usually 
this method is applied to gauge-invariant product of currents, and involves
only gauge invariant quantities (for a recent review see~\cite{Ioffe:2002ee}).
However it can be extended to gauge-dependent operators (like QCD propagators) and 
involve gauge-variant condensates (\cite{Lavelle:1992yh},\cite{Ahlbach:1991ws}).
We do not discuss here the subtile question of the renormalisation of condensates
and of calculation of their anomalous dimensions. On the lattice the MOM-type renormalisation 
process is non-ambiguous (\cite{Boucaud:2000nd},\cite{Boucaud:2001st}), 
because the non-perturbative value for the l.h.s in (\ref{OPE_example}) is available. This allows to define 
the condensates at fixed ultraviolet cut-off. Then one can apply a MOM renormalisation
prescription on the both sides of (\ref{OPE_example}) and thus renormalise 
the condensates $\left\langle \mathcal{O}^{[2n]}_i \right\rangle$.

In the following paragraph we will discuss the dominant power corrections, and corresponding 
condensates, in the case of the gluon and ghost correlators.

%
\subsection{The dominant OPE power correction for the gluon propagator}
%

\begin{figure}[ht]
\begin{center}
\begin{tabular}{cc}
\includegraphics[width=7cm]{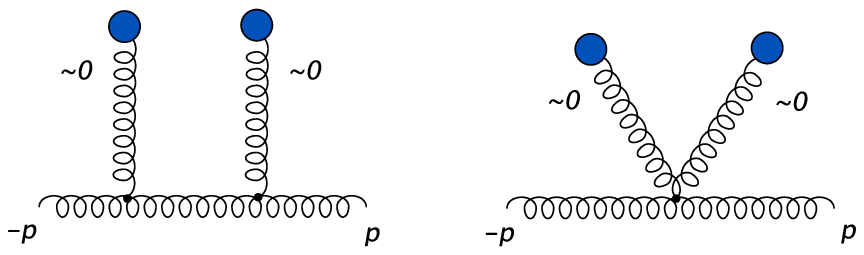}\hspace{0.35cm} &
\hspace{0.35cm} \includegraphics[width=7cm]{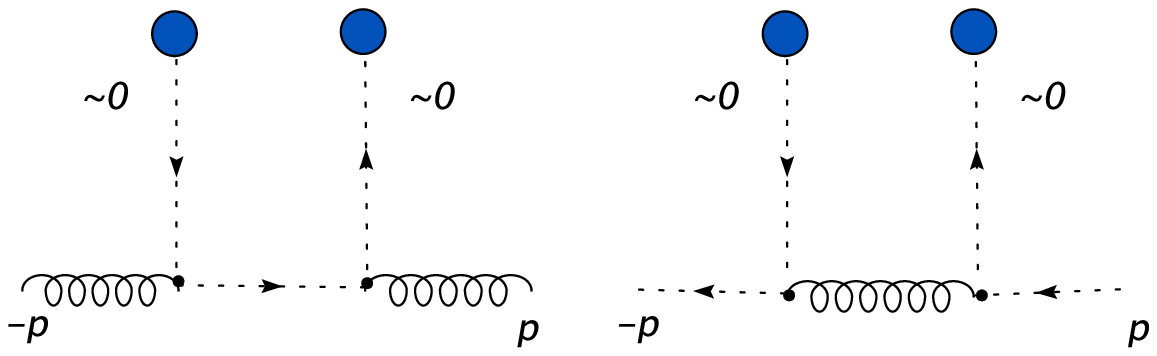}
\\ 
(a) & (b)
\end{tabular}
\end{center}
\caption{\footnotesize\it
(a) Contribution of the gluon $A^2-$condensate (represented as soft external gluons) to the gluon two-point function
(b) Contribution of the ghost $\bar{c}c$ condensate (represented as soft external ghosts) to the gluon and ghost  two-point functions. These contributions vanish because 
they are proportional to ($\sim$zero) momentum of the outgoing ghost is the ghost-gluon vertex.}
\label{CONDENSAT_DIAGRAMS}
\end{figure}

The basis of operators in the pure Yang-Mills case is 
\begin{equation}
\underline{~\id_{~} }
\quad
A_\mu^a
\quad
c^a
\quad
\d_\mu A^a_\nu
\quad
\bar{c}^a c^b
\quad
A^a_\mu c^b 
\quad
\underline{A^a_\mu A^b_\nu}
\quad
\d_\mu c^a
\quad 
\bar{c}^a\bar{c}^b
\quad
c^a c^b
\quad
\ldots
\end{equation}
At the leading order (a $\propto 1/p^2$ power 
correction compared to perturbation theory) only underlined operators contribute~\cite{Boucaud:2000nd} to the gluon propagator, because 
operators with an odd number of fields cannot satisfy colour and Lorentz invariance and $\bar{c} c$ does not contribute because of the particular
structure of the ghost-gluon vertex (cf. Figure \ref{CONDENSAT_DIAGRAMS}(b)). 
%
%
We write then for the gluon propagator:
\begin{equation}
\begin{split}
\left( \widetilde{G}^{(2)} \right)^{ab}_{\mu\nu}(p^2) & \equiv  
\left\langle 
T\left( \widetilde{A}^a_\mu(-p) \widetilde{A}^b_\nu(p)\right) 
\right\rangle 
= \\ & =
\left( V_0 \right)^{ab}_{\mu\nu}(p^2) + 
\left( V_2 \right)^{ab a^\prime b^\prime}_{\mu\nu \mu^\prime \nu^\prime}(p^2) 
\delta^{a^{\prime} b^{\prime} }
\delta_{\mu^{\prime} \nu^{\prime}  }
\frac{\left\langle : A^c_\rho(0)A^c_\rho(0):  \right\rangle}{4(N_c^2-1)}+\ldots,
\end{split}
\end{equation}
where $\langle \bullet \rangle$ is a v.e.v with respect to the non-perturbative vacuum and $:A^c_\rho(0)A^c_\rho(0):$ is
a free-field normal product. In the perturbative vacuum the v.e.v. of all the normal 
products give zero, and thus only $V_0$ is non-vanishing. Hence
\begin{equation}
\left( V_0 \right)^{ab}_{\mu\nu}(p^2) = \left( \widetilde{G}^{(2)}_{\text{pert}} \right)^{ab}_{\mu\nu} (p^2).
\end{equation}
The coefficient $V_2$ is obtained at the tree-level order from
\begin{equation}
\langle g\arrowvert  
: A^c_\rho(0)A^c_\rho(0):  
\arrowvert g\rangle = 2 + O(\alpha_s)
\end{equation}
and
\begin{equation}
\langle g\arrowvert  
T\left( \widetilde{A}^a_\mu(-p) \widetilde{A}^b_\nu(p)\right) 
\arrowvert g\rangle_{\text{connected}} = 
\left( V_2 \right)^{ab a^\prime b^\prime}_{\mu\nu \mu^\prime \nu^\prime}(p^2) 
\langle g\arrowvert  
: \widetilde{A}^{a^\prime}_{\mu^\prime}(0) \widetilde{A}^{b^\prime}_{\nu^\prime}(0) : 
\arrowvert g\rangle,
\end{equation}
where $\arrowvert g\rangle$ is a soft gluon state. So, using the LSZ rule to cut the soft external gluons, we obtain
\begin{equation}
\label{wilson_coeff_gluon}
\left( V_2 \right)^{ab a^\prime b^\prime}_{\mu\nu \mu^\prime \nu^\prime}(p^2)  = 
\frac{1}{2}
\frac{
	\left\langle 
		\widetilde{A}^{t}_{\tau}(0)  
		\widetilde{A}^{a}_{\mu}(-p)
		\widetilde{A}^{b}_{\nu}(p)
		\widetilde{A}^{s}_{\sigma}(0)
	\right\rangle}{ \left( G^{(2)}_{\text{pert}}(0) \right)^{t a^\prime}_{\tau\mu^\prime}   \left( G^{(2)}_{\text{pert}} (0) \right)^{s b^\prime}_{\sigma\nu^\prime} },
\end{equation}
which can be computed in perturbation theory (cf. Figure \ref{CONDENSAT_DIAGRAMS}(a)). Finally,
\begin{equation}
\left( \widetilde{G}^{(2)} \right)^{ab}_{\mu\nu}(p^2) = 
\frac{1}{p^2} \left( \delta_{\mu\nu} - \frac{p_\mu p_\nu}{p^2}\right)
\left( 
	p^2 \widetilde{G}^{(2)}_{\text{pert}} (p^2) 
	+ N_C\frac{g_0^2 \langle A^2 \rangle}{4\left(N_C^2 -1\right)}\frac{1}{p^2} + O(g^4,p^{-4})
\right).
\end{equation}
A MOM-type renormalisation prescription may be defined non-perturbatively. This allows an easy renormalisation 
procedure for the $A^2-$condensate~\cite{Boucaud:2000nd}. Here we do not include the 
effects of the anomalous dimension of the $A^2$ operator~\cite{Boucaud:2002jt} and hence 
we apply the MOM prescription by imposing the tree-level value to the Wilson coefficient at the 
renormalisation point $p^2=\mu^2$ for the last equation. This allows to factorise the perturbative gluon propagator giving finally
\begin{equation}
\label{Z3gluon}
{Z_3}(\mu^2) \ = \ {Z_{\rm 3,pert}}(\mu^2) \
\left(  1 + \frac{N_C}{\mu^2} \frac{g^2_R \langle A^2 \rangle_R} {4 \left(N_C^2-1\right)} + O(g_R^4,\mu^{-4}) \right).
\end{equation}
%

%
\subsection{The dominant OPE power correction for the ghost propagator}
\label{OPEsection}
%

\begin{figure}[ht]
\begin{center}
\includegraphics[scale=1]{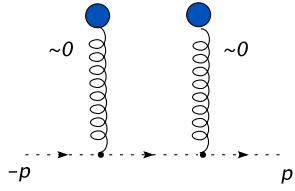}
\end{center}
\caption{\footnotesize\it Contribution of the gluon $A^2-$condensate (external soft gluons) to the ghost propagator} 
\label{CONDENSAT_DIAGRAM_A2_GHOST_PROPAGATOR}
\end{figure}

In the case of the ghost propagator the set of basic operators is the same, the ghost condensate $\bar{c}c$ does not contribute for the same reasons as for the
gluon propagator (cf. Figure~\ref{CONDENSAT_DIAGRAMS}(b) ). Thus, applying the OPE to the ghost two-point function, we obtain:
\beq\label{OPE1}
\widetilde{F}^{(2)a b}(p^2) &=& (\widetilde{V}_0)^{a b}(p^2) \ + \ \left( \widetilde{V}_2 \right)^{a b \sigma \tau}_{s t}(p^2)
\langle : A_\sigma^s(0) A_\tau^t(0): \rangle \ + \ \dots \nonumber \\ 
&=& F^{(2)a b}_{\rm pert}(p^2) \ + \ 
w^{a b} \ \frac{\langle A^2 \rangle}{4 (N_c^2-1)} \ + \ \dots 
\eeq
where, in analogy with (\ref{wilson_coeff_gluon}), the Wilson coefficient reads
\begin{small}
\beq\label{OPE3}
w^{a b} \ &=& \ \left( \widetilde{V}_2 \right)^{a b \sigma \tau}_{s t} \delta^{s t} \delta_{\sigma \tau} \ = \ 
\frac 1 2 \ \delta^{s t} \delta_{\sigma \tau} \frac{\int d^4x e^{i p \cdot x} \
\langle \Am{\tau'}{t'}{0} \ T\left( c^a \overline{c^b} \right) \ \Am{\sigma'}{s'}{0} \rangle_{\rm connected}}
{{G^{(2)}}_{\sigma \sigma'}^{s s'}(0) {G^{(2)}}_{\tau \tau'}^{t t'}(0) }
\eeq
\end{small}
which is equal to twice the diagram represented on the Figure~\ref{CONDENSAT_DIAGRAM_A2_GHOST_PROPAGATOR}
%
%
that describes the coupling of  the ghost propagator to the gluon $A^2-$condensate. Hence
\begin{equation}
w^{a b} = \frac{1}{2}\delta^{st}\delta_{\sigma\tau}
\cdot 2 \frac{\delta^{aa_1}}{p^2}\left( i g_0 f^{a_2 t a_1}\right)
\frac{\delta^{a_2 a_3}}{p^2}
\left( ig_0 f^{a_4 s a_3}\right)
\frac{\delta^{a_4 b} }{p^2} = N_C \frac{g_0^2}{p^2}\widetilde{F}^{(2)ab}_{\text{tree}}(p^2).
\end{equation}
This gives the leading non-perturbative contribution, because the first Wilson coefficient 
trivially gives the perturbative ghost propagator. Finally,
\beq\label{Fin1}
\widetilde{F}^{(2)a b}(p^2) \ = \ \widetilde{F}_{\rm pert}^{(2)a b}(p^2) \
\left( 1 + \frac{N_C}{q^2} \ 
\frac{g^2_0 \langle A^2 \rangle} {4 \left(N_C^2-1\right)}  + \ {\cal O}\left(g_0^4,q^{-4} \right) \right) \
\eeq
where all quantities are bare. Performing the MOM renormalisation 
we obtain for the renormalisation factor:
\beq\label{Z3fantome}
\widetilde{Z_3}(\mu^2) \ = \ \widetilde{Z_{\rm 3,pert}}(\mu^2) \
\left(  1 + \frac{N_C}{\mu^2} \frac{g^2_R \langle A^2 \rangle_R} {4 \left(N_c^2-1\right)}  
+ O(g_R^4,\mu^{-4}) \right) \ ,
\eeq
where the $A^2$-condensate is renormalised as in the case of the gluon propagator.
We see that the dominant multiplicative correction to the 
perturbative $\widetilde{Z_{\rm 3,pert}}$  is identical to the one obtained 
in the previous section for the gluon propagator (\ref{Z3gluon}).

%
\subsection{Constraints on the Wilson coefficients from the Slavnov-Taylor identity}
%
\label{subsection_on_Wilson_coeff_from_ST}

The gauge-dependent power corrections due to the $\langle A^2\rangle$-condensate
are obviously absent in gauge-invariant quantities. Because of this the Wilson
coefficients for the $\langle A^2\rangle$-condensate in different Green functions
are not independent. Some relations may be obtained from 
the Slavnov-Taylor identity (\ref{STid}) but their role in the MOM renormalisation
constants is not obvious.
%
%
\begin{figure}[!h]
\begin{center}
\includegraphics[width=0.3\linewidth]{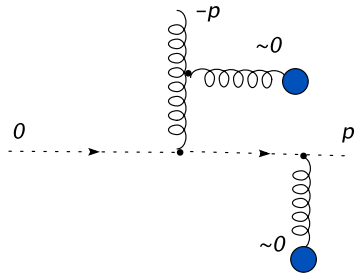}
\end{center}
\caption{\footnotesize\it The $\langle A^2 \rangle$ contribution to the ghost-gluon vertex with 
vanishing entering momentum. The above diagram is zero in Landau gauge because of the projector in the gluon propagator.}
\label{CONDENSAT_DIAGRAM_A2_GHOST_GLUON_VERTEX_1PI}
\end{figure}
%
\begin{figure}[!h]
\begin{center}
\includegraphics[width=0.3\linewidth]{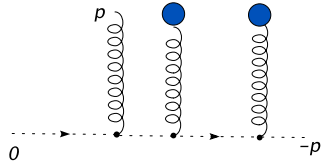}
\end{center}
\caption{\footnotesize\it Non-zero dominant $\langle A^2 \rangle$ contribution to the ghost-gluon vertex with 
vanishing entering momentum. This term contribute to the external ghost propagator.}
\label{CONDENSAT_DIAGRAM_A2_GHOST_GLUON_VERTEX}
\end{figure}
%

It is interesting to know if there are any power corrections
to this vertex, because perturbation theory predicts the non-renormalisation of
this vertex, i.e. it is equal to $1$ to all orders (\cite{Taylor:1971ff},\cite{Chetyrkin:2000dq}).
If the $\langle A^2\rangle$ power corrections are present they will constitute
the main contribution at low energies. One can directly evaluate the Wilson coefficient
to the ghost-gluon vertex $\widetilde{\Gamma}_{\mu}^{abc}(-p,0;p)$ with 
vanishing entering ghost momentum.

The only non-zero correction to the ghost-gluon vertex with one vanishing ghost momentum is
the one that contributes to the external ghost propagator (see Figure~\ref{CONDENSAT_DIAGRAM_A2_GHOST_GLUON_VERTEX}).
But all the diagrams with the condensate interaction attached to different external legs
are zero in Landau gauge (see Figure~\ref{CONDENSAT_DIAGRAM_A2_GHOST_GLUON_VERTEX_1PI}).
Thus the ghost-gluon vertex in this particular kinematic configuration does not contain
the $\frac{1}{p^2}$ power-corrections, and thus the non-renormalisation theorem holds at this order.
However, this it is not true if the external ghost momentum is not \emph{exactly} zero.

%
%
\section{Data analysis}
%
%

We calculated, using the techniques described in the chapter~\ref{chapter_Lattice_Green_functions},
the ghost and the gluon propagators of the Landau gauge $SU(3)$ gauge theory at different lattice volumes and different values of 
the $\beta$ parameter (cf. Table~\ref{tab:simulation})~\cite{Boucaud:2005np}.
\begin{table}[h]
  \centering
\begin{tabular}[h]{rcccc}
  \hline
  $\beta$ & $V$ &  $a^{-1}$ (GeV) & $V_{\text{phys}}$ ($\text{fm}^4$) & \# Configurations \\
  \hline
  $\rightarrow6.0$   &  $16^{4}$ & $1.96$ & $6.73$ & $1000$ \\
  $6.0$   &  $24^{4}$ & $1.96$ & $33.17$ & $500$ \\
  $\rightarrow6.2$   &  $24^{4}$ & $2.75$ & $8.43$ & $500$  \\
  $\rightarrow6.4$   &  $32^{4}$ & $3.66$ & $8.85$ & $250$ \\
  \hline
\end{tabular}
\caption{\footnotesize\it Lattice setup parameters. The lattice spacings are taken from Table~3 in \cite{Bali:1992ru} 
with a physical unit normalised by $\sqrt{\sigma}=445$ MeV. The lattices marked by the ``$\rightarrow$" symbol
have similar physical volume. }
\label{tab:simulation}
\end{table}
The lattices marked by the ``$\rightarrow$" symbol correspond to similar physical volume. The produced data allow
us to study the propagators in the momentum range $[\approx 2\text{GeV},\, \approx 6.5\text{GeV}]$.

This section is organised in the following way. In the first subsection we 
present the fits of the ghost and the gluon propagators separately,
and compare the fitted values for $\Lqcd$. Thus we test the self-consistency of the method. 
Non-perturbative effects are quite important in the energy interval accessible to us.
This is why another motivation is to study the asymptoticity of the perturbative series.
The latter is done by comparing the results in different renormalisation schemes 
($R=\ms,\widetilde{\text{MOM}},\widetilde{\text{MOM}_c},\widetilde{\text{MOM}_{c0}}$) 
and at different orders (from two to four loops). In the second subsection
we use the analytical result of the previous section namely that the
dominant non-perturbative effects are the same for the ghost and for the gluon propagators,
and hence the ratio of the gluon and the ghost dressing functions is 
better described by perturbation theory at low energies. We shall see that lattice data support this claim.

%
\subsection{Fitting the gluon and the ghost propagators}
%

\begin{figure}[h]
\centering
\psfig{figure=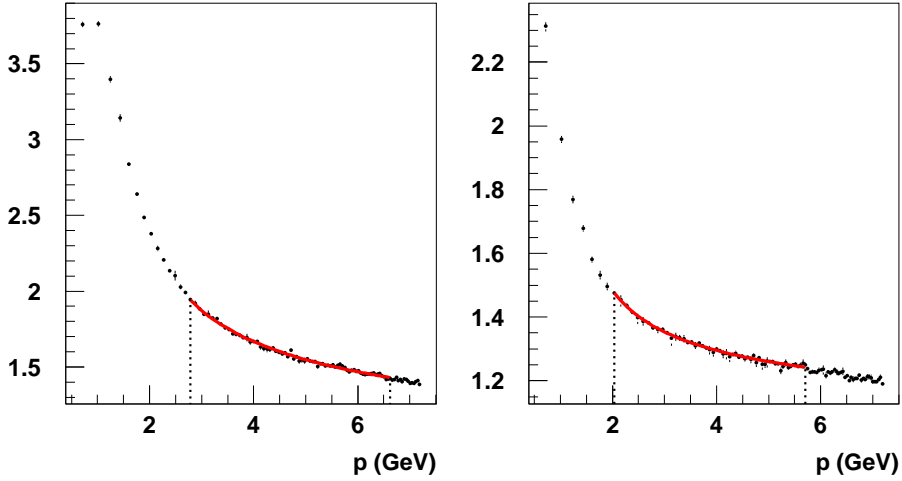, width=0.9\linewidth}
\caption{\footnotesize\it Extrapolated lattice data at $\beta=6.4$ for $G(p)$ (left)
and $F(p)$ (right). The solid line is the fit at
four-loop order in the $\ms$ scheme. The vertical dotted lines
delimit the window of each fit.}
\label{exemple_fits_Zn}
\end{figure}

We extracted $\Lqcd$ from the dressing functions of our lattice propagators by fitting them to the 
formula (\ref{eq:Zmom}) (with $h$ given by (\ref{betainvert}) ) in different MOM renormalisation schemes.
There are two parameters of the fit - the wanted $\Lqcd$ and the integration constant $Z_0$.
An example of such a fit if presented at Figure~\ref{exemple_fits_Zn}. 
The obtained value of $\Lambda_{R}$ in a scheme $R$ is converted to 
the $\ms$ scheme using the exact~\footnote{A relation between the values of $\Lqcd$ in 
different schemes $\textit{A}$ and $\textit{B}$ reads 
\begin{equation}
\label{conversion_asympt_a_une_boucle}
\frac{\Lambda^{\textit{B}}}{\Lambda^{\textit{A}}} = \exp{\left[\frac{1}{2\beta_0} 
\left( \frac{1}{g_{\textit{A}}^2(\mu^2)} -\frac{1}{g_{\textit{B}}^2(\mu^2)} \right) + O\left(g^2(\mu^2)\right) \right]}.
\end{equation}
If $g^2_{\textit{B}}=g^2_{\textit{A}}\left(1+ \zeta\,\frac{g^2_{\textit{A}}}{4\pi}+\ldots\right)$ then the asymptotic freedom
gives $\Lambda^{\textit{B}}=\Lambda^{\textit{A}}e^{\frac{\zeta}{2\beta_0}}$. Thus the exact conversion coefficient
is given by an one-loop calculation~\cite{Celmaster:1979km}. } asymptotic formulae
\begin{equation}
\label{conversion}
\begin{array}{l}
\Lambda_{\ms} = 0.346\, \Lambda_{\widetilde{\text{MOM}} }
\\
\Lambda_{\ms} = 0.429\, \Lambda_{\widetilde{\text{MOM}_c} }
\\
\Lambda_{\ms} = 0.428\, \Lambda_{\widetilde{\text{MOM}_{c0}} }
\end{array}
\end{equation}
The results in $\ms$ are given in Tables~\ref{tab:ms_values},\ref{tab:mom_tilde_values1},\ref{tab:mom_tilde_values2}. 
The errors include the statistical error, extrapolation errors and the bias due to the choice of the fit window.
%
\begin{table}[!h]
\centering
\begin{tabular}{c||c;{0.4pt/1pt}c|c;{0.4pt/1pt}c}
\hline
$\beta$ & $\Lambda^{(3)}_{\ms,gluon}$ & $\Lambda^{(3)}_{\ms,ghost}$ &
$\Lambda^{(4)}_{\ms,gluon}$ & $\Lambda^{(4)}_{\ms,ghost}$ \\
\hline
6.0 &  $519(6)^{+12}_{-4~}$ & $551(12)^{+33}_{-16}$ & $441(4)^{+8~}_{-4~}$ & 
$461(10)^{+29}_{-14}$ \\
6.2 &  $509(6)^{+17}_{-27}$ & $550(8)^{+27}_{-63}~$ & $435(6)^{+11}_{-19}$ & 
$465(8)^{+33}_{-36}~$ \\
6.4 &  $476(7)^{+44}_{-40}$ & $549(7)^{+55}_{-51}~$ & $410(4)^{+33}_{-29}$ & 
$468(7)^{+48}_{-40}~$ \\
\hline
\end{tabular}
\caption{\footnotesize\it Three-loop and four-loop physical values of $\Lambda_{\ms}$ in MeV
extracted from fits in the $\ms$ scheme.}
\label{tab:ms_values}
\end{table}
%
\begin{table}[!h]
\centering
\begin{tabular}{c||c;{0.4pt/1pt}c|c;{0.4pt/1pt}c}
	\hline
	$\beta$ & $\Lambda^{(3)}_{\ms,gluon}$ & $\Lambda^{(3)}_{\ms,ghost}$ & $\Lambda^{(4)}_{\ms,gluon}$ & $\Lambda^{(4)}_{\ms,ghost}$ \\
	\hline
	6.0 &  324(2)$^{+2~}_{-5~}$ & 322(8)$^{+20}_{-16}$ & --- & --- \\
	6.2 &  320(2)$^{+8~}_{-14}$ & 326(5)$^{+26}_{-33}$ & --- & 331(8)$^{+21}_{-16}$ \\
	6.4 &  312(1)$^{+9~}_{-25}$ & 331(4)$^{+42}_{-35}$ & 320(4)$^{+6~}_{-4~}$ & 353(9)$^{+17}_{-38}$ \\
	\hline
\end{tabular}
\caption{\footnotesize\it Three-loop physical values of $\Lambda_{\ms}$ in MeV
converted from fits in the $\widetilde{\text{MOM}}$ scheme.
}
\label{tab:mom_tilde_values1}
\end{table}
\begin{table}[!h]
\centering
\begin{tabular}{c||c;{0.4pt/1pt}c|c;{0.4pt/1pt}c}
	\hline
	$\beta$ & $\Lambda^{(3)}_{\ms,gluon}$ & $\Lambda^{(3)}_{\ms,ghost}$ & $\Lambda^{(4)}_{\ms,gluon}$ & $\Lambda^{(4)}_{\ms,ghost}$ \\
	\hline
	6.0 &  345(3)$^{+4~}_{-4~}$ & 369(9)$^{+3~}_{-2~}$ & --- & --- \\
	6.2 &  341(2)$^{+6~}_{-7~}$ & 364(8)$^{+11}_{-19}$ & 344(4)$^{+9~}_{-6~}$ & 357(10)$^{+8~}_{-16}$ \\
	6.4 &  323(2)$^{+17}_{-11}$ & 354(8)$^{+28}_{-20}$ & 332(2)$^{+14}_{-30}$ & 351(8)$^{+23}_{-25}$ \\
	\hline
\end{tabular}
\caption{\footnotesize\it Three-loop physical values of $\Lambda_{\ms}$ in MeV
converted from fits in the $\widetilde{\text{MOM}}_c$ scheme.
}
\label{tab:mom_tilde_values2}
\end{table}
%
We see from these tables that at a given order and in a given renormalisation scheme the
values obtained from the gluon and ghost propagators are consistent within the error bars, and
are quite independent of the ultraviolet cut-off. The results from a direct fit in the $\ms$ scheme
(Table~\ref{tab:ms_values}) confirm the old claim that we are still far from asymtoticity 
in the considered momentum interval in this scheme~\cite{Becirevic:1999hj}.
\begin{figure}[ht]
\centering
\psfig{figure=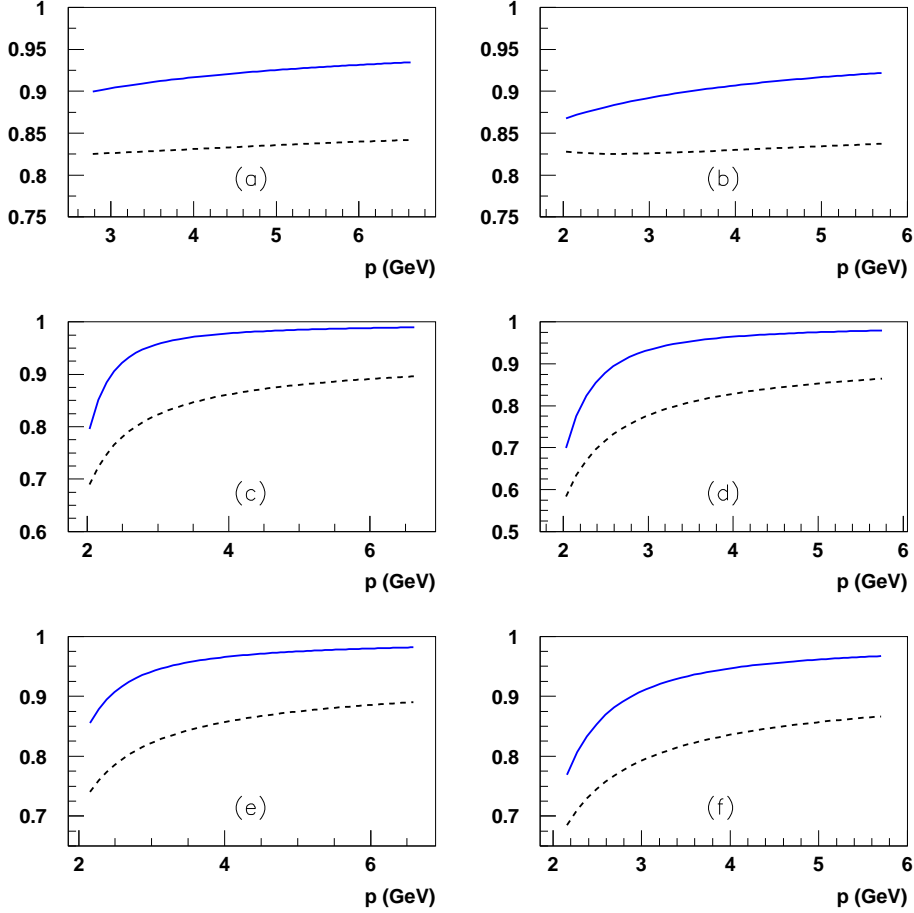, width=0.9\linewidth}
\caption{\footnotesize\it $\frac{\Lambda^{(n+1)}_{R}}{\Lambda^{(n)}_{R}}$ for
$n=2$ (dashed lines) and $n=3$ (solid lines), for the
gluon propagator in the $\ms$ scheme (a), $\widetilde{\text{MOM}}$ scheme (c) and
$\widetilde{\text{MOM}}_c$ scheme (e), and for the ghost propagator in the $\ms$
scheme (b), $\widetilde{\text{MOM}}$ scheme (d) and $\widetilde{\text{MOM}}_c$ scheme (f).}
\label{asympt_lambda_ratio}
\end{figure}
Our analysis suggests that the perturbative series 
become asymptotic at the NNLO in the case of $\widetilde{\text{MOM}}$ and $\widetilde{\text{MOM}}_c$ 
renormalisation schemes. However, the property of asymptoticity is only 
approximate at considered momenta. To see this one can use the perturbative expression 
(analogue to (\ref{Lambda_pert_generic})) for $\Lambda_{R}$ in terms of the coupling $h_R$ 
to the order four
\begin{equation}
\label{Lambda_n_n_plus_1}
2\ln\Lambda^{(4)}_{R} = \ln\mu^{2} - \frac{1}{\beta_{0}h_{R}} - 
\frac{\beta_{1}}{\beta_{0}^{2}}\ln(\beta_{0}h_{R}) - 
\frac{\beta_{0}\beta_{2}-\beta_{1}^{2}}{\beta_{0}^{3}}h_{R} -
\frac{\beta_{0}^{2}\beta_{3} - 2\beta_{0}\beta_{1}\beta_{2} +
\beta_{1}^{3}}{2\beta_{0}^{4}}h_{R}^{2},
\end{equation}
and plot the ratio (Figure~\ref{asympt_lambda_ratio}) of the 
consecutive orders $\frac{\Lambda^{(n+1)}_{R}}{\Lambda^{(n)}_{R}}$.
There is a qualitative agreement between the ratios presented at 
Figure~\ref{asympt_lambda_ratio} and our results 
(see Tables~5-10 in \cite{Boucaud:2005np}). The influence of truncation,
responsible for the differences between different orders and renormalisation schemes,
is mostly due to the large value of the effective coupling at considered energies~\cite{Boucaud:2005np}.
In fact, as shown in~\cite{Boucaud:2000ey}, the real value of the coupling constant may
be smaller, because of the power correction discussed in the section~\ref{Green_OPE_section}. 
Indeed, according to the OPE analysis the effective coupling constant is modified by a factor 
\begin{equation}
\alpha_s\rightarrow\alpha_s\left(  1 + \text{const} \cdot \frac{\langle A^2 \rangle}{p^2} \right).
\end{equation}
According to the results of the section~\ref{Green_OPE_section}, one can eliminate the dominant power correction 
by considering the ratio of the propagators. In this case one expects a better
behaviour of perturbative series at low momenta. We discuss this strategy in the following subsection.

%
%
\subsection{Fit of the ratio}
%

\begin{figure}[ht]
\begin{center}
\includegraphics[width=0.6\linewidth]{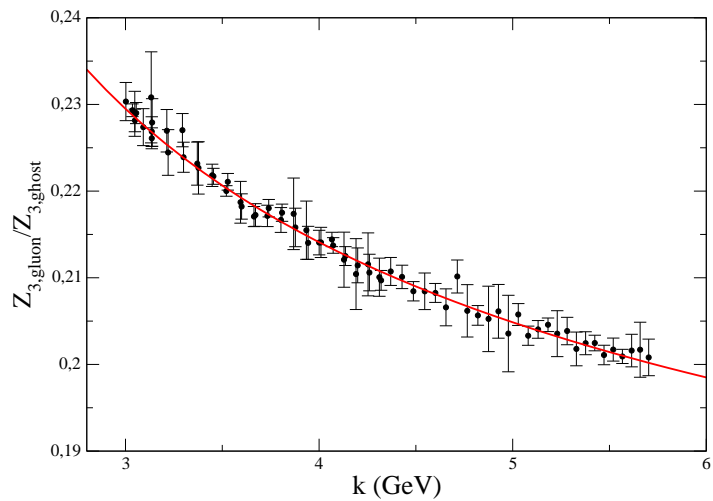}
\end{center}
\caption{\footnotesize\it Plot (in the $\widetilde{\text{MOM}}$ scheme) of the $\frac{Z_3(p^2)}{\widetilde{Z}_3(p^2)}$ for the best fit parameter $\Lambda_{\ms}=269(5)$ MeV. }
\label{ratio}
\end{figure}

\begin{figure}[h]
\vspace*{0.51cm}
\begin{center}
\includegraphics[width=0.6\linewidth]{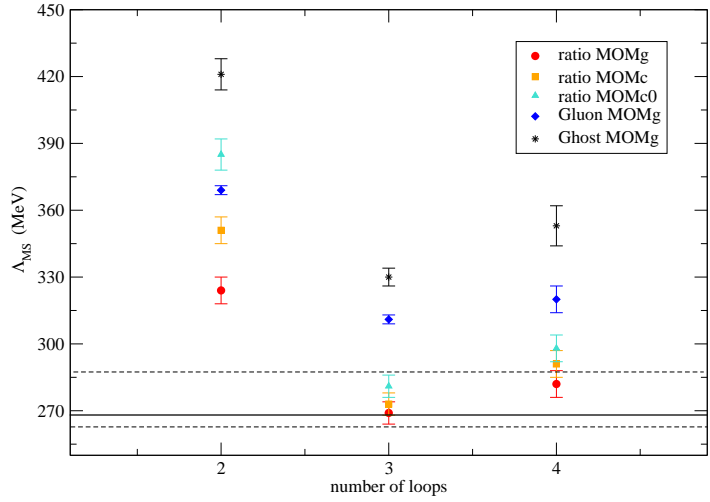}
\end{center}
\caption{\footnotesize\it Evolution of the parameter $\Lambda_{\ms}$, extracted from fits 
of the ratio \ref{ratioNP} and propagators alone (rhombus and star markers, extracted from 
Tables~\ref{tab:mom_tilde_values1},\ref{tab:mom_tilde_values2}~\cite{Boucaud:2005np}) to perturbative 
formulae, as function of the order of perturbation theory. The solid line corresponds to the value (\ref{Lambda-final}). 
Only statistical error is quoted.} 
\label{loops_a}
\end{figure}

\begin{figure}[h]
\begin{center}
\includegraphics[width=0.4\linewidth]{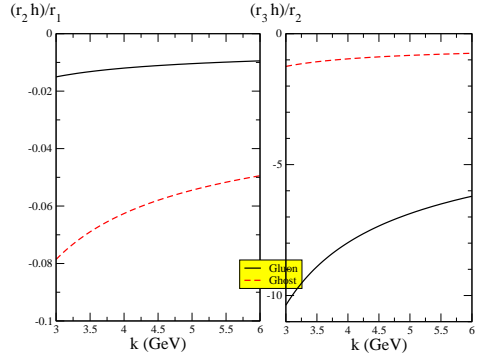}
\end{center}
\caption{\footnotesize\it (b) Ratio of four-loop to three-loop 
contributions (and of three-loop to two-loops for the sake of comparison) for the perturbative 
expansion of $\log{Z_3}$ and $\log{\widetilde{Z_3}}$ (in $\widetilde{\text{MOM}}$) in \ref{LnZ-loops},
plotted versus the momenta inside our fitting window.} 
\label{loops_b}
\end{figure}

Given that at the leading order the non-perturbative power 
corrections factorise (\ref{Z3gluon}),(\ref{Z3fantome}) and are identical 
in the case of the ghost and gluon propagators, we can fit the ratio
\beq
\label{ratioNP}
\frac{\widetilde{Z_3}(q^2,\Lambda_{R},\langle A^2 \rangle)}
{Z_3(q^2,\Lambda_{R},\langle A^2 \rangle)} = 
\frac{\widetilde{Z_{3,\text{pert}}}(q^2,\Lambda_{R})}{Z_{3,\text{pert}}(q^2,\Lambda_{R})},
\eeq
to the ratio of \emph{perturbative} formulae in scheme $R$ given by (\ref{eq:Zmom}), and
then convert $\Lambda_{R}$ to $\Lambda_{\ms}$ using (\ref{conversion}). It is interesting to notice 
that non-perturbative corrections cancel out in this ratio even in
the unquenched case with $n_f \neq 0$ flavours of dynamical quarks. The $\Lambda_{\rm QCD}$-parameter extracted from this ratio 
is free from non-perturbative power corrections up to contributions related to the operators of dimension four.
In Table~\ref{best-fits} the best-fit parameters for the three schemes are presented and we plot in 
Figure~\ref{ratio} the lattice data and the $\widetilde{\text{MOM}}$ best-fit 
curve for the ratio (\ref{ratioNP}).
%
%
\begin{figure}[ht]
\begin{center}
\includegraphics[width=0.6\linewidth]{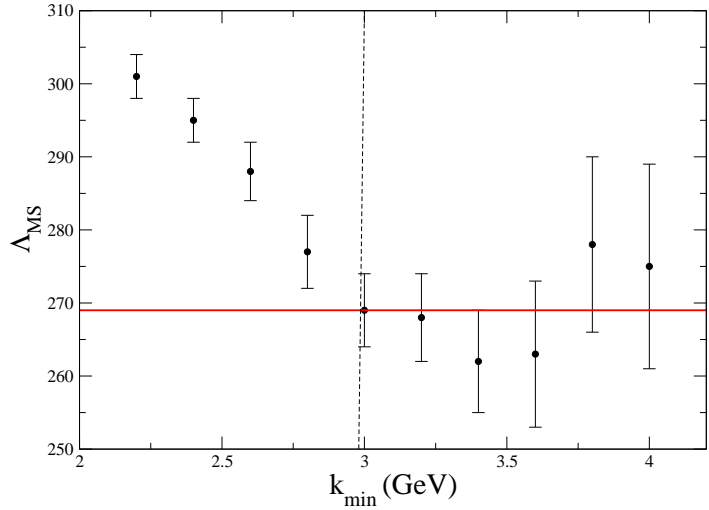}
\end{center}
\caption{\footnotesize\it The determination of the optimal window fit (from 3 GeV to $k_{\rm max} a \le \pi/2$) 
results from the search for 
some ``plateau'' of $\Lambda_{\ms}$ when one changes the low bound of the fit window. Fits are done in the 
$\widetilde{\text{MOM}}$ scheme.}
\label{kmin}
\end{figure}
%
%
\begin{table}[ht]
\begin{center}
\begin{tabular}{c|c;{0.4pt/1pt}c|c;{0.4pt/1pt}c|c;{0.4pt/1pt}c}
\hline 
scheme & $\Lambda_{\ms}^{(2)}$ & $\chi^2$/n.d.f & $\Lambda_{\ms}^{(3)}$ & $\chi^2$/n.d.f & $\Lambda_{\ms}^{\text{4 loops}}$ & $\chi^2$/n.d.f 
\\ \hline
$\widetilde{\text{MOM}}$ \rule[0cm]{0cm}{0.5cm} &  $324(6)$ & $0.33$ & $269(5)$ & $0.34$ & $282(6)$ & $0.34$
 \\
$\widetilde{\text{MOM}}_c$ \rule[0cm]{0cm}{0.5cm} &  $351(6)$ & $0.33$ & $273(5)$ & $0.34$ & $291(6)$ & $0.33$
 \\
$\widetilde{\text{MOM}}_{c0}$  \rule[0cm]{0cm}{0.5cm} &  $385(7)$ & $0.33$ & $281(5)$ & $0.34$ & $298(6)$ & $0.33$
 \\ 
\hline 
\end{tabular}
\end{center}
\caption{\footnotesize\it The best-fitted values of $\Lambda_{\ms}$ for the three considered renormalisation 
schemes. As discussed in the text, $\momc$ seems to be 
the one showing the best asymptotic behaviour.}
\label{best-fits}
\end{table}
%
In Figure~\ref{loops_a} we show the evolution of the fitted parameter $\Lambda_{\ms}$ 
when changing the order of perturbation theory used in the fitting formula.
One can conclude from Figure~\ref{loops_a} that the $\widetilde{\text{MOM}}$ scheme 
at three loops gives the most stable result for $\Lambda_{\ms}$. It can also be seen from
the ratio of four to three loops contributions (see Figure~\ref{loops_b}) 
for the perturbative expansion of $\ln{Z_3}$,
\beq\label{LnZ-loops}
\ln{Z_3} \ = \ r_0 \ln{h_R} + \sum_{i=1} r_i h_R^i \ ,
\eeq
where the coefficients $r_i$ are to be computed from those 
in equations (\ref{LnZ}-\ref{betainvert}) using the Table~\ref{betacoefs}. 
The same is done for $\ln{\widetilde{Z_3}}$. 

According to our analysis, and in agreement with the result of the separate fit, three loops seems to 
be the optimal order for the asymptoticity~\footnote{Note that the the asymptoticity property is better verified in
the case of the ratio, see Figure~\ref{loops_a}.}. Indeed, the values of $\Lambda_{\ms}$ for the three considered 
renormalisation schemes practically match each other at three loops (see Figure~\ref{loops_a}). 
The approximate value
\beq
\label{Lambda-final}
\Lambda_{\ms}=269(5)^{+12}_{-9}
\eeq
could be presented as the result for the fits of the ratio of dressing functions to perturbative 
formulae.

The results of the previous subsection and~\cite{Boucaud:2005np} suggest that our present 
systematic uncertainty may be underestimated (narrowness of the momentum interval, truncation of the
perturbative series, etc.), that is why we prefer simply to quote $\Lambda_{\ms} \approx 270 \text{ MeV}$
for future reference. This value is pretty smaller than the 
value of $\approx330\text{ MeV}$ obtained by independent fits of dressing 
functions (see Tables~\ref{tab:mom_tilde_values1},\ref{tab:mom_tilde_values2}). In light of our
OPE analysis and previous results~\cite{Boucaud:2001st}, this argues in favour of presence of low-order
non-perturbative corrections to the ghost and gluon propagators in the momentum range $[2\text{ GeV},6\text{ GeV}]$.
%
%
%
%

%
%
\subsection{Comparing the results}
%
%

We showed that perturbation theory is quite successfull in describing (up to NNNLO)
lattice propagators in the momentum range $[2\,\text{GeV},6\,\text{GeV}]$, yielding
compatible values of $\Lqcd$. The separate fit of the ghost and gluon propagators,
and the fit of their ratio favours the existence of $\propto \frac{1}{p^2}$ power
corrections and validates the OPE approach in the case of ghost and gluon propagators.
The difference with previous approaches is that we have not introduced any additional fit parameter, and have only used
perturbation theory. Our method can also be used to calculate $\Lqcd$ from propagators alone 
in the unquenched case. In principle, it can be used to estimate the value of the 
$\langle A^2 \rangle$ condensate.

The main limitation of the application of perturbation theory to lattice 
Green functions in the accessible energy interval is the lack of asymptoticity
and the truncation of the series. In fact, even the conversion formula 
(\ref{conversion_asympt_a_une_boucle}) is not exact at considered momenta.
We estimate the accuracy of our results at around $10\%$. It can be improved by
performing the simulations at $\beta=6.6, 6.8$ on the lattices of sizes 
$48^4,64^4$, respectively. The choice of parameters is motivated by
the necessity to have the same physical volume of the lattice in order
to control the finite-size effects.

%
%
%
\chapter{The infrared behaviour of Green functions}
%
%
%
\label{chapter_IR}

There are compelling reasons to think that confinement is a property of QCD, and 
does not result from some other theory. One of such indications
is a non-zero value of the string tension found in the lattice simulations.
As a matter of fact lattice simulations give access to many non-perturbative 
quantities. We are particularly interested in knowing the Landau gauge Green functions at low momenta,
i.e. at energies of order and below than $\Lqcd$. No free quarks 
or gluons exist at very small momenta because of confinement.
So, a study of gluonic correlation functions in the deep-infrared
domain may seem useless. However, in order to study the property of confinement 
from the first principles one has to understand the change in behaviour
of Green functions found at low momenta. Knowing these functions exactly
would be a great support for the future development of the theory, because many 
confinement scenarii (for example the Gribov-Zwanziger scenario) give predictions for 
low energies momentum dependence of the Green functions in Landau gauge.
Lattice simulations allow to test these predictions.

Lattice results for different Green functions of QCD have been 
successfully tested at large momenta by perturbation theory up to 
NNNLO (see chapter~\ref{chapter_UV_behaviour} and~\cite{Boucaud:2005np},\cite{Boucaud:2005xn}). We shall see below that 
lattice Green functions also satisfy the complete ghost Schwi\-nger-Dyson 
equation (see Figure~\ref{SDcheck}). Thus lattice approach 
gives consistent results non only in the ultraviolet domain but also 
in the infrared one. Of course numerical methods could never
give us complete Green functions for all possible values of momenta. 
Nevertheless, lattice gives a \emph{quasi-unique method for testing different 
analytical approaches}, like study of truncated system of Schwinger-Dyson equations,
renormalisation group flow equations and other non-perturbative 
relations like Slavnov-Taylor identities.

Most of analytical predictions are done for the infrared exponents $\alpha_F$ and $\alpha_G$
that describe power-law deviations from free propagators when $p\rightarrow 0$
\begin{equation}
\label{PowerLawParametrization}
\begin{array}{l}
p^2 G^{(2)}(p^2) \equiv G(p^2)\propto \left(\frac{p^2}{\lambda_G^2}\right)^{\alpha_G} \\
p^2 F^{(2)}(p^2) \equiv F(p^2)\propto \left(\frac{p^2}{\lambda_F^2}\right)^{\alpha_F} \\
\ldots 
\end{array}
\end{equation}
where $\lambda_{G,F}$ are some fixed parameters of dimension one. When $p$ is large, the functions
$F(p^2)$ and $G(p^2)$ are logarithmic functions of momentum, and thus $\alpha_{F,G}=0$. But at low momenta 
it may not be true. In fact, power-law behaviour is the crudest approximation, allowing to exhibit the most general features
of the momentum dependence of the Green functions in the infrared. 
The real law governing the infrared gluodynamics might be much more complicated. 
In the following section we review (very briefly) different predictions 
for the exponents $\alpha_{G}$ and $\alpha_{F}$ of, respectively, the gluon and ghost
two-point functions. Next we present our analysis of the Slavnov-Taylor identities 
imposing some limits on these exponents. After this we turn to the study of
the ghost Schwinger-Dyson equation and test the
widely accepted relation (\ref{R}) between the exponents $\alpha_{G}$ and $\alpha_{F}$.
Our conclusion (supported by numerical simulations) is that this relation is not valid.
We revisit the usual proof and conclude that either the ghost-gluon vertex behaves
unexpectedly  in the infrared, or that $\alpha_{F}=0$. At the end of the chapter 
we discuss the results of direct fits of two-point functions, and compare 
our results with other lattice collaborations.

%
%
\section{Review of today's analytical results}
%
%
\label{Review_of_today_s_analytical_results}

In this section we quote the main analytical results regarding the infrared 
exponents (\ref{PowerLawParametrization}). We start with the Zwanziger's 
prediction obtained for gluonic correlation functions. Next we present the
results of other analytical approaches.

%
\subsection{Zwanziger's prediction}
%
\label{subsection_Zwanziges_prediction}

Zwanziger suggested in~\cite{Zwanziger:1990by},\cite{Zwanziger:1991gz} that the (Landau gauge) gluon propagator 
vanishes at zero momentum in the infinite volume limit. The argument is the following.
The Faddeev-Popov operator (\ref{MFPlat}) is positive definite at a local minimum of the functional 
(\ref{LatticeGaugeMinimisingFunctional})
\begin{equation}
\label{poz}
(\omega,\MFPlat[U]\omega)\ge 0.
\end{equation}
Choosing a test vector
\begin{equation}
\omega^a(x)=\frac{\exp{i\frac{2\pi\emu}{L} x}}{\sqrt{V}}\chi^a,
\end{equation}
where the normalised colour vector $\chi^a$ is an eigenfunction of the ``angular momentum" operator
$\left(J^b\right)_{ac} = if^{abc}$, one obtains from (\ref{MFPlat}) and (\ref{poz}) 
the following limit on the mean colour spin
\begin{equation}
\label{Zwanziger_estimate}
\left\arrowvert\frac{1}{V}\sum_{x} A_\mu(x)\right\arrowvert\le 2\tan{\frac{\pi}{L} }.
\end{equation}
Introducing an external colour field $H^a_\mu$ source (independent of $x$), 
one obtains from (\ref{Zwanziger_estimate}) an estimate for the generating functional
\begin{equation}
\frac{Z(H)}{Z(0)}=\frac{1}{Z(0)} \int [\mathcal{DA}] e^{-S[\mathcal{A}] + H^a_\mu \sum_{x}A^a_\mu(x)} \le e^{2V\sum_{\mu}
\left\arrowvert H_\mu \right\arrowvert\tan{\frac{\pi}{L}}}.
\end{equation}
The free energy density $w(H)=\frac{1}{V}\log{Z[H]}$ is convex and bounded from below ($w(0)=0$), thus
one has 
\begin{equation}
0\le w(H)\le2\sum_\mu\arrowvert H_\mu \arrowvert\tan{ \frac{\pi}{L} }.
\end{equation}
All connected gluonic Green functions can be obtained by calculating the variations of the free energy 
with respect to the external sources $H_\mu^a(x)$. The last inequality suggest that in the infinite
volume limit all Green functions vanish at zero momentum. However, the inversion of derivation and 
thermodynamic limit is not supported by a rigorous proof.

%
\subsection{Study of truncated SD and ERG equations}
%

Diverse analytical approaches (study of truncated Schwinger-Dyson equations
and of renormalisation group equation, see Table~\ref{Predictions_analytiques})
agree that the infrared divergence of the ghost propagator is enhanced,
i.e. $\alpha_F \leq 0$; while they predict different values for $\alpha_G$,
mostly around $\alpha_G\approx 1.2$. This means that the gluon
propagator is suppressed in the infrared. However, some groups obtain $\alpha_G\leq 1$, i.e.
an infrared-divergent gluon propagator. 
Lattice simulations confirmed the prediction for the ghost propagator, 
whereas the lattice gluon propagator seems to remain finite and non-zero 
in the infrared, i.e. $\alpha_G = 1$. We discuss this question in details in the section~\ref{section_direct_fits}.
%
\begin{table}[!h]
\centering
\begin{small}
\begin{tabular}{l|lcc}
\hline\hline
{\bf Reference} &{\bf Method} & \boldmath$\alpha_G$\unboldmath & \boldmath$\alpha_F$\unboldmath
 \\ \hline
Zwanziger~\cite{Zwanziger:1990by} & see subsection~\ref{subsection_Zwanziges_prediction} & $>1$ & no
\\ \hdashline[0.4pt/1pt]
Bloch~\cite{Bloch:2003yu}  & SD truncation + perturbation theory  & $[0.34,1.06]$ & $[-0.53,-0.17]$
\\ 
von Smekal et al.~\cite{vonSmekal:1997is} & SD truncation & $1.84$ & $-0.92$
\\
Zwanziger~\cite{Zwanziger:2001kw} & SD truncation + Zwanziger condition & $2$ or $1.19$ & $-1$ or $-0.595$
\\ 
Aguilar et al.~\cite{Aguilar:2004sw} & SD equation & $0.98$ & $-0.04$
\\ \hdashline[0.4pt/1pt]
Kato~\cite{Kato:2004ry}  & ERGE & $0.292$ & $-0.146$
\\ 
Pawlowski et al.~\cite{Pawlowski:2003hq} & ERGE & $1.19$ & $-0.595$
\\
Fischer et al.~\cite{Fischer:2004uk} & ERGE & $1.02$ & $-0.52$
\\
\hline\hline
\end{tabular}
\caption{\footnotesize\it Summary of various analytical predictions}
\label{Predictions_analytiques}
\end{small}
\end{table}
%

%
%
\section{Constraints on the infrared exponents and the Slavnov-Taylor identity}
%
%
\label{section_Slavnov_Taylor_IR}

Let us consider the Slavnov-Taylor identity (\ref{STid}) relating the three-gluon vertex
$\Gamma_{\lambda \mu \nu}$, the ghost-gluon vertex $\widetilde{\Gamma}_{\lambda\mu}(p,q;r)$
and the ghost and gluon propagators:
\begin{equation}
\label{ST}
p^\lambda\Gamma_{\lambda \mu \nu} (p, q, r)  =
\frac{F(p^2)}{G(r^2)} (\delta_{\lambda\nu} r^2 - r_\lambda r_\nu) \widetilde{\Gamma}_{\lambda\mu}(r,p;q) -
\frac{F(p^2)}{G(q^2)} (\delta_{\lambda\mu} q^2 - q_\lambda q_\mu) \widetilde{\Gamma}_{\lambda\nu}(q,p;r).
\end{equation}
Taking the limit $r \rightarrow 0$ keeping $q$ and $p$ finite, and using 
the parametrisation $G(r^2) \simeq \left( r^2\right)^{\alpha_G}$ valid for $r^2\ll \Lqcd^2$, 
one finds the following limits on the infrared exponents
\begin{equation}
\label{STpredictions}
\left\{
\begin{array}{ll}
\alpha_G < 1  & \text{\small gluon propagator \emph{diverges} in the infrared, and} \\
\alpha_F \le 0 & \text{\small the divergence of the ghost propagator is \emph{unchanged} or \emph{enhanced} in the infrared }
\end{array}
\right.
\end{equation}

\noindent Let us discuss in details the origin of the limits (\ref{STpredictions}). The ghost-gluon 
vertex $\widetilde{\Gamma}_{\mu\nu}(p,k;q)$ may be parametrised~\cite{Ball:1980ax} in the most general way as
\bea
\widetilde{\Gamma}_{\mu}^{abc}(p,k;q) & = & f^{abc} (-i p_\nu) g_0
\widetilde{\Gamma}_{\nu\mu}(p,k;q) \label{vertghost}  \\
&=& f^{abc} (-i p_\nu) g_0 
\cdot  \left[
\delta_{\nu\mu} a(p,k;q) - q_\nu k_\mu b(p,k;q) + p_\nu q_\mu c(p,k;q)  + \nonumber 
\right. \\ &&\qquad\qquad\qquad\quad + q_\nu p_\mu d(p,k;q) + \left.  p_\nu p_\mu e(p,k;q)\right] 
\label{vertgluon}
\eea
We recall that in this formula $-p$ is the momentum of the outgoing ghost, $k$ is the momentum of the 
incoming one and $q=-p-k$ the momentum of the gluon (all momenta are taken as entering).
For some particular kinematic configurations we use the following dense notations
\begin{equation}
\begin{array}{l}
a_3(p^2) = a(-p,p; 0 ) \\
a_1(p^2) = a(0, -p; p), \quad b_1(p^2) = b(0, -p; p), \quad d_1(p^2) = d(0, -p; p).
\end{array}
\end{equation}
The limit $r^2\rightarrow 0 $ leads to an asymmetric kinematic configuration for the three-gluon
vertex in the l.h.s. of (\ref{ST}). This particular configuration allows a general parametrisation
~\cite{Chetyrkin:2000dq}
\begin{small}
\begin{equation}
\Gamma_{\mu\nu\rho}(p,-p,0) = 
\left( 2\delta_{\mu\nu}p_\rho - \delta_{\mu\rho}p_\nu - \delta_{\rho\nu}p_\mu\right) T_1(p^2) - 
\left(  \delta_{\mu\nu} - \frac{p_\mu p_\nu}{p^2} \right) p_{\rho}T_2(p^2) +
p_\mu p_\nu p_\rho T_3 (p^2).
\end{equation}
\end{small}
with functions $T_{1,2,3}(p^2)$. The scalar function $T_1(p^2)$ is proportional to the
gauge coupling in the $\widetilde{\text{MOM}}$ renormalisation scheme. 
Now, exhibiting the dominant part of each term in (\ref{ST}), we obtain
%
\begin{eqnarray}
\label{STlimr0}
& T_1(q^2)\left(q_\mu q_\nu - q^2 \delta_{\mu\nu}\right)+q^2 T_3(q^2)q_\mu q_\nu +\eta_{1\mu\nu}(q,r)= \nonumber \\
& \frac{F\left((q+r)^2\right)}{G(r^2)}\Big[(a_1(q^2)+r_1(q,r))\left(\delta_{\mu\nu}r^2-r_\mu r_\nu\right)
+ \left(b_1(q^2)+r_2(q,r)\right)
q_\mu\left(r^2 q_\nu-(q\cdot r) r_\nu\right) + \nonumber 
\\& +(b_1(q^2)+d_1(q^2)+r_3(q,r))r_\mu(r^2 q_\nu - (q\cdot r) r_\nu)\Big] + \nonumber 
\\
& +\frac{F\left((q+r\right)^2)}{G(q^2)}\Big[a_3(q^2)(q_\mu q_\nu - q^2 \delta_{\mu\nu}) +\eta_{2\mu\nu}(q,r)\Big]
\end{eqnarray}
%
with $r_{1,2,3}$ and $\eta_{1,2}$ satisfying
\begin{align}
\lim_{r\rightarrow 0 }r_1(q,r) = \lim_{r\rightarrow 0 }r_2(q,r) = \lim_{r\rightarrow 0 }r_3(q,r) = 0 
\nonumber
\\
\lim_{r\rightarrow 0 }\eta_{1\mu\nu}(q,r) = \lim_{r\rightarrow 0 } \eta_{2\mu\nu}(q,r) = 0 
\end{align}
Identifying the leading terms of the scalar factors multiplying the tensors $q_\mu q_\nu$ and
$\left(q_\mu q_\nu - q^2 \delta_{\mu\nu}\right)$ we obtain the usual relations (\cite{Chetyrkin:2000dq}):
\begin{equation}
\label{WIhab}
\begin{array}{l}
T_1(q^2) = \frac{F(q^2)}{G(q^2)}a_3(q^2) \\
T_3(q^2) = 0.
\end{array}
\end{equation}
Using these relations in (\ref{STlimr0}) we get
\begin{equation}
\lim_{r\rightarrow 0}\, \frac{F(p^2)}{G(r^2)}\Big[a_1(q^2)\left(r^2\delta_{\mu\nu}-r_\mu r_\nu\right)+b_1(q^2) 
(r^2 q_\mu q_\nu -(r\cdot q) q_\mu r_\nu)\Big] = 0.
\end{equation}
Thus one sees that if 
\begin{equation}
\label{cond_a1}
a_1(q^2)\neq 0\quad\text{or}\quad b_1 \neq 0
\end{equation}
then (\ref{ST}) can only be compatible with the parametrisation~(\ref{PowerLawParametrization}) if 
\begin{equation}
\alpha_G < 1. 
\end{equation}
The condition (\ref{cond_a1}) is satisfied because at large momentum one has to 
all orders $a_1(p^2)$ = 1 ( because of the non-renormalisation 
theorem ~\cite{Taylor:1971ff},\cite{Chetyrkin:2000dq}).

We can also, instead of  letting $r \to 0$,   take the limit $p\rightarrow 0$ of (\ref{STid}) 
as is done in \cite{Chetyrkin:2000dq}. The dominant part of the l.h.s. of (\ref{STid}) is
\begin{equation}
\left(2\delta_{\mu\nu}(p\cdot q)-p_\mu q_\nu-p_\nu q_\mu \right)T_1(q^2)-
\left(\delta_{\mu\nu}-\frac{q_\mu q_\nu}{q^2}\right)(p\cdot q) T_2(q^2)
+(p\cdot q) q_\mu q_\nu T_3(q^2) 
\end{equation}
The r.h.s. is the product of $F(p^2)$ with an expression of at least first order in p.  $T_1$ and
$T_2$ being different from zero we can conclude in this case that $\alpha_F \le 0$.

Let us repeat here that all these considerations are valid only if all scalar factors
of the ghost-ghost-gluon and three-gluons vertices are regular functions when one momentum 
goes to zero while the others remain finite. Under those quite reasonable hypotheses 
one obtains important constraints on the gluon and ghost propagators - namely that they are both divergent in 
the zero momentum limit, and the divergence of the ghost propagator is enhanced.

Let us stress that the limit (\ref{STpredictions}) on $\alpha_G$ disagrees 
with many other analytical predictions quoted in the section~\ref{Review_of_today_s_analytical_results}.

%
\section{Relation between the infrared exponents}
%
\label{section_Relation_between_the_infrared_exponents}

The Schwinger-Dyson equation for the two-point correlation function (and for the
quark propagator, but we consider only pure Yang-Mills case here) has the simplest
form among other non-perturbative relations between Green functions. It has 
been used to constrain the the infrared exponents. Even more, there is a
a commonly accepted relation between the infrared exponents
\begin{equation}
\label{R}
2\alpha_F + \alpha_G = 0.
\end{equation}
which we shall discus now. The origin of this relation is the dimensional analysis of 
the Schwinger-Dyson equation for the ghost propagator
\begin{equation}
\label{SDghost_IR}
\frac{1}{F(k)}  = 1 + g_0^2 N_c \int \frac{d^4 q}{(2\pi)^4} 
\left( \rule[0cm]{0cm}{0.8cm}
\frac{F(q^2)G((q-k)^2)}{q^2 (q-k)^2} 
\left[  
\frac{(k\cdot q)^2 - k^2q^2}{k^2(q-k)^2}  
\right]
\  H_1(q,k) 
\right),
\end{equation}
where $H_1(q,k)$ is one of the scalar functions defining the ghost-gluon vertex:
\begin{equation}
\label{H12_definition}
q_{\nu'} \widetilde{\Gamma}_{\nu'\nu}(-q,k;q-k) =  q_\nu H_1(q,k) + (q-k)_\nu H_2(q,k),
\end{equation}
where $H_{1,2}$ are functions of the factors $a,b,c,d,e$ (\ref{vertghost}).
The large momentum behaviour (\cite{Chetyrkin:2000dq},\cite{Taylor:1971ff}) 
of this vertex depends on the kinematic configuration:
\begin{equation}
\begin{array}{l}
\frac{p_{\mu}p_{\nu}}{p^2} \cdot \widetilde{\Gamma}_{\mu\nu}^{\overline{\text{\tiny MS}}}(-p,0;p) = 1\quad\text{to \emph{all} orders}
\\
\\
\frac{p_{\mu}p_{\nu}}{p^2} \cdot \widetilde{\Gamma}_{\mu\nu}^{\overline{\text{\tiny MS}}}(-p,p;0) = 1 + \frac{9}{16\pi}\alpha_s(p^2) + \ldots
\end{array}
\end{equation}
Note that in the case of the vanishing momentum of the out-going ghost (and only in this case) 
the non-renormalisation theorem is applicable~\cite{Taylor:1971ff} and hence
\begin{equation}
\label{NonRenormalization}
H_1(q,0) + H_2(q,0)=1.
\end{equation}
Let us now consider two infrared scales $k_1\equiv k$ and $k_2\equiv \kappa
k$.  Calculating the difference of the Schwinger-Dyson equation (\ref{SDghost_IR}) 
taken at scales $k_1$ and $k_2$ and supposing for the moment that $\alpha_F\neq 0$ one obtains
\begin{equation}
\label{k1k2}
\begin{split}
\frac{1}{F(k)} - \frac{1}{F(\kappa k)} & \propto
(1-\kappa^{-2\alpha_F} )  (k^{2})^{-\alpha_F}
=  g_0^2 N_c \int \frac{d^4 q}{(2\pi)^4} 
\left( \rule[0cm]{0cm}{0.8cm}
\frac{F(q^2)}{q^2} \left(\frac{(k\cdot q)^2}{k^2}-q^2\right) \right.
\times \\ & \left. \times
\left[ \rule[0cm]{0cm}{0.6cm}
\frac{G((q-k)^2)H_1(q,k)}{\left((q-k)^2\right)^2} -  
\frac{G((q-\kappa k)^2)H_1(q,\kappa k)}{\left((q-\kappa k)^2\right)^2}
\rule[0cm]{0cm}{0.6cm} \right]
\rule[0cm]{0cm}{0.8cm} \right).
\end{split}
\end{equation}
This integral equation, as well as the initial equation (\ref{SDghost_IR}),
is written in terms of bare Green functions, and the integral may contain 
ultraviolet divergences. It can be cast into a well-defined renormalised form by 
multiplying (in (\ref{SDghost})) $G^{(2)}$ (resp. $F^{(2)}$) by $Z_3^{-1}$ (resp. $\widetilde Z_3^{-1}$) 
and the bare coupling $g_0^2$ by $Z_g^{-2} = Z_3  \widetilde Z_3^2$, 
and finally multiplying  the $k^2$ term by $\widetilde Z_3$.
However, in the subtracted equation (\ref{k1k2}) all ultraviolet
divergences are cancelled, as well as the $\widetilde Z_3 k^2$ term. 
Thus the subtracted Schwinger-Dyson equation holds both 
in terms of bare and renormalised Green functions without 
any explicit renormalisation factors.

We now  make the hypothesis that there exists a scale $q_0$ below which the
power-law parametrisation is valid
\begin{equation}
\label{def_q0}
G(q^2) \ \sim \ (q^2)^{\alpha_G}, \quad
F(q^2) \ \sim \ (q^2)^{\alpha_F}  , \quad \mbox{\rm for}  \quad q^2 \le q_{0}^2.
\end{equation}
The equation (\ref{NonRenormalization}) suggests that if \emph{both} functions $H_{1,2}$
are non-singular then one can suppose $H_1(q,k)\simeq 1$ in (\ref{k1k2}),
and (\ref{R}) is straightforward by a dimensional analysis. However, we have a priori no reason
to think that the scalar functions $H_1(q,k)$ and $H_2(q,k)$ are \emph{separately} non-singular 
for all $q,k$. Writing for example~\footnote{In fact there are many possible parametrisation. We
choose (\ref{parametrization_H1}) in order to illustrate the argument that follows.}
\begin{equation}
\label{parametrization_H1}
H_1(q,k) \ \sim \ (q^2)^{ \alpha_\Gamma } \ h_1\left(\frac{q\cdot k}{q^2},\frac{k^2}{q^2} \right),
\end{equation}
with a non-singular function $h_1$, 
we keep all the generality of the argument admitting a singular behaviour of
the scalar factor $H_1(q,k)$. Doing the dimensional analysis of the equation (\ref{k1k2}) 
\emph{without} putting $H_1(q,k)\simeq 1$, we obtain that the relation (\ref{R}) 
is satisfied if and only if the following triple condition is verified~\cite{Boucaud:2005ce}:
\begin{equation}
\label{conditions}
2\alpha_F + \alpha_G = 0 \quad \longleftrightarrow \quad
\left\{ 
\begin{array}{l}
\alpha_F \ne 0 \\
\alpha_\Gamma=0 \\
\alpha_F +\alpha_G < 1
\end{array}
\right.
\end{equation}
All possible cases and limits obtained from the integral convergence 
conditions are given in Table~\ref{tabledescas}. As we shall see the case $2$ is excluded
by lattice simulations. The case $4$ is particularly interesting,
it corresponds to the situation when the power-law infrared behaviour of the ghost propagator is the same as 
in the free case, and no relation between the infrared exponents follows from the Schwinger - Dyson equation.
We shall return to this the discussion of this case in the section~\ref{section_direct_fits}.
%
\begin{table}[!h]
\centering
\begin{small}
\begin{tabular}{l|c|c|c|c}
\hline\hline
\textbf{case} & \textbf{1} & \textbf{2} & \textbf{3} & \textbf{4} \\
\hdashline[0.4pt/1pt]
 & $\alpha_F \ne 0$&$\alpha_F \ne 0$&$\alpha_F = 0$&$\alpha_F = 0$\\
&$\alpha_F +\alpha_G +\alpha_\Gamma < 1$&$\alpha_F +\alpha_G +\alpha_\Gamma \ge1$&$\alpha_G +\alpha_\Gamma < 1$&$\alpha_G +\alpha_\Gamma \ge1$\\
\hline
\textbf{IR} &&&&\\
\textbf{convergence}&$\alpha_F  +\alpha_\Gamma > -2$&$\alpha_F  +\alpha_\Gamma> -2$ &$\alpha_\Gamma> -2$&$  \alpha_\Gamma> -2$\\
\textbf{conditions}&$\alpha_G +\alpha_\Gamma> -1$&$\alpha_G +\alpha_\Gamma> -1$&$ \alpha_G +\alpha_\Gamma> -1$&$\alpha_G +\alpha_\Gamma> -1$\\
\hline
\textbf{SD}&&&&\\
\textbf{constraints}& $2\alpha_F + \alpha_G +\alpha_\Gamma 
= 0$& $\alpha_F =-1$&excluded&none\\
\hline\hline
\end{tabular}
\caption{\footnotesize\it Summary of the various cases regarding the $\alpha$ coefficients}
\label{tabledescas}
\end{small}
\end{table}
%

The first and the last conditions (\ref{conditions}) are compatible with limits coming from the analysis 
of the Slavnon-Taylor identity (\ref{STpredictions}), and are also consistent with 
lattice simulations (see section~\ref{section_direct_fits}, \cite{Boucaud:2005ce}). 
If one of the conditions (\ref{conditions}) is not verified then, according to the Table~\ref{tabledescas}, 
(\ref{R}) should be replaced by
\begin{equation}
\label{R_Gamma}
2\alpha_F + \alpha_G + \alpha_\Gamma = 0.
\end{equation}
In the following section we present the results of a numerical test of the relation (\ref{R}),
and thus we probe the validity of the condition on $\alpha_\Gamma$.

One remark regarding the power-law parametrisation is in order. Suppose for the moment that this
parametrisation is exact below the scale $q_0$ defined in (\ref{def_q0}). 
Then one can differentiate (\ref{k1k2}) $n$ times with respect to $\kappa$, keeping $q,k$ finite. We obtain
\begin{equation}
\left( k^2 \right)^{-2\alpha_F}\left(-2\alpha_F\right)\cdot\ldots\cdot
\left( -2\alpha_F -n \right) \kappa^{-2\alpha_F -n} \propto \int d^4 q \,
\frac{d^n}{d^n \kappa} \left( \frac{G((q-\kappa k)^2)H_1(q,\kappa k)}{\left((q-\kappa k)^2\right)^2} \right).
\end{equation}
The r.h.s of the last equation is not equal to zero for finite $k$, and thus one immediately
has 
\begin{equation}
\alpha_F \neq -\frac{n}{2},\qquad n=1,2,\ldots.
\end{equation}
Thus any half-integer predictions for $\alpha_F$ should be considered as an indication of
incompleteness of the power-law parametrisation (\ref{PowerLawParametrization}).

%
\section{Lattice study of the ghost Schwinger-Dyson equation}
%

%
\subsection{Complete ghost Schwinger-Dyson equation in the lattice formulation}
%

In order to derive the discretized version of the ghost Schwinger-Dyson equation
we repeat the same steps as in the continuum case (\ref{SD_simple_begin} - \ref{SD_simple_end})
but for the lattice version of the  Faddeev-Popov operator (\ref{MFPlat}).
We define the covariant Laplacian
\begin{equation}\label{DU}
\begin{array}{l}
\Delta_U^{ab} = \sum_\mu  \left(G_\mu^{ab}(x) 
\left( 
        \delta_{x,y} - \delta_{y,x+\emu}
\right)
-
G_\mu^{ab}(x-\emu) 
\left( 
        \delta_{y,x-\emu} - \delta_{y,x}
\right)
\right).
\end{array}
\end{equation}
\noindent The appearance of $\Delta_U$ in (\ref{MFPlat}) is due to the appropriate 
discretisation of the usual Laplacian operator $\Delta$, dictated by
the non-locality of derivatives in the lattice formulation, i.e.
replacement of the $\nabla$ operator by its covariant version. 

Multiplying (\ref{MFPlat}) by $F^{(2)}(x,y)$ from the right, one obtains
\begin{equation}
\label{SD_L1}
\begin{split}
& \frac{1}{N_c^2 -1} \Delta_U^{ab} (y,z) F_{\text{1conf}}^{(2)ba}(U;z,x)   = \delta_{y,u} -
 \\ -
 \frac{f^{abc}}{2(N_c^2 -1)} &
\left[ 
A_\mu^c(y) F_{\text{1conf}}^{(2)ba}(U;y+\emu,x)  - A_\mu^c(y-\emu) F_{\text{1conf}}^{(2)ba}(U;y-\emu,x)
\right].
\end{split}
\end{equation}
This is an exact mathematical identity for each gauge configuration $U$, 
and thus the consequences that can be derived from this relation are free of 
any ambiguity originating from the presence of Gribov copies. 
Performing an averaging $\langle \bullet \rangle$ over the configurations $U$ one gets
\begin{equation}
\label{SD_L1_av}
\begin{split}
 & \frac{1}{N_c^2 -1} \tr \left\langle\Delta_U(y,z) F_{\text{1conf}}^{(2)}(z,x)\right\rangle  = \delta_{y,x} -
\\ &   - 
 \frac{f^{abc}}{2(N_c^2 -1)} 
\left\langle A_\mu^c(y) F_{\text{1conf}}^{(2)ba}(U,y+\emu,x)  - A_\mu^c(y-\emu) F_{\text{1conf}}^{(2)ba}(U,y-\emu,x)
\right\rangle
\end{split}
\end{equation}
This averaging  procedure depends on the way chosen to treat the Gribov problem:
the particular set of configurations over which it is performed depends on the prescription which is adopted 
(fc/bc procedures on the lattice, restriction to  the fundamental modular region; 
see the subsection~\ref{subsection_Lattice_Green_functions_and_Gribov} for details). 
Consequently, the Green functions may vary but they must in any case satisfy the above equation, even 
when the volume of the lattice is finite. 

Like in the continuum case, we perform a Fourier transform and obtain:
\begin{small}
\begin{equation}
\label{SD_L1_av2}
\begin{split}
& \frac{1}{N_c^2 -1} \tr \sum_{x} e^{i p\cdot x} \left\langle\Delta_U(0,z) F_{\text{1conf}}^{(2)}(U,z,x)
\right\rangle  = 1
- i \sin(p_\mu)
\frac{f^{abc}}{(N_c^2 -1)} 
\left\langle A_\mu^c(0) \tilde{F}_{\text{1conf}}^{(2)ba}(U,p) \right\rangle
\end{split}
\end{equation}
\end{small}
Although the equations (\ref{SD_L1}) and (\ref{SD_L1_av}) have to be exactly verified by lattice data,
the relation (\ref{SD_L1_av2}) does only approximately (within statistical errors) since it 
relies on translational invariance, which could be guaranteed only if we used an infinite
number of Monte-Carlo configurations.

\begin{figure}[!h]
\begin{center}
\includegraphics[width=0.3\linewidth]{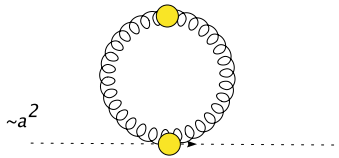} 
\end{center}
\caption{\footnotesize\it Example of the terms in the Schwinger-Dyson 
equation on the lattice.}
\label{tadpole}
\end{figure}

The presence of $\Delta_U$ in the last equation is due to non-zero lattice spacing effects. 
Indeed, lattice perturbation theory possesses an infinite number of ghost-gluon vertices 
depending on the lattice spacing $a$, giving tadpole contributions like the one presented at
the Figure~\ref{tadpole}.
%
%
Such tadpole contributions may be estimated by a mean field method~\cite{Lepage:1992xa}.
Using the average plaquette  $\langle P \rangle$ (for $\beta=6.0$ $\langle P\rangle \simeq 0.5937$) one predicts 
a tadpole correction factor $\propto \langle P \rangle^{-(1/4)} \simeq 1.14$. 
These terms disappear in the continuum limit, but they do so only very slowly : the
tadpole corrections (1 - plaquette) vanish  only as an inverse  logarithm with the
lattice spacing. This is to be contrasted with the corrections arising in the r.h.s 
which are expected to be of order $a^2$. Our lattice calculation~\cite{Boucaud:2005ce} gives
\begin{equation}
\Delta_U \simeq \Delta/\left(1.16 \pm 0.01 \right),
\end{equation}
almost independently of the momentum. This is in good agreement with
the  correction factor $1.14$ quoted above.

\begin{figure}[!h]
\begin{center}
\includegraphics[angle=-90, width=0.7\linewidth]{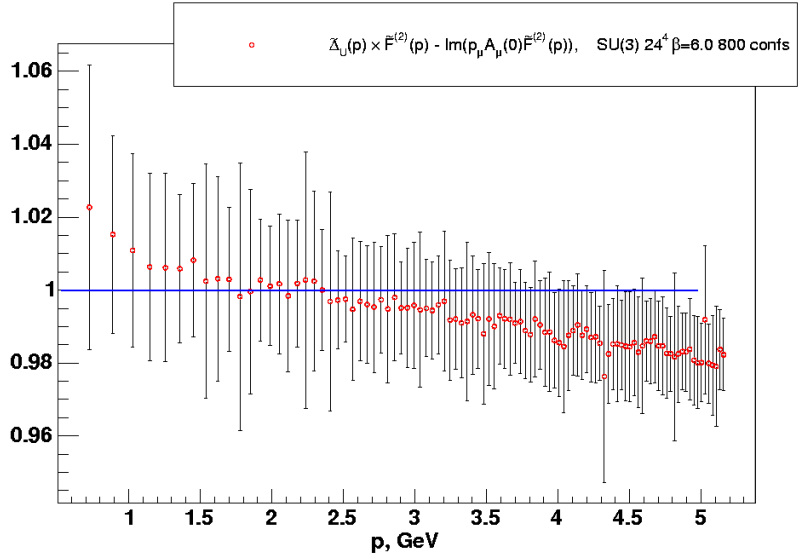}
\end{center}
\caption{\footnotesize\it Checking that lattice Green functions satisfy 
the ghost SD equation (\ref{SD_L1_av2}). We plot $\frac{1}{1.16}\widetilde{F}(p^2) 
-  g_0 \frac{p_\mu}{N_c^2 -1} f^{abc}\langle A^c_\mu(0)\,\widetilde{F}_{\text{1conf}}^{(2)ba}
(\mathcal{A},p)\rangle $ compared to $1$.}
\label{SDcheck}
\end{figure}

We see from Figure~\ref{SDcheck} that the lattice Green functions 
match pretty well the SD equation (\ref{SD_L1_av2}) in both the ultraviolet and
infrared regions.
%
%
Lattice propagators were successfully checked by the perturbation theory at large 
momentum, and they satisfy the ghost Schwinger-Dyson equation. This means that
lattice approach gives consistent results also in the infrared.

%
\subsection{Checking the validity of the tree-level approximation for the ghost-gluon vertex}
%

The simplest approximation (used by many authors, see section~\ref{Review_of_today_s_analytical_results}) 
of the ghost Schwinger - Dyson equation~(\ref{SDghost_IR}) corresponds to the case 
\begin{equation}
H_1(q,k)=1\quad \forall q,k,
\end{equation}
an approximation motivated by the non-renormalisation theorem (\ref{NonRenormalization}) 
valid for the sum $H_1(q,k)+H_2(q,k)$ when $k=0$. This gives
\begin{equation}
\label{trunc_SD_cont}
\frac{1}{F(k)} = 1 + \frac{g_0^2 N_c}{k^2} \int \frac{d^4 q}{(2\pi)^4} 
\Big(\frac{F(q^2)G((k-q)^2)}{q^2 (k-q)^2} 
\frac{(k\cdot q)^2 - k^2 q^2}{(q-k)^2}
\,\cdot 1 \, \Big).
\end{equation}
Strictly speaking this equation, written in this way, is meaningless since it involves 
UV-divergent quantities. However it is well defined at fixed ultraviolet cut-off.

We want to check whether lattice propagators satisfy it. According to perturbation theory,
it should be approximately true at large $k$. Lattice propagators are discrete functions of momentum and 
thus one has to handle the problem of the numerical evaluation of the loop 
integral $I$ in (\ref{trunc_SD_cont}). Let us express the integrand solely 
in terms of $q^2$ and $(k-q)^2$
\begin{equation}
\label{I_is}
I =\int \frac{d^4 q}{(2\pi)^4} \,
\frac{F^2(q)G(k-q)}{q^2 (k-q)^2 } 
\Big[ \frac{(k-q)^2}{4} + \frac{(k^2)^2+(q^2)^2 - 2k^2 q^2}{4(k-q)^2} 
-\frac{q^2+k^2}{2} \Big].
\end{equation}
Then we write
\begin{equation}
I=I_1+I_2+I_3+I_4+I_5+I_6,
\end{equation}
each $I_{i}$ corresponds to one term in (\ref{I_is}). All these integrals have the form 
\begin{equation}
I_i = C_i(k) \int \frac{d^4 q}{(2\pi)^4} f_i (q) h_i(k-q).
\end{equation}
The convolution in the r.h.s. is just the Fourier transform of the product at the same point
in configuration space:
\begin{equation}
\int \frac{d^4 q}{(2\pi)^4} \, f_i (q) h_i(k-q) = \text{F}_{+}
\Big(  \text{F}_{-}(f_i)[x] \text{F}_{-}(h_i)[x]  \Big)(k),
\end{equation}
where $\text{F}_{-}(\hat{f})(x)$ is an inverse and $\text{F}_{+}(f)(k)$ a direct Fourier transform. 
Thus, in order to calculate the integral $I$ from discrete lattice propagators one proceeds as follows:
\begin{enumerate}
\item calculate $\{f_i\}(p)$  and $\{h_i\}(p)$ as functions of $F(p), G(p), p^2$ for all $i$
\item apply the inverse Fourier transform $\text{F}_{-}$ to all these functions and get $f_i(x)$ and $h_i(x)$
\item compute the product at the same point $f_i(x)\cdot h_i(x)$
\item apply the direct Fourier transform $\text{F}_{+}$ to $f_i(x)\cdot h_i(x)$
\end{enumerate}
The integrands in (\ref{I_is}) depend only on the squared norms $q^2$ and $(k-q)^2$, 
and thus the angular part may be integrated out, giving the four-dimensional 
Hankel transformation
\begin{align}
& \widehat{f(\arrowvert x \arrowvert)}[p]=
\frac{1}{\arrowvert p \arrowvert}
\int_{0}^{\infty}
J_1(\arrowvert p \arrowvert r)r^{2}
f(r)dr\nonumber
\\ &
f(r)=\frac{1}{(2\pi)^2}
\frac{1}{r} 
\int_{0}^{\infty}
J_{1}(\rho r)\rho^{2}
\widehat{f(\arrowvert x \arrowvert)}[\rho] d\rho.
\end{align}
These integrals are evaluated numerically by means of the Riemann sum
\begin{equation}
\label{HT}
f(r) = (2\pi)^{-2} \arrowvert r \arrowvert^{-1}
\sum_{i=1}^{N} J_{1}(r \rho_i ) \rho_i^{2} \frac{\hat{f}[\rho_i] + \hat{f}[\rho_{i-1}]}{2} (\rho_i-\rho_{i-1}),\qquad \rho_0 = 0,
\end{equation}
where $N$ is the number of data points. The inverse transformation is 
done in the similar way. In  practice, because of the lattice artifacts 
(see subsection~\ref{subsection_discretisation_errors}) which become
important at large $\rho$ the summation has to be 
restricted to $\rho < \rho_{max} \simeq 2.2$ instead  of the maximal value $ 2 \pi$.

\begin{figure}[!htb]
\begin{center}
	\includegraphics[angle=0, width=0.8\linewidth]{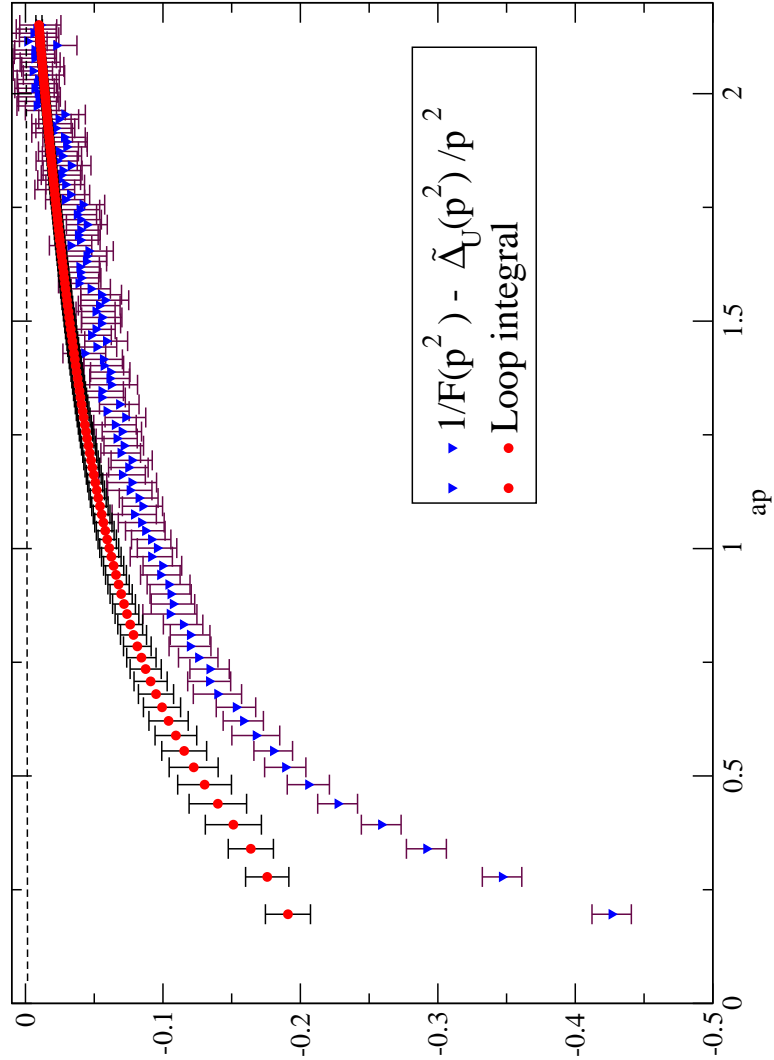}  
\end{center}
\caption{\footnotesize\it Checking whether lattice Green functions satisfy the ghost SD equation (\ref{SDghost_IR}) with
an assumption $H_1(q,k)=1$. The upper line(circles) correspond to the loop integral in (\ref{SDghost_IR}), 
and the down line (triangles) corresponds to $1/F(p^2)-1$. In this plot $a^{-1}\approx 3.6$ GeV. }
\label{figure_SDcheckH1}
\end{figure}

Now we are ready to check the approximate equation~(\ref{trunc_SD_cont}) on the lattice.
We still have to face the same problem we have already encountered in the previous subsection,
namely that the lattice Faddeev-Popov operator involves the non trivial discretisation $\Delta_U$ 
of the Laplacian operator.  This is taken into account by means of the substitution of $\widetilde{\Delta_U}(p^2)/p^2$ to the ``1''  term in the l.h.s of equation~(\ref{trunc_SD_cont})
We present on Figure~\ref{figure_SDcheckH1} the result of the numerical integration 
described above. We have chosen for this purpose the data set from the simulation
with the gauge group $SU(3)$ at $\beta=6.4, V=32^4, a^{-1}\approx 3.6\text{ GeV}$. 
One sees that the equality is achieved  at large momenta, but in the infrared 
the naive approximation of the ghost Schwinger-Dyson equation fails.
%
%
The errors on Figure~\ref{figure_SDcheckH1} include statistical Monte-Carlo errors 
for $F(q^2)$ and $G(q^2)$ and the bias coming from the UV cut-off of the integral $I$.

We see that at small momenta (below $\approx3$ GeV) the ghost Schwinger-Dyson equation 
with the assumption $H_1(q,k)=1$ is not satisfied. However, it is quite difficult to establish 
whether this disagreement is due to the infrared or ultraviolet dependencies of 
$H_1(q,k)$. To check this one has to know $H_1(q,k)$ for all values of $q,k$.
Unfortunately this information is not available. Thus the main conclusion of the
present subsection is that the scalar function $H_1(q,k)$ plays an important 
role in the infrared gluodynamics, and it cannot be set to one.

%
%
\section{Direct fits of infrared exponents}
%
%
\label{section_direct_fits}

We have seen in the previous section that lattice simulations give consistent 
results for the Green functions at all momenta. Another interesting feature 
that we have established is the important role 
of the scalar factor $H_1(q,k)$ coming from the complete ghost-gluon 
vertex (\ref{H12_definition}). In this section we present numerical 
results allowing to check the relation (\ref{R}). After this we
present our results for direct fits of the exponents $\alpha_F$ and $\alpha_G$,
and compare them to the results of other lattice collaborations.
%
\begin{figure}[!thb]
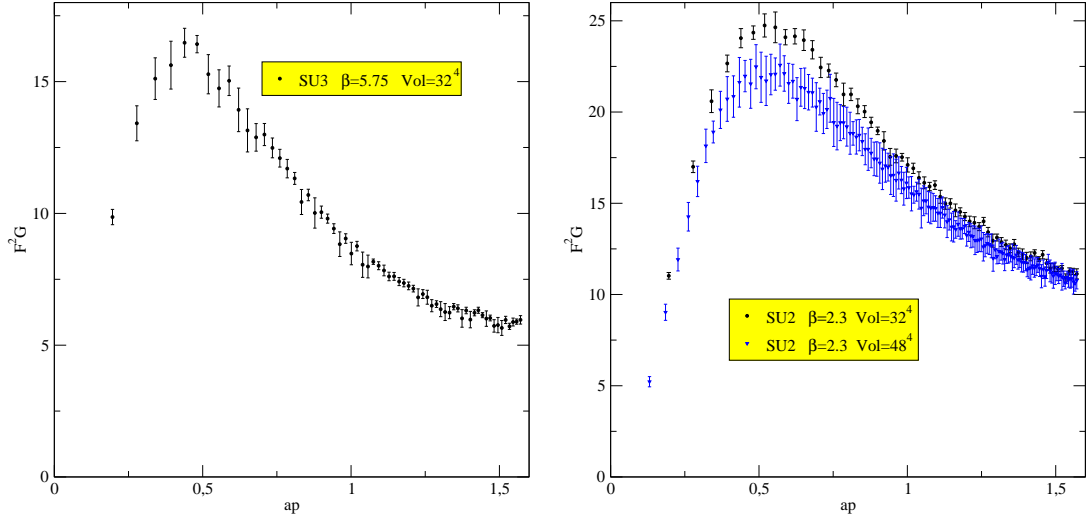

\vspace*{0.02\paperheight}
\centering
\hspace*{-0.1cm}
\begin{tabular}{lr}
\includegraphics[width=0.45\linewidth]{EPS/SU3_F2G}&\includegraphics[width=0.45\linewidth]{EPS/SU2_F2G}
\end{tabular}
\caption{\footnotesize\it Direct test of the relation $2\alpha_F + \alpha_G=0$. If the last is true $F^2G$ has to be
constant in the infrared. We see that it is clearly not the case. In these plots  $a^{-1}\approx 1.2$ GeV, so the peak is located at $\approx 600$ MeV.}
\label{F2G}
\end{figure}
%

%
\subsection{Testing the relation $2\alpha_F + \alpha_G =0$.}
%

In order to test the relation (\ref{R}) we plot at Figure~\ref{F2G} the quantity $F^2(p^2) G(p^2)$.
If all the conditions (\ref{conditions}) are satisfied this quantity should be constant 
in the infrared (or slightly varying). We see from Figure~\ref{F2G} that in the infrared (below $\approx 600$ MeV) 
the quantity $F^2 G$ is not constant, and thus one of the conditions (\ref{conditions})
is not verified. We have seen that the conditions $\alpha_F \neq 0 $ and $\alpha_F+\alpha_G < 1$
are consistent with the limits (\ref{STpredictions}) from the Slavnov-Taylor
identity (\ref{ST}). We have also seen (cf. Figures~\ref{SDcheck} and \ref{figure_SDcheckH1})
that neglecting the momentum dependence of the vertex is not possible in the infrared, because in this case
the ghost Schwinger-Dyson equation is no longer satisfied by lattice propagators.
Thus the only possibility is to admit that $H_1(q,k)$ plays an important role,
and that the relation (\ref{R}) is not verified.  If $\alpha_F\neq 0$ then the 
modified form (\ref{R_Gamma}) that takes in account the singularity of $H_1(q,k)$ should be
considered (according to Fig.\ref{figure_SDcheckH1}), with $\alpha_\Gamma < 0$
in our parametrisation. This singularity is probably related to the non-perturbative
power corrections to the vertex discussed in the subsection~\ref{subsection_on_Wilson_coeff_from_ST}.
%
\begin{figure}[!h]
\vspace*{0.02\paperheight}
\begin{center}
\includegraphics[width=0.6\linewidth]{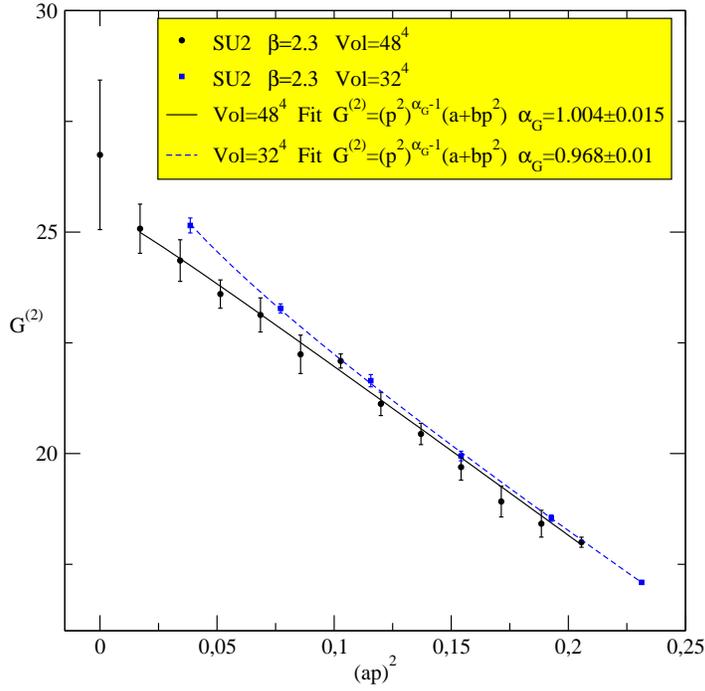}
\end{center}
\caption{\footnotesize\it $G^{(2)}(p^2)$ from lattice simulation for  $SU(2)$ (left).
$\beta_{\text{SU(2)}}=2.3$ and $\beta_{\text{SU(3)}}=5.75$.
The  volumes are $32^4$ and $48^4$ for $SU(2)$. 
In these plots $a^{-1}\approx 1.2$ GeV.}
\label{plotSU2}
\end{figure}
%

Another reason to think that the relation (\ref{R}) is not exact 
is the dependence of $\alpha_F$ and $\alpha_G$ on the 
choice of the Gribov copy. We have seen in the section~\ref{section_Gribov_copies_sur_reseau}
that the low-momentum dependence of the gluon propagator 
is not sensitive to the bc/fc choice while the infrared behaviour
of the ghost propagator depends on it. But the Schwinger - Dyson equation
for the ghost propagator is independent of the choice of the copy, because it is valid
exactly for every gauge configuration, even on a finite lattice 
(see equations (\ref{SD_simple_end}) and (\ref{SD_L1_av})). Hence if there is a relation 
between the infrared exponents $\alpha_F$ and $\alpha_G$ 
resulting from the ghost Schwinger - Dyson equation then 
it could not depend on the choice of the copy. Thus it is 
not possible to have a relation with $\alpha_F$ and $\alpha_G$ alone.
This above argument is not directly applicable in the case $\alpha_F = 0$.

According to the analysis performed in the section~\ref{section_Relation_between_the_infrared_exponents},
and given (see next subsection) that the case $2$ of the Table~\ref{tabledescas} is excluded by lattice simulations 
the following explanations of the non-validity of the relation $2\alpha_F+\alpha_G=0$ are possible:
\begin{enumerate}
\item The ghost-gluon vertex contains scalar factors that are singular in the infrared, i.e. $\alpha_\Gamma\neq 0$ in 
the equation \ref{parametrization_H1}.

\item The case $4$ of the Table~\ref{tabledescas} is realised~\cite{Boucaud:2006if} and hence there exists \emph{no} relation between the infrared exponents. 
Let us recall that in the above case one has $\alpha_F=0$ and $\alpha_G+\alpha_\Gamma\geq 1$. If the ghost-gluon vertex is regular in the 
infrared then one has
\begin{equation}
\label{case_4_vertex_regulier}
\left\{
\begin{array}{l}
\alpha_F=0 \\
\alpha_G \geq 1.
\end{array}
\right.
\end{equation}
\end{enumerate}
%

%
\subsection{Lattice fits for $\alpha_F$ and  $\alpha_G$.}
%

\begin{figure}[!h]
\begin{center}
\includegraphics[ angle=-90 ,width=0.75\linewidth]{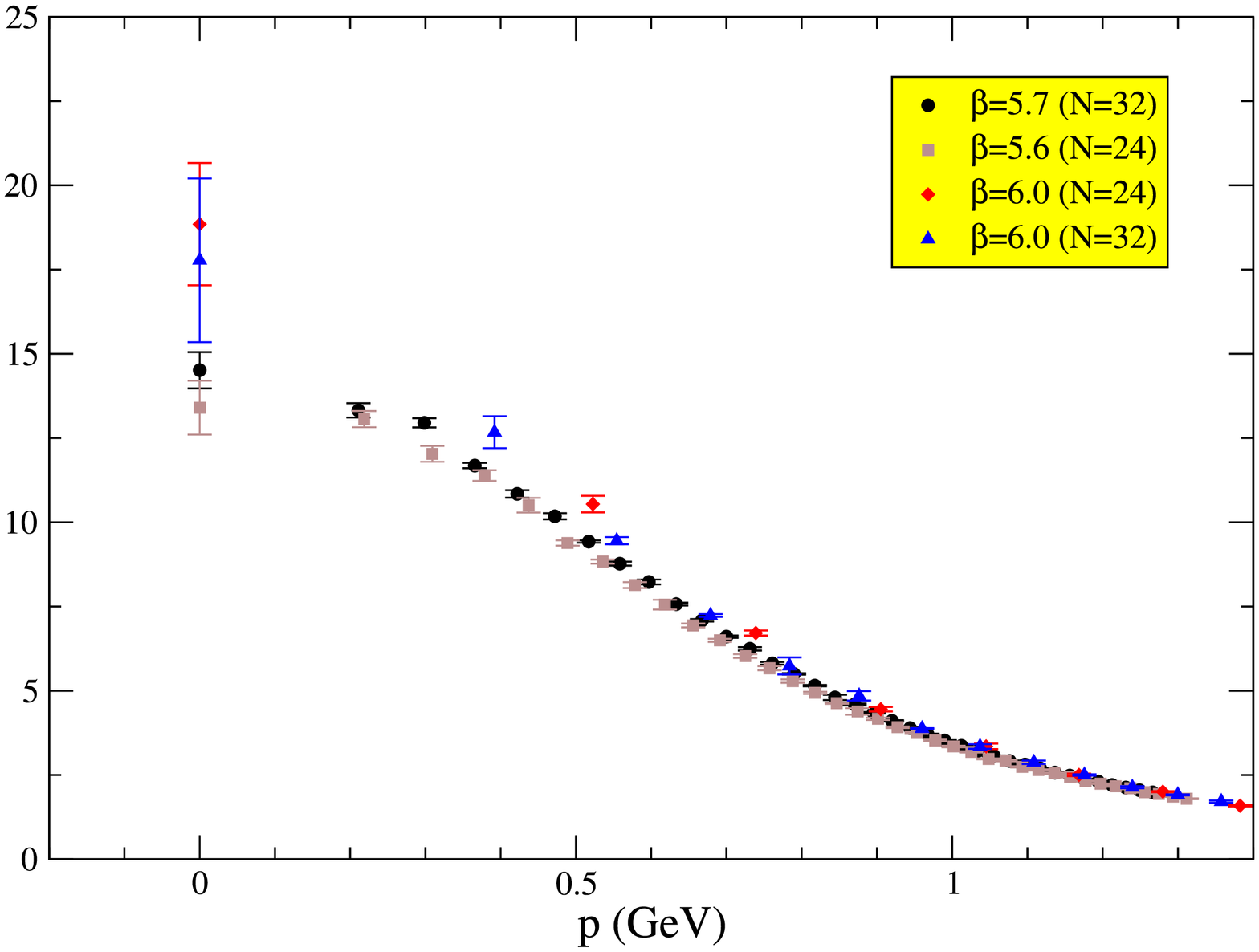}
\end{center}
\caption{\footnotesize\it The continuity of the lattice gluon propagator in the infrared.} 
\label{GdPAll}
\end{figure}

Let us now discuss the direct fits for the infrared exponents $\alpha_G$ and $\alpha_F$.
The examples of such fits of lattice data are presented on Figure~\ref{plotSU2}.
The errors are quite large, leading to an instability in the fit results. 
That is why we fit both propagators in the infrared to the formula
\begin{equation}
(q^2)^\alpha (\lambda + \mu q^2)
\end{equation}
where we added an additional term of the form $\mu q^2$ in order to describe a situation 
like the one at Figure~\ref{plotSU2}(left) where $G^{(2)}(p^2)$ seems 
to go  to a finite limit when $p$ goes to zero.
The obtained values for $\alpha_{F,G}$ are summarised 
in Table~\ref{tablalpha}. 
%
\begin{table}[h]
\begin{center}
\begin{tabular}{c|c|c|cc}
\hline\hline
\text{Group}&\text{Volume}&$\beta$&$\alpha_G$&$\alpha_F$\\
\hline
$SU(2)$ & $48^4$ & $2.3$  & $1.004(15)$ &$-0.087(15)$\\
\hdashline[0.4pt/1pt]
$SU(2)$ & $32^4$ & $2.3$  & $0.968(11)$ &$-0.109(14)$\\
\hline\hline
\end{tabular}
\caption{Summary of the fit results for the $F$ and $G$ functions}
\label{tablalpha}
\end{center}
\label{default}
\end{table}%
%
For $SU(2)$ and the larger lattice volume the value obtained for $\alpha_G$ is compatible 
with 1.
We also take into account our experience from previous studies of the 
gluon propagator where we have always observed that the gluon 
propagator goes continuously to a finite limit in the infrared region (see Figure~\ref{GdPAll}).
%
%
However, the fits are quite instable, and depend a lot on the choice of the fit formula
that can considerably change the result. The main problem is the lack of data points at low momenta.

Regarding the gluon propagator another strategy may be taken. It consist in
extrapolating the available data to the infinite volume limit.
A very detailed study of the gluon dressing function and specially of 
its volume dependence at $k=0$  has already been performed in~\cite{Bonnet:2001uh}. 
This study shows that a value $\alpha_G = 1$ is compatible with the 
data (the dressing function shows no signal of discontinuity in the neighbourhood 
of zero) and that no pathology shows up as the volume goes to infinity.
Let us compare all available lattice results for the point $G^{(2)}(0)$ 
and check whether there is an agreement between the data. 
Following~\cite{Bonnet:2001uh} we renormalise the gluon propagator in the MOM scheme 
at $4~\text{GeV}$ and use the suggested extrapolation formula
\begin{equation}
G_R^{(2)}(0,\mu=4 \,{\rm GeV}) =  G_{R\,\infty}^{(2)}(0,\mu=4 \,{\rm GeV})
+ \frac{c}{V}.
\end{equation}
We compare the results of~\cite{Bonnet:2001uh},\cite{Oliveira:2005hg} and our data from the 
Table~\ref{Gzero_OLD_data}. The results for the fit parameters $G_{R\,\infty}^{(2)}(0)$ and $c$ 
are presented in the Table~\ref{Gde0}.
%
\begin{table}[h]
\centering
\begin{tabular}{c;{0.4pt/1pt}c;{0.4pt/1pt}c;{0.4pt/1pt}c}
\hline\hline
$\beta$ &  $V$ \text{in units of} $a$    
     &\text{bare propagator} $G^{(2)}(p)$  & $1/aL$ \text{in GeV}
\\ \hline 
5.7  &$16^4 $&$16.81\pm0.13$   & 0.0672 \\
5.7  &$24^4 $ &$15.06\pm0.29$  & 0.0448 \\
5.8  &$16^4 $ &$19.12\pm0.16$  & 0.0841 \\
5.9  &$24^4 $ &$18.12\pm0.30$  & 0.0685 \\
6.0  &$32^4 $ &$17.70\pm0.59$  &0.0615  \\
6.0  &$24^4$  &$19.67\pm0.35$  &0.0821  \\ 
\hline\hline
\end{tabular}
\caption{\footnotesize\it Physical lattice sizes and raw data for the gluon propagator at zero momentum
$G^{(2)}(p)$ from our old data.} 
\label{Gzero_OLD_data}
\end{table}
%
%
\begin{table}[h]
\centering
\begin{tabular}{c|c|c|c}
\hline\hline
\text{reference} &
$G_R^{(2)}(0,\mu=4 \,{\text{GeV}})$ \text{in GeV}$^{-2}$ &  $c$ \text{in GeV}$^{-2}$\text{ fm}$^4$   
      & \text{max vol in  fm}$^4$  
\\ \hline   \cite{Bonnet:2001uh} &
$7.95\pm 0.13$           &  $245\pm 22$ & 2000
\\  Table \ref{Gzero_OLD_data} &
$9.1\pm 0.3$          & $140\pm 50 $   & 90
\\ \cite{Oliveira:2005hg} &
$10.9-11.3$          & $47-65$   & 110s
\\ \hline\hline
\end{tabular}
\caption{\footnotesize\it Summary of the infinite volume zero momentum 
propagator and its slope in terms of $1/V$ for three different simulations.
 The largest volume used in the fit  is also indicated. The statistical error is not
 quoted in~\cite{Oliveira:2005hg}.}
\label{Gde0}
\end{table}
%
We are aware that not all systematic errors are taken into account: 
$O(a)$ effects, effect due to different lattice shape, insufficiently large volumes 
(for the second and third lines),  uncertainty in the estimate of
the lattice spacing in physical units, etc. However, it seems that not only
there is a clear indication in favour of a finite non vanishing zero momentum
gluon propagator, but that different lattice collaborations agree on the value. 
Of course a more extensive study is necessary to check this statement.
The other free parameter of the fit - the slope $c$ - is clearly different, but
still all the values are in agreement in the order of magnitude.

We conclude~\cite{Boucaud:2006pc} thus that all available numerical results point towards a 
finite non-vanishing and zero momentum renormalised lattice gluon propagator in 
the infinite volume limit. This suggests that $\alpha_G = 1$.
An additional study at much larger lattices is needed to get a reliable
result for this infrared exponent.

This last result is in conflict with the limits found from the study of the
Slavnov - Taylor identity (\ref{STpredictions}), and contradicts the Zwanzigers's 
prediction that the gluon propagator is infrared suppressed. However, it is very close to
the results presented in~\cite{Aguilar:2004sw}. 

\begin{figure}[!h]
\begin{center}
\includegraphics[width=0.75\linewidth]{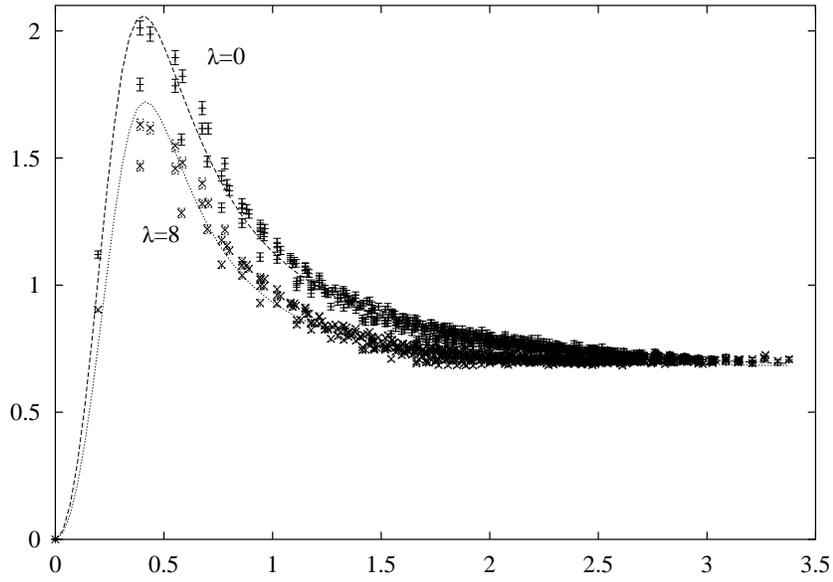}
\end{center}
\caption{\footnotesize\it Transverse part of the gluon propagator $p^2 G^{(2)}(p)$ in
covariant gauges as a function of $p$. The two sets of data refer to ($\xi=\lambda$) $\xi=0$ (Landau gauge) and $\xi=8$,
221 thermalized $SU(3)$  configurations at $\beta=6.0$ with a volume $V\times T=16^3\times 32$.
Extracted from~\cite{Giusti:2000yc}.}
\label{gluon_gauge_xi}
\end{figure}

Let us finally discuss the gauge-dependence of the parameter $\alpha_G$. This is still an open
question. However, the results of works~\cite{Giusti:2000yc},\cite{Giusti:2001kr} suggest 
that the value of $\alpha_G$ does not change drastically (see Figure~\ref{gluon_gauge_xi}) when
changing the gauge parameter $\xi$. This question deserves a separate study.
%
\begin{figure}[!h]
\begin{center}
\includegraphics[width=0.55\linewidth]{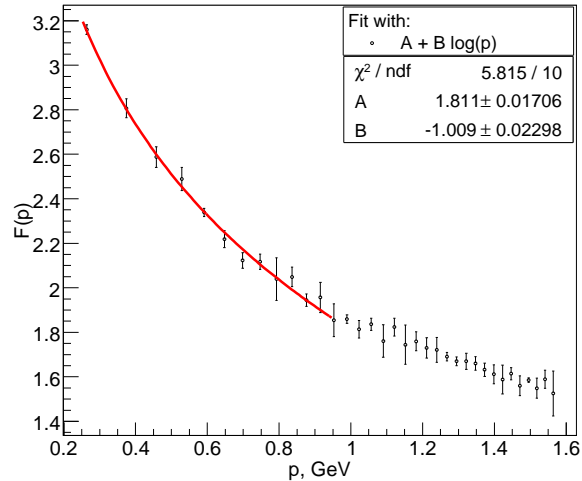}
\end{center}
\caption{\footnotesize\it The fit of the ghost scalar factor $F(p)$ to the formula $A+B\log{p}$. It suggests that
the infrared divergence of $F(p)$ is very slow. Hence $\alpha_F$ is close to zero~\cite{Boucaud:2006if}. The simulations was performed on
a $V=32^4$ lattice at $\beta=5.8$. The Landau gauge was fixed using the f.c. choice for the Gribov copies.}
\label{ghost_log_fit}
\end{figure}
%

To finish this chapter, let us summarise the lattice results. We have found 
that $\alpha_F$ is very close to 0 (see Figure~\ref{ghost_log_fit}), $\alpha_G$ is close to 1 and the widely 
used relation $2\alpha_F + \alpha_G = 0$ is not true. Going back to the possibilities
given in the Table~\ref{tabledescas} we find that the cases $2$ and $3$ are not realised.
We are left with the cases $1$ and $4$, and for the moment lattice simulations cannot
say which possibility is true. However, all numerical results are better explained 
by the possibility (\ref{case_4_vertex_regulier}) corresponding to the 
case $4$ of the Table~\ref{tabledescas} supplied with an hypothesis of the regularity of the
scalar factors entering the ghost-gluon vertex. We recall that in this case one has:
\begin{equation}
\left\{
\begin{array}{l}
\alpha_F=0 \\
\alpha_G \geq 1. \\
\text{no relation between $\alpha_F$ and $\alpha_G$ follows from the ghost SD equation}.
\end{array}
\right.
\end{equation}
%
%
%
Note that it is still in conflict with the constraints coming from the Slavnov - Taylor identity (\ref{STpredictions}).
Thus the essential question today is to understand whether $\alpha_F=0$ or not~\cite{Boucaud:2006if}. And, of course, a study on
larger lattices is necessary to perform better fits of the infrared exponents.
\chapter{Conclusions}

In this chapter we discuss the conclusions of the present dissertation. Lattice 
simulations is a great tool to study the non-perturbative effects in QCD.
The main goal of this dissertation is to exhibit how these effects influence the momentum behaviour of 
different Green functions in Landau gauge. Our hope is that the knowledge of the change in momentum behaviour
at low momenta can help in the understanding of one of the most difficult puzzles of QCD -
the mechanism of confinement.

\vspace*{1cm}
\noindent
In the chapter~\ref{chapter_UV_behaviour} the large momentum behaviour of the ghost 
and the gluon propagator of a pure Yang-Mills theory in Landau gauge is investigated.
The main parameter under study is the scale $\Lqcd$. We show that the values 
of $\Lqcd$ fitted from the ghost and the gluon propagator are consistent. 
However, the available momentum range (from $\approx 2$ to $\approx 6.5$ GeV) is
situated in the zone where non-perturbative effects cannot be neglected. So at first 
glance the agreement between $\Lqcd$ extracted from different Green function may seem 
strange. An explanation of this fact comes from the OPE analysis allowing to estimate
the influence of dominant non-perturbative power corrections. We found that these
corrections are the same in the case of the ghost and gluon propagators, that is why
the values of $\Lqcd$ are compatible. According to the OPE calculation, the value of
$\Lqcd$ that is extracted from the propagators is modified by the non-perturbative effects.
We used the fact that the equivalence of the leading power corrections implies that
their ratio is free of power corrections at the considered order. Thus the ratio 
of the ghost and gluon propagators is a quantity which is better described by perturbation
theory in the considered energy interval than the propagators themselves. 
Indeed, our analysis of the lattice data showed
that the fit of the ratio gives a smaller value for $\Lqcd$ ($\approx 270\text{ MeV}$ in 
the $\ms$ renormalisation scheme) compared to the value obtained from the separate fits
of the propagators ($\Lambda_{\ms}\approx 330\text{ MeV}$). Both fits are performed on the
same data samples. This speaks in favour of a presence of non-perturbative power corrections
in the interval $[2\text{ GeV},6.5\text{ GeV}]$, in agreement with the OPE predictions and the
results of previous investigations.

One can use the Slavnov - Taylor identities in order to find other relations between the Wilson coefficients in different Green functions.
For example we found that the dominant $\propto\frac{1}{p^2}$ power corrections 
are the same in the case of the three gluon vertex and the ghost-gluon vertex in asymmetric
kinematic configurations (with the gluon momentum set to zero). We have also shown that 
the power correction to the ghost-gluon vertex with vanishing momentum of the entering ghost 
is equal to zero (in Landau gauge). However, this is not true if the momentum of the entering ghost 
is not \emph{exactly} equal to zero. In this case the vertex has an $\propto \frac{1}{p^2}$
power correction that becomes quite important at low momenta.

As a partial conclusion we stress the attention of the reader on the fact that
lattice simulations are very successfull in describing the Green functions at large
momentum, the results are consistent with the predictions of perturbation theory,
completed with the OPE calculation of the power corrections, up to NNNLO.

\vspace*{1cm}
\noindent
In the chapter~\ref{chapter_IR} we turn to the study of very low momentum behaviour 
(of order and below $\Lqcd$) of the Green functions. One of the very interesting 
puzzles in the infrared is the problem of the Gribov ambiguity.
The lattice method has an advantage to explicitly perform the Gribov quantisation. However, the gauge
is not fixed in a unique way and there are Gribov copies on the lattice. We showed
that the probability to find a secondary Gribov copy possess the property of scaling.
This probability increases significantly when the physical volume of the lattice excesses
some critical volume (of around $(2.75/\sqrt{\sigma})^4$ in the case of $SU(2)$ gauge theory).
Our conclusion is that in order to study the non-perturbative effects one has to
work at lattices with physical volume larger then the critical one that we found.

Our first step in the study of low-momentum behaviour of Green functions is to
check that lattice simulations can give reliable results in the infrared. 
For this purpose we verified (numerically, see Figure~\ref{SDcheck}) that lattice Green functions
satisfy the complete ghost Schwinger - Dyson equation (\ref{SDghost}) for all considered momenta. These tests allow 
us to conclude that numerical simulation on the lattice give relevant results not only 
in the ultraviolet domain, but also in the infrared one.

The quantitative parameters we are interested in are the infrared exponents $\alpha_F$ 
and $\alpha_G$ describing the power-law deviation (this is a crudest approximation) 
from free propagators (ghost and gluon respectively) in the deep infrared. 
Our analysis of the Slavnov-Taylor identity relating the three-gluon vertex, 
the ghost-gluon vertex and the propagators showed that the power-law infrared 
divergence of the ghost propagator is unchanged or enhanced in the infrared
(compared to the free case), and that the gluon propagator must diverge in the infrared.
The latter limit is in conflict with today's lattice results yelding a finite non-zero gluon propagator at
zero momentum, and with most present analytical estimations that we
quote in the section~\ref{Review_of_today_s_analytical_results}, that
support a vanishing gluon propagator in the infrared.

Another analytical relation imposing constraints on the infrared exponents is the ghost 
Schwinger-Dyson equation. We revisited the commonly accepted relation (\ref{R})
between these exponents saying that $2\alpha_F + \alpha_G=0$.
According to our analysis this relation is true only if the ghost-gluon vertex 
contains no singularity, in none of the scalar functions defining the vertex. 
Our numerical studies showed that the relation in question 
is not valid, because $F^2 G$ is infrared suppressed, and hence $2\alpha_F + \alpha_G > 0$.
This statement is supported by the fact that lattice propagators do not match the reduced
Schwinger - Dyson equation (see Figure~\ref{figure_SDcheckH1}), whereas the complete one is perfectly
verified, Figure~\ref{SDcheck}. There are even more reasons to think that the relation (\ref{R}) is not true.
First, we have seen that the non-perturbative ghost-gluon vertex contains
singular contributions from the $\langle A^2 \rangle$ condensate for most kinematic
configurations. However, it is difficult to estimate the role of such contributions at very low momenta.
Second, we have seen that the infrared behaviour of the ghost and gluon propagator seem to
vary differently with the choice of the Gribov copy (bc or fc). We have seen that
the form of the complete ghost Schwinger - Dyson equation does not depend on the choice
of the Gribov copy. If $\alpha_F\neq 0$ and it depends on the choice of the copy,
it is impossible to have an exact relation between $\alpha_F$ and $\alpha_G$ alone, and
to satisfy the condition of independence of the choice of the copy.

The direct fit of the propagators in the infrared supports the prediction that
the infrared behaviour of the ghost propagator is enhanced in the infrared. But
this enhancement is very slight. The gluon propagator is found to be infrared finite. 
This last result is in conflict with the limit found from the analysis 
of the Slavnov - Taylor identity. For the moment
we have no explanation regarding this disagreement. 

Summarising the numerical results, we found that the gluon propagator is finite 
in the infrared ($\alpha_G \approx 1$), that the infrared divergence of the ghost 
propagator is almost the same as in the free case ($\alpha_F \approx 0$) and that the commonly accepted relation 
$2\alpha_F + \alpha_G = 0$ is not true. Going back to the analysis of the ghost 
Schwinger - Dyson equation (see Table~\ref{tabledescas}), two solutions are possible:
\begin{enumerate}
\item The infrared exponent of the ghost propagator $\alpha_F$ is \emph{strictly} equal to zero, i.e.
the power-law infrared dependence is the same as at large momenta. This implies that \emph{no}
relation between $\alpha_F$ and $\alpha_G$ follows from the ghost Schwinger - Dyson equation. 
If we now suppose that the ghost-gluon vertex contains no (infrared) singular components then
all our lattice results are perfectly described.

\item The infrared exponent of the ghost propagator $\alpha_F \neq 0$, then there is a relation
between $\alpha_F$, $\alpha_G$ and $\alpha_\Gamma$ following from the ghost Schwinger - Dyson equation.
The fact that the relation $2\alpha_F + \alpha_G = 0$ is not verified on the lattice suggests that there is
a singularity in one of the scalar factors defining the ghost-gluon vertex i.e. $\alpha_\Gamma \neq 0$.
\end{enumerate}
The today's lattice results speak in favour of the first possibility, but calculations at larger lattices
are necessary in order to conclude.

\vspace*{1cm}

\addcontentsline{toc}{chapter}{References}

\begin{thebibliography}{10}

\bibitem{Gribov:1977wm}
V.~N. Gribov.
\newblock Quantization of non-abelian gauge theories.
\newblock {\em Nucl. Phys.}, B139:1, 1978.

\bibitem{Maas:2005qt}
Axel Maas.
\newblock On the spectrum of the Faddeev-Popov operator in topological
  background fields.
\newblock 2005, hep-th/0511307.

\bibitem{Franke_Semenov_Tyan_Shanskii}
M.~Semenov-Tyan-Shanskii and V.~Franke.
\newblock All gauge orbits and some Gribov copies encompassed by the Gribov
  horizon.
\newblock {\em Zap. Nauch. Sem. Leningrad. Otdeleniya Matematicheskogo
  Instituta im. V. A. Steklova, AN SSSR}, vol. 120:159, 1982.
\newblock (English translation:New York, Plenum Press 1986).

\bibitem{vanBaal:1991zw}
Pierre van Baal.
\newblock More (thoughts on) Gribov copies.
\newblock {\em Nucl. Phys.}, B369:259--275, 1992.

\bibitem{Zwanziger:2003cf}
Daniel Zwanziger.
\newblock Non-perturbative Faddeev-Popov formula and infrared limit of QCD.
\newblock {\em Phys. Rev.}, D69:016002, 2004, hep-ph/0303028.

\bibitem{Alkofer:2000wg}
Reinhard Alkofer and Lorenz von Smekal.
\newblock The infrared behavior of QCD Green's functions: Confinement,
  dynamical symmetry breaking, and hadrons as relativistic bound states.
\newblock {\em Phys. Rept.}, 353:281, 2001, hep-ph/0007355.

\bibitem{Slavnov:1972fg}
A.~A. Slavnov.
\newblock Ward identities in gauge theories.
\newblock {\em Theor. Math. Phys.}, 10:99--107, 1972.

\bibitem{Taylor:1971ff}
J.~C. Taylor.
\newblock Ward identities and charge renormalization of the Yang-Mills field.
\newblock {\em Nucl. Phys.}, B33:436--444, 1971.

\bibitem{Taylor:1976ru}
J.~C. Taylor.
\newblock Gauge theories of weak interactions.
\newblock Cambridge 1976, 167p.

\bibitem{Bagnuls:2000ae}
C.~Bagnuls and C.~Bervillier.
\newblock Exact renormalization group equations: An introductory review.
\newblock {\em Phys. Rept.}, 348:91, 2001, hep-th/0002034.

\bibitem{Ellwanger:1996wy}
Ulrich Ellwanger, Manfred Hirsch, and Axel Weber.
\newblock The heavy quark potential from Wilson's exact renormalization group.
\newblock {\em Eur. Phys. J.}, C1:563--578, 1998, hep-ph/9606468.

\bibitem{Wilson:1974sk}
Kenneth~G. Wilson.
\newblock Confinement of quarks.
\newblock {\em Phys. Rev.}, D10:2445--2459, 1974.

\bibitem{Montvay:1994cy}
I.~Montvay and G.~Munster.
\newblock Quantum fields on a lattice.
\newblock Cambridge, UK: Univ. Pr. (1994) 491 p. (Cambridge monographs on
  mathematical physics).

\bibitem{Smit:2002ug}
J.~Smit.
\newblock Introduction to Quantum Fields on a lattice: A robust mate.
\newblock {\em Cambridge Lect. Notes Phys.}, 15:1--271, 2002.

\bibitem{Osterwalder:1977pc}
K.~Osterwalder and E.~Seiler.
\newblock Gauge field theories on the lattice.
\newblock {\em Ann. Phys.}, 110:440, 1978.

\bibitem{Itzykson:1980fz}
C.~Itzykson, Michael~E. Peskin, and J.~B. Zuber.
\newblock Roughening of Wilson's surface.
\newblock {\em Phys. Lett.}, B95:259, 1980.

\bibitem{Adler:1987ce}
Stephen~L. Adler.
\newblock Overrelaxation algorithms for lattice field theories.
\newblock {\em Phys. Rev.}, D37:458, 1988.

\bibitem{Lokhov:2005ra}
A.~Y. Lokhov, O.~Pene, and C.~Roiesnel.
\newblock Scaling properties of the probability distribution of lattice Gribov
  copies.
\newblock 2005, hep-lat/0511049.

\bibitem{Boucaud:1998bq}
P.~Boucaud, J.~P. Leroy, J.~Micheli, O.~Pene, and C.~Roiesnel.
\newblock Lattice calculation of alpha(s) in momentum scheme.
\newblock {\em JHEP}, 10:017, 1998, hep-ph/9810322.

\bibitem{Alles:1996ka}
B.~Alles et~al.
\newblock $\alpha_s$ from the nonperturbatively renormalised lattice three
  gluon vertex.
\newblock {\em Nucl. Phys.}, B502:325--342, 1997, hep-lat/9605033.

\bibitem{Boucaud:2005np}
Ph. Boucaud et~al.
\newblock Large momentum behavior of the ghost propagator in SU(3) lattice
  gauge theory.
\newblock {\em Phys. Rev.}, D72:114503, 2005, hep-lat/0506031.

\bibitem{Saad}
Yousef Saad.
\newblock Iterative methods for sparse linear systems.
\newblock PWS Publishing, New York, 1996.

\bibitem{Jackknife1}
B.~Efron.
\newblock Bootstrap methods: Another look at the jackknife.
\newblock {\em The Annals of Statistics}, 7:1--26, 1979.

\bibitem{Jackknife2}
A.~C. Davison and D.~V. Hinkley.
\newblock Bootstrap methods and their applications.
\newblock {\em Monographs on Statistics and Applied Probability 57, Chapman and
  Hall/CRC}, (1994).

\bibitem{Sternbeck:2005vs}
A.~Sternbeck, E.~M. Ilgenfritz, and M.~Mueller-Preussker.
\newblock Spectral properties of the Landau gauge Faddeev-Popov operator in
  lattice gluodynamics.
\newblock{\em Phys.\ Rev.}, D73 (2006) 014502,  hep-lat/0510109.

\bibitem{Bloch:2003sk}
Jacques C.~R. Bloch, Attilio Cucchieri, Kurt Langfeld, and Tereza Mendes.
\newblock Propagators and running coupling from SU(2) lattice gauge theory.
\newblock {\em Nucl. Phys.}, B687:76--100, 2004, hep-lat/0312036.

\bibitem{Fingberg:1992ju}
J.~Fingberg, Urs~M. Heller, and F.~Karsch.
\newblock Scaling and asymptotic scaling in the SU(2) gauge theory.
\newblock {\em Nucl. Phys.}, B392:493--517, 1993, hep-lat/9208012.

\bibitem{Cucchieri:1997dx}
Attilio Cucchieri.
\newblock Gribov copies in the Minimal Landau gauge: The influence on gluon and
  ghost propagators.
\newblock {\em Nucl. Phys.}, B508:353--370, 1997, hep-lat/9705005.

\bibitem{Bakeev:2003rr}
T.~D. Bakeev, Ernst-Michael Ilgenfritz, V.~K. Mitrjushkin, and
  M.~Mueller-Preussker.
\newblock On practical problems to compute the ghost propagator in SU(2)
  lattice gauge theory.
\newblock {\em Phys. Rev.}, D69:074507, 2004, hep-lat/0311041.

\bibitem{Sternbeck:2004qk}
A.~Sternbeck, E.~M. Ilgenfritz, M.~Muller-Preussker, and A.~Schiller.
\newblock The influence of Gribov copies on the gluon and ghost propagator.
\newblock {\em AIP Conf. Proc.}, 756:284--286, 2005, hep-lat/0412011.

\bibitem{Sternbeck:2004xr}
A.~Sternbeck, E.~M. Ilgenfritz, M.~Muller-Preussker, and A.~Schiller.
\newblock The gluon and ghost propagator and the influence of Gribov copies.
\newblock {\em Nucl. Phys. Proc. Suppl.}, 140:653--655, 2005, hep-lat/0409125.

\bibitem{Silva:2004bv}
P.~J. Silva and O.~Oliveira.
\newblock Gribov copies, lattice QCD and the gluon propagator.
\newblock {\em Nucl. Phys.}, B690:177--198, 2004, hep-lat/0403026.

\bibitem{Sternbeck:2005tk}
A.~Sternbeck, E.~M. Ilgenfritz, M.~Mueller-Preussker, and A.~Schiller.
\newblock Towards the infrared limit in SU(3) Landau gauge lattice
  gluodynamics.
\newblock {\em Phys. Rev.}, D72:014507, 2005, hep-lat/0506007.

\bibitem{Chetyrkin:2000dq}
K.~G. Chetyrkin and A.~Retey.
\newblock Three-loop three-linear vertices and four-loop mom beta functions in
  massless QCD.
\newblock 2000, hep-ph/0007088.

\bibitem{Chetyrkin:2004mf}
K.~G. Chetyrkin.
\newblock Four-loop renormalization of QCD: Full set of renormalization
  constants and anomalous dimensions.
\newblock {\em Nucl. Phys.}, B710:499--510, 2005, hep-ph/0405193.

\bibitem{vanRitbergen:1997va}
T.~van Ritbergen, J.~A.~M. Vermaseren, and S.~A. Larin.
\newblock The four-loop beta function in Quantum chromodynamics.
\newblock {\em Phys. Lett.}, B400:379--384, 1997, hep-ph/9701390.

\bibitem{Boucaud:2000ey}
P.~Boucaud et~al.
\newblock Lattice calculation of $1/p^2$ corrections to $\alpha_S$ and of
  $\Lqcd$ in the $\widetilde{MOM}$ scheme.
\newblock {\em JHEP}, 04:006, 2000, hep-ph/0003020.

\bibitem{Boucaud:2001qz}
P.~Boucaud et~al.
\newblock Preliminary calculation of $\alpha_S$ from Green functions with
  dynamical quarks.
\newblock {\em JHEP}, 01:046, 2002, hep-ph/0107278.

\bibitem{Wilson:1969zs}
Kenneth~G. Wilson.
\newblock Nonlagrangian models of current algebra.
\newblock {\em Phys. Rev.}, 179:1499--1512, 1969.

\bibitem{Ioffe:2002ee}
B.~L. Ioffe.
\newblock Condensates in Quantum chromodynamics.
\newblock {\em Phys. Atom. Nucl.}, 66:30--43, 2003, hep-ph/0207191.

\bibitem{Lavelle:1992yh}
Martin Lavelle and Michael Oleszczuk.
\newblock The Operator Product Expansion of the QCD propagators.
\newblock {\em Mod. Phys. Lett.}, A7:3617--3630, 1992.

\bibitem{Ahlbach:1991ws}
Jorg Ahlbach, Martin Lavelle, Martin Schaden, and Andreas Streibl.
\newblock Propagators and four-dimensional condensates in pure QCD.
\newblock {\em Phys. Lett.}, B275:124--128, 1992.

\bibitem{Boucaud:2000nd}
P.~Boucaud et~al.
\newblock Consistent OPE description of gluon two point and three point green
  function?
\newblock {\em Phys. Lett.}, B493:315--324, 2000, hep-ph/0008043.

\bibitem{Boucaud:2001st}
Ph. Boucaud et~al.
\newblock Testing Landau gauge ope on the lattice with a condensate.
\newblock {\em Phys. Rev.}, D63:114003, 2001, hep-ph/0101302.

\bibitem{Boucaud:2002jt}
P.~Boucaud et~al.
\newblock A transparent expression of the $\langle A^2 \rangle$-condensate's renormalization.
\newblock {\em Phys. Rev.}, D67:074027, 2003, hep-ph/0208008.

\bibitem{Bali:1992ru}
Gunnar~S. Bali and Klaus Schilling.
\newblock Running coupling and the lambda parameter from SU(3) lattice
  simulations.
\newblock {\em Phys. Rev.}, D47:661--672, 1993, hep-lat/9208028.

\bibitem{Celmaster:1979km}
William Celmaster and Richard~J. Gonsalves.
\newblock The renormalization prescription dependence of the QCD coupling
  constant.
\newblock {\em Phys. Rev.}, D20:1420, 1979.

\bibitem{Becirevic:1999hj}
D.~Becirevic et~al.
\newblock Asymptotic scaling of the gluon propagator on the lattice.
\newblock {\em Phys. Rev.}, D61:114508, 2000, hep-ph/9910204.

\bibitem{Boucaud:2005xn}
Ph. Boucaud et~al.
\newblock Non-perturbative power corrections to ghost and gluon propagators.
\newblock {\em JHEP}, 01:037, 2006, hep-lat/0507005.

\bibitem{Zwanziger:1990by}
Daniel Zwanziger.
\newblock Vanishing color magnetization in lattice Landau and Coulomb gauges.
\newblock {\em Phys. Lett.}, B257:168--172, 1991.

\bibitem{Zwanziger:1991gz}
D.~Zwanziger.
\newblock Vanishing of zero momentum lattice gluon propagator and color
  confinement.
\newblock {\em Nucl. Phys.}, B364:127--161, 1991.

\bibitem{Bloch:2003yu}
J.~C.~R. Bloch.
\newblock Two-loop improved truncation of the ghost-gluon Dyson-Schwinger
  equations: Multiplicatively renormalizable propagators and nonperturbative
  running coupling.
\newblock {\em Few Body Syst.}, 33:111--152, 2003, hep-ph/0303125.

\bibitem{vonSmekal:1997is}
Lorenz von Smekal, Andreas Hauck, and Reinhard Alkofer.
\newblock A solution to coupled Dyson-Schwinger equations for gluons and ghosts
  in Landau gauge.
\newblock {\em Ann. Phys.}, 267:1, 1998, hep-ph/9707327.

\bibitem{Zwanziger:2001kw}
Daniel Zwanziger.
\newblock Non-perturbative Landau gauge and infrared critical exponents in QCD.
\newblock {\em Phys. Rev.}, D65:094039, 2002, hep-th/0109224.

\bibitem{Aguilar:2004sw}
A.~C. Aguilar and A.~A. Natale.
\newblock A dynamical gluon mass solution in a coupled system of the
  Schwinger-Dyson equations.
\newblock {\em JHEP}, 08:057, 2004, hep-ph/0408254.

\bibitem{Kato:2004ry}
Junya Kato.
\newblock Infrared non-perturbative propagators of gluon and ghost via exact
  renormalization group.
\newblock 2004, hep-th/0401068.

\bibitem{Pawlowski:2003hq}
Jan~M. Pawlowski, Daniel~F. Litim, Sergei Nedelko, and Lorenz von Smekal.
\newblock Infrared behaviour and fixed points in Landau gauge QCD.
\newblock {\em Phys. Rev. Lett.}, 93:152002, 2004, hep-th/0312324.

\bibitem{Fischer:2004uk}
Christian~S. Fischer and Holger Gies.
\newblock Renormalization flow of Yang-Mills propagators.
\newblock {\em JHEP}, 10:048, 2004, hep-ph/0408089.

\bibitem{Ball:1980ax}
James~S. Ball and Ting-Wai Chiu.
\newblock Analytic properties of the vertex function in gauge theories. 2.
\newblock {\em Phys. Rev.}, D22:2550, 1980.

\bibitem{Boucaud:2005ce}
Ph. Boucaud et~al.
\newblock The infrared behaviour of the pure Yang-Mills Green functions.
\newblock 2005, hep-ph/0507104.

\bibitem{Lepage:1992xa}
G.~Peter Lepage and Paul~B. Mackenzie.
\newblock On the viability of lattice perturbation theory.
\newblock {\em Phys. Rev.}, D48:2250--2264, 1993, hep-lat/9209022.

\bibitem{Boucaud:2006if}
J.P. Leroy A. Le Yaouanc A.Y. Lokhov J. Micheli O. Pene J. Rodriguez-Quintero
  Ph.~Boucaud, Th.~Bruntjen.
\newblock Is the QCD ghost dressing function finite at zero momentum?
\newblock {\em JHEP}{ 0606}, 001 (2006), hep-ph/0604056.

\bibitem{Bonnet:2001uh}
Frederic D.~R. Bonnet, Patrick~O. Bowman, Derek~B. Leinweber, Anthony~G.
  Williams, and James~M. Zanotti.
\newblock Infinite volume and continuum limits of the Landau-gauge gluon
  propagator.
\newblock {\em Phys. Rev.}, D64:034501, 2001, hep-lat/0101013.

\bibitem{Oliveira:2005hg}
Orlando Oliveira and Paulo~J. Silva.
\newblock Finite volume effects in the gluon propagator.
\newblock {\em PoS}, LAT2005:287, 2005, hep-lat/0509037.

\bibitem{Boucaud:2006pc}
Ph. Boucaud et~al.
\newblock Short comment about the lattice gluon propagator at vanishing
  momentum.
\newblock 2006, hep-lat/0602006.

\bibitem{Giusti:2000yc}
L.~Giusti, M.~L. Paciello, S.~Petrarca, C.~Rebbi, and B.~Taglienti.
\newblock Results on the gluon propagator in lattice covariant gauges.
\newblock {\em Nucl. Phys. Proc. Suppl.}, 94:805--808, 2001, hep-lat/0010080.

\bibitem{Giusti:2001kr}
L.~Giusti, M.~L. Paciello, S.~Petrarca, B.~Taglienti, and N.~Tantalo.
\newblock Quark and gluon propagators in covariant gauges.
\newblock {\em Nucl. Phys. Proc. Suppl.}, 106:995--997, 2002, hep-lat/0110040.

\end{thebibliography}
\end{document}